\documentclass[12pt]{JHEP3}

\usepackage{amsmath,amssymb,amscd,amsfonts,xspace,mathrsfs,amsthm,dsfont}
\usepackage{bbm}
\usepackage{braket}
\usepackage{caption}
\usepackage{color}
\let\normalcolor\relax
    \definecolor{darkgreen}{rgb}{0,0.5,0}
    \definecolor{darkred}{rgb}{0.5,0,0}
    \definecolor{darkblue}{rgb}{0,0,0.6}
    \definecolor{purple}{rgb}{0.4,.2,0.7}

\usepackage[mathcal]{eucal}
\usepackage{float}
\usepackage{graphicx}
\usepackage{mathtools}
\usepackage{pdfsync}
\usepackage{slashed}
\usepackage[normalem]{ulem}
\usepackage{upgreek}
\usepackage{url}
\allowdisplaybreaks

\usepackage[ragged]{footmisc}
\def\twofig#1#2#3#4#5#6#7#8{
\begin{figure}[t]
 \centerline{\includegraphics[width=#3]{#2}\hspace{#4}\includegraphics[width=#6]{#5}}
 \vspace{#7}
  \caption{#1 \label{#8}}
 \end{figure}
}

\newcommand{\ben}{\begin{eqnarray}\displaystyle}
\newcommand{\een}{\end{eqnarray}}
\def\bea{\begin{eqnarray}}
\def\eea{\end{eqnarray}}
\def\be{\begin{equation}}
\def\ee{\end{equation}}
\def\ba{\begin{align}}
\def\ea{\end{align}}
\def\bse{\begin{subequations}}
\def\ese{\end{subequations}}

\numberwithin{equation}{section}

\def\det{\,{\rm det}\, }

\def\sign{{\rm sgn}}

\def\Im{\,{\rm Im}\,}
\def\Re{\,{\rm Re}\,}

\newcommand{\p}{\partial}

\renewcommand{\tilde}{\widetilde}

\def\({\left(}
\def\){\right)}
\def\[{\left[}
\def\]{\right]}
\def\<{\left\langle}
\def\>{\right\rangle}
\def\hf{{1\over 2}}

\newcommand{\cA}{\mathcal{A}}

\newcommand{\cD}{\mathcal{D}}

\newcommand{\cF}{\mathcal{F}}

\newcommand{\cI}{\mathcal{I}}
\newcommand{\cJ}{\mathcal{J}}

\newcommand{\cO}{\mathcal{O}}

\newcommand{\cS}{\mathcal{S}}

\newcommand{\cU}{\mathcal{U}}
\newcommand{\cV}{\mathcal{V}}

\newcommand{\IR}{\mathds{R}}
\newcommand{\IC}{\mathds{C}}
\newcommand{\IZ}{\mathds{Z}}

\newcommand{\IN}{\mathds{N}}

\newcommand{\Zint}{\mathds{Z}}

\newcommand\scX{\mathscr{X}}

\newcommand{\nn}{\nonumber}
\newcommand{\non}{\nonumber}
\newcommand{\refb}{\eqref}

\newcommand\eps{\epsilon}
\newcommand\vp{\varphi}
\newcommand\vpz{\varphi^{(0)}}

\def\hmu{\hat\mu}
\def\hym{\hat y_-}

\def\tX{\tilde X}
\def\tR{\tilde R}

\def\tcV{\tilde \cV}
\def\tcJ{\tilde\cJ}

\def\Xp{X_{+}}
\def\Xm{X_{-}}
\def\Xpm{X_{\pm}}
\def\Xmp{X_{\mp}}

\def\xp{x_{+}}
\def\xm{x_{-}}
\def\xpm{x_{\pm}}
\def\xmp{x_{\mp}}

\def\pxp{x'_{+}}
\def\pxm{x'_{-}}
\def\pxpm{x'_{\pm}}

\def\yp{y_{+}}
\def\ym{y_{-}}
\def\ypm{y_{\pm}}

\def\sind{\nu}
\def\bind{\bar\nu}

\def\thpm{\theta^{\pm}}

\def\tauip#1{\tau_{+,\sind}^{(#1)}}
\def\tauim#1{\tau_{-,\sind}^{(#1)}}
\def\tauipm#1{\tau_{\pm,\sind}^{(#1)}}

\def\thpm{\theta^{0}}
\def\thpm{\theta^{\sind}}

\def\Pe{\Psi^{_{E}}}

\def\pse{ \psi^{_{E}} }

\def\Pep{\Pe_+}
\def\Pem{\Pe_-}
\def\Pepm{\Pe_\pm}

\def\psep{\pse_+}
\def\psem{\pse_-}
\def\psepm{\pse_\pm}

\def\lamcr{\lambda^{\rm cr}}
\def\lam{\tilde\lambda_k}
\def\lamm{\tilde\lambda_{2k}}

\def\mU{\mathfrak{U}}
\def\ZZZ{{\mathbb Z}}
\def\kmax{k_{\rm max}}
\def\str{s}
\def\repla{\tilde\lambda}

\def\upert{u^{(\rm p)}}
\def\wpert{w^{(\rm p)}}
\def\uinst{u^{(\rm in)}}
\def\winst{w^{(\rm in)}}
\def\uinstd#1{\smash{u^{(\rm in)}_{#1}}}
\def\winstd#1{\smash{w^{(\rm in)}_{#1}}}
\def\vin{v^{({\rm in})}}

\def\Scz{\boldsymbol{S}}
\def\Acz{\boldsymbol{A}}
\def\Fcz{\boldsymbol{\cF}}
\def\mucz{\boldsymbol{\mu}}
\def\rcz{\boldsymbol{r}}

\def\km{{k_{\max}}}
\def\bS{{\bf S}}
\def\hlm{\hat\lambda_0}

\title{Instantons in sine-Liouville theory}

\author{Sergei Alexandrov$^1$, Raghu Mahajan$^2$, Ashoke Sen$^3$
\\
$^1$ {\it
Laboratoire Charles Coulomb (L2C), Universit\'e de Montpellier,
CNRS, F-34095, Montpellier, France}\\

$^2$ {\it Department of Physics, Stanford University, Stanford, CA 94305-4060, USA}\\

$^3$ {\it International Centre for Theoretical Sciences - TIFR  \\ Bengaluru - 560089, India}\\

\vspace*{2mm} {\tt e-mail:
\email{sergey.alexandrov@umontpellier.fr},
\email{raghumahajan@stanford.edu},
\email{ashoke.sen@icts.res.in}
}

\vspace*{-3mm}

}

\abstract{
We compute instanton corrections to the partition function of sine-Liouville (SL) theory, which provides a worldsheet description of two-dimensional string theory in a non-trivial tachyon background.
We derive these corrections using a matrix model formulation based on a chiral representation of matrix quantum mechanics and using string theory methods.
In both cases we restrict to the leading and subleading orders in the string coupling expansion.
Then the CFT technique is used to compute two orders of the expansion in the SL perturbation parameter $\lambda$, while the matrix model gives results which are non-perturbative in $\lambda$.
The matrix model results perfectly match those of string theory in the small $\lambda$ expansion.
We also generalize our findings to the case of perturbation by several tachyon vertex operators carrying different momenta, and obtain interesting analytic predictions for the disk two-point and annulus one-point functions with ZZ boundary condition.
}

\begin{document}

\baselineskip 16.pt
\setlength{\parskip}{0.15cm}

\section{Introduction}
\label{sec-intro}

Bosonic two-dimensional string theory is a fruitful playground to test various ideas and techniques of critical string theory; see \cite{Klebanov:1991qa,Ginsparg:1993is,Jevicki:1993qn,Polchinski:1994mb} for reviews.
In its spectrum, it contains only one field-theoretic degree of freedom --- a massless scalar field, which is commonly called ``tachyon" because this is the same field that has negative mass-squared in higher dimensions.
In the simplest linear dilaton background, the background value of this tachyon field is taken to be time independent and exponentially dependent on the spatial coordinate.
In this case the theory is well understood, being an exactly solvable model.
On the worldsheet, it is described by Liouville theory at the central charge $c=25$ coupled to $c=1$ matter\footnote{Throughout this paper we set $\alpha'=1$ and we use the convention that $d^2z = dx dy$ in terms of Cartesian coordinates $x,y$. We have omitted the linear term in $\phi$ from the action since we have specialized to a patch with flat background metric. Of course, the worldsheet CFT also contains the $bc$ ghosts.}
\be
S_0=\frac{1}{\pi}\int d^2 z\, \Bigl(\p \phi \bar\p\phi +\pi \mu e^{2\phi}
+\p X \bar\p X\Bigr),
\label{CFTc=1}
\ee
where $\mu$ is the Liouville coupling (also known as 2d cosmological constant) which is inversely proportional to the string coupling $g_\str$.

It is much more challenging however to describe 2d strings in non-trivial backgrounds which can be obtained, for example, by introducing the so called ``tachyon condensate", i.e. choosing a background tachyon field different from the Liouville potential.
There are two main reasons to investigate such backgrounds:
\begin{itemize}
\item
In Lorentzian signature, such a tachyon condensate typically implies a time-dependent background
\cite{Alexandrov:2003uh,Karczmarek:2004ph,Mukhopadhyay:2004ff}
(see also, e.g., \cite{Karczmarek:2004yc,Das:2004aq,Karczmarek:2007ag}).
Understanding time-dependence in string theory is a notoriously difficult problem and getting a handle on it in the simplified setting of two-dimensional theory can hopefully bring new insights on how to deal with time-dependent backgrounds in more complicated setups. Other approaches to constructing time-dependent backgrounds in two-dimensional
string theory can be found in \cite{Rodriguez:2023kkl,Rodriguez:2023wun,Collier:2023cyw}

\item
In Euclidean signature, the ``time" direction can be compactified on a circle of radius $R$.
Then after T-duality, the tachyon condensate turns to a condensate of string winding modes.
If this condensate is chosen in its simplest form corresponding to the so-called sine-Liouville (SL) theory obtained by adding
\be
\lambda \int d^2 z \, \tcV(z,\bar z),
\qquad
\tcV\sim \cos(\tR\tX)\, e^{(2-\tR)\phi}
\label{SLwinding}
\ee
to the worldsheet action \eqref{CFTc=1}, the resulting model was conjectured to be dual to 2d string theory in a black hole background \cite{FZZ} that has the famous ``cigar" geometry \cite{Callan:1992rs} and is described by $[SL(2,\IR)]_k/U(1)$ coset WZW model \cite{Witten:1991yr, Mandal:1991tz}.\footnote{More precisely, the duality is supposed to hold at $\mu=0$ and $\tR=\sqrt{k}$.
The string theory condition $c=26$ further fixes these parameters to $\tR=3/2$ and $k=9/4$.}
Here $\tR=1/R$ is the dual radius and $\tX=X_R-X_L$ is the dual field.
(The conjecture was proven later in \cite{Hikida:2008pe}.)
Thus, analyzing non-trivial tachyon backgrounds, one can try to access various important issues of the black hole physics.
\end{itemize}

However, the CFT technique is not powerful enough to solve the theory.
Typically, it allows to compute only correlation functions on sphere and torus with a very few insertions of vertex operators.
Fortunately, 2d string theory is known to have a dual description in terms of Matrix Quantum Mechanics (MQM) \cite{Brezin:1989ss,Ginsparg:1990as,Gross:1990ay}.
This matrix model not only provides an exact solution to 2d string theory in the simplest background \eqref{CFTc=1}, but also reveals an integrable structure underlying a large class of backgrounds with a non-trivial tachyon condensate, including in particular the SL theory relevant for the 2d black hole
(see \cite{Alexandrov:2003ut} for review).

More precisely, if we are not interested in the winding modes of compactified Euclidean theory, MQM can be reduced to a system of free fermions moving in the inverse oscillator potential $V(x)=-\hf\, x^2$.
Different backgrounds of string theory then correspond to different states of this system \cite{Das:1990kaa}.
In particular, the background described in \eqref{CFTc=1} is realized as the vacuum state where the fermions fill all energy levels up to $E=-\mu$ on one side of the potential, with $\mu$ being the same parameter as in \eqref{CFTc=1}.
How to generate states corresponding to more general backgrounds was shown in \cite{Alexandrov:2002fh}.
The construction heavily relies on a chiral or light-cone representation of MQM where the scattering of fermions off the potential turns out to be encoded in the Fourier transform relating the left and right representations.
Using this formalism, it was shown that the tachyon backgrounds described by the worldsheet action
\be
S=S_0+\sum_n \lambda_{n} \int d^2 z \, \cV_{n/R}(z,\bar z),
\qquad
\cV_\omega \sim \cos(\omega X)\, e^{\(2-\omega\)\phi},
\label{genCFT}
\ee
are all governed by the Toda integrable hierarchy, consistently with the previous findings of \cite{Dijkgraaf:1992hk, Imbimbo:1995np}, which allows, at least in principle, to compute the corresponding partition and correlation functions order by order in perturbation theory and exactly in the parameters $\lambda_n$.

Remarkably, it turns out to be possible also to go beyond the perturbative description.
Both CFT and MQM formulations of 2d string theory have instanton effects which go like
\be
A \, e^{-\bS_{\rm inst}},
\qquad \bS_{\rm inst}\sim g_\str^{-1}.
\label{nonpert}
\ee
In CFT, they arise due to the existence of the so-called ZZ boundary conditions in Liouville theory \cite{Zamolodchikov:2001ah}.
Then $\bS_{\rm inst}$ is given by the disk amplitude representing an open string ending on a ZZ-brane, while $A$ is related to the annulus amplitude with the same boundary conditions.
On the other hand, in MQM the instanton effects are related to the tunneling of fermions through the inverse oscillator potential.

Starting from 2003, these non-perturbative effects have been extensively studied (see e.g. \cite{McGreevy:2003kb,Martinec:2003ka,Klebanov:2003km,McGreevy:2003ep,Alexandrov:2003nn,Alexandrov:2003un,Alexandrov:2004ks}) and it was demonstrated that, at the level of the instanton action, the two formulations of 2d string theory perfectly agree.
In particular, in MQM the quantity $\bS_{\rm inst}$ has been computed for the SL theory and its first terms in the small $\lambda$ expansion have been shown to reproduce the disk correlation functions in the worldsheet CFT with $(1,n)$ ZZ boundary conditions \cite{Alexandrov:2003nn,Alexandrov:2003un}.\footnote{For generic central charge, there is a two-parameter family of ZZ boundary conditions labelled by $(m,n)\in\IN^2$ \cite{Zamolodchikov:2001ah}.
However, it was argued in \cite{Alexandrov:2003un,Alexandrov:2004ks}, and recently confirmed by the analysis of \cite{Balthazar:2019ypi}, that only $(1,n)$ branes survive in the $c=1$ limit.}

At the subleading order in the $g_\str$-expansion, in particular for the quantity $A$ above, the situation however was more complicated.
On one hand, the CFT annulus amplitude is formally divergent which blocked the computation of the factor $A$ from the string theory side until recent works \cite{Balthazar:2019ypi, Balthazar:2019rnh, Sen:2019qqg, Sen:2021qdk}.
On the other hand, the factor $A$ was computed in the dual MQM in \cite{Alexandrov:2004cg}.
Moreover, this was done by two methods: first, by solving a linearized Toda equation and, second, by analyzing quasi-classical fermion wave functions in the chiral representation.
Both methods produced the same very non-trivial function of the SL parameter $\lambda$, see \eqref{resAnSL}.
However, the first method is not able to fix the overall normalization constant.
The second method can do it, but led to a strange result $A\sim 1/\log\Lambda$ where $\Lambda$ is a MQM cut-off supposed to be taken to infinity.
Such a result cannot be physical which implies that something must have been missed in the analysis of \cite{Alexandrov:2004cg}.
Another puzzle related to the function \eqref{resAnSL} found in \cite{Alexandrov:2004cg} is that at small $\lambda$ it behaves as $\lambda^{-1/2}$ and thus diverges, while one could expect that it should reproduce the well-known instanton corrections to the partition function in the linear dilaton background.

In this paper we return to these problems with the new insights coming from recent progress in calculation of D-instanton amplitudes.
Using string field theory, it was understood how to properly regularize the divergences appearing in the annulus amplitudes with D-brane boundary conditions evaluated in the naive CFT approach \cite{Sen:2020cef,Sen:2020eck,Sen:2021qdk}.
The recipe resulting from this understanding has then been successfully applied in ten-dimensional type IIB string theory \cite{Sen:2021tpp,Sen:2021jbr}, in Calabi-Yau compactifications of critical strings \cite{Alexandrov:2021shf,Alexandrov:2021dyl,Alexandrov:2022mmy}, in $c<1$ non-critical models \cite{Eniceicu:2022nay,Eniceicu:2022dru,Eniceicu:2022xvk}, and for S-matrix of 2d string theory in the background described in \eqref{CFTc=1}
\cite{Sen:2020eck,Sen:2021qdk,Chakravarty:2022cgj}.
Here we apply it to 2d string theory in non-trivial tachyon backgrounds.

Before doing this, however, we revise the MQM analysis that was done in \cite{Alexandrov:2004cg} and led to a cut-off dependent normalization factor.
It was based on the observation \cite{Alexandrov:2004ks} that the instanton contributions to the partition function of the perturbed theory \eqref{genCFT}
correspond to the so-called double points of the MQM complex curve.
While this is true, we realize that the analysis of \cite{Alexandrov:2004ks} did not take into account the existence of a large class of double points.
Fixing this problem, it turns out to be possible to remove the cut-off dependence.
Nevertheless, a close inspection reveals a few other subtle issues in the analysis so that it is not really trustworthy.
Instead, we redo the calculation in a way that avoids these issues and yields a well-defined result for the instanton corrections \eqref{nonpert} in the SL theory.
Furthermore, we generalize this result to a generic set of parameters in \eqref{genCFT} and make it explicit for the case of two non-vanishing parameters, $\lambda_k$ and $\lambda_{2k}$, which we dub `double sine-Liouville' (dSL).
In particular, in this case we find a new set of non-perturbative effects which is absent in the usual SL theory.

Next, we move on to the string theory analysis of non-perturbative effects in the SL and dSL theories.
First, we compute the instanton action $\bS_{\rm inst}$ up to second order in the parameters $\lambda_n$.
In our approach, it is obtained by extremizing certain `tachyon potential' determined by one and two-point disk amplitudes with ZZ boundary conditions.
While the one-point amplitude is easily computed analytically, this is impossible for the two-point function given by a highly complicated integral.
Moreover, it depends on a free parameter reflecting an inherent ambiguity of string field theory needed to get a finite amplitude from a naive divergent CFT expression.
Fortunately, at extrema, the potential depends only on combinations of the two-point functions where the ambiguity is cancelled, and it can be checked numerically that these combinations are equal to simple elementary functions of external momenta, precisely such that the resulting instanton action is identical to the one in MQM.

Finally, we compute the subleading contribution $A$ and its first order correction in $\lambda$.
First, we show that in the small $\lambda$ limit, $A\sim \lambda^{-1/2}$ as predicted by MQM.
This behavior arises due to the lifting of an instanton zero mode by the SL perturbation.
A careful analysis also allows to match the precise normalization factor of the non-perturbative effects in MQM and string theory.
Second, we derive the next order contribution in the $\lambda$-expansion, which turns out to be given by two-point disk and one-point annulus amplitudes.
Like the former, the latter cannot be computed analytically and depends on the same ambiguous parameter of string field theory.
But again in the final result the ambiguity is cancelled and one can show numerically that the two functions combine into an elementary function reproducing predictions of MQM.
Thus, our analysis establishes a perfect agreement at the non-perturbative level of the two descriptions of 2d string theory in the presence of a non-trivial tachyon background up to several orders in the SL parameters and string coupling expansion.

The organization of the paper is the following.
In section \ref{sec-MQM} we review the description of the tachyon perturbations in MQM and the known results on their perturbative and non-perturbative free energy.
In section \ref{sec-MQMSL}, using this description, we provide a derivation of the instanton effects in the SL theory.
In section \ref{sec-MQMdSL} we generalize these results to generic perturbations and  analyze in detail the case of two non-vanishing perturbation parameters.
Then in section \ref{sec-instact} we compute the instanton action in the SL theory up to second order in the perturbation parameter using string theory methods.
In section \ref{sec-sublead} the same is done for the subleading factor $A$ in the instanton contribution, and section \ref{sec-dSL} repeats these calculations for the case of the double SL theory.
Finally, section \ref{sec-cocl} contains our conclusions.
Several appendices provide various consistency checks, technical calculations and other useful information.

\section{Tachyon backgrounds in MQM}
\label{sec-MQM}

In this section we review the MQM description of non-trivial tachyon backgrounds of 2d string theory as well as perturbative and non-perturbative results about free energy associated to these backgrounds.
While the results presented in subsections \ref{subsec-MQM}-\ref{subsec-free} will be used below in section \ref{sec-MQMSL}, those given in subsections \ref{subsec-integr} and \ref{subsec-instantons} will be rederived by following an alternative approach that is better suited for studying instanton effects.
Nevertheless, we have reviewed them here for comparison.

\subsection{MQM in the chiral representation} \label{subsec-MQM}

We are interested in the quantum mechanical system where the degrees of freedom consist of a single time-dependent hermitian $N\times N$ matrix, restricted to the singlet sector of the $U(N$) symmetry group.
The matrix can be diagonalized by a unitary transformation $M\to \Omega^\dagger M\,\Omega$, and upon integrating out the degrees of freedom encoded by $\Omega$,
the system reduces to $N$ free non-relativistic fermions moving in a potential \cite{Brezin:1977sv}.
This system becomes dual to 2d string theory in the so-called double scaling limit where $N\to \infty$, while a parameter of the potential is tuned in a correlated way to its critical value at which the Fermi sea reaches a local maximum of the potential \cite{Brezin:1989ss,Ginsparg:1990as,Gross:1990ay}.
In this limit, the form of the original potential becomes irrelevant, and we end up with a system of free fermions in the inverse oscillator potential $V(x)=-\hf\, x^2$.
The resulting system is parametrized by a scaling parameter $\mu$, which is equal to the difference between the maximum of $V$ and the Fermi level, and turns out to be proportional to the inverse of the string coupling.

A particularly simple way to describe the dynamics of these fermions is to introduce chiral coordinates in phase space
\be
\xpm=\frac{x\pm p}{\sqrt{2}}\, ,
\qquad
\{ \xm,\xp\}=1\, ,
\label{lcx}
\ee
so that $x_-$ and $x_+$ are canonically conjugate variables.
The simplifications come from the fact that in the chiral representation the one-fermion Hamiltonian $H_0=\hf(p^2-x^2)$ becomes a {\it first order} linear differential operator
\be
\hat H_0^\pm=-\hf\(\hat x_+\hat x_- +\hat x_-\hat x_+\)=\mp i\( \xpm\frac{\p}{\p\xpm}+\hf\),
\label{chiralH}
\ee
so that its eigenfunctions take a very simple form
\be
\psepm(\xpm)= \frac{e^{i\phi_\pm(E)}}{\sqrt{2\pi }}\,  \xpm^{ \pm i E-{1\over 2}}\, ,
\qquad E\in \IR,
\label{enekk1}
\ee
where $\phi_\pm(E)$ are any constant phases.
The non-trivial dynamics, which is usually encoded in the parabolic cylinder functions solving the eigenvalue problem in the $x$-representation \cite{Moore:1991zv}, in this case is hidden in the relationship between the two chiral representations.
Namely, since $\psep(\xp)$ and $\psem(\xm)$ represent the same physical state in conjugate representations, they must be related by a Fourier transform\footnote{In most of the previous literature, the phase factor in \eqref{enekk1} was not included into the definition of the wave functions and instead appeared as an explicit factor on the right hand side of \eqref{ennormcond1}.}
\be
\hat S  [\psep](\xm)\equiv
\frac{1}{ \sqrt{2\pi}}\int d\xp \, e^{ i\xp\xm} \psep(\xp)=\psem(\xm).
\label{ennormcond1}
\ee
If one sets $\phi_\pm(E) =\mp\hf\, \phi_0(E)$ by choosing the overall phase of $\psi_\pm$, this condition can be used to determine $\phi_0(E)$.
The result depends on the choice of the integration interval, but this choice affects only the non-perturbative completion of the theory.
If one integrates in \eqref{ennormcond1} over the positive half-line, one obtains\footnote{
If we introduce a new variable $u$ via $x_+ = (-E/x_-) \, e^u$, then the integral reduces to $(-E/x_-)^{iE+{1\over 2}} \int_{-\infty}^\infty du \, e^{u/2} e^{-iE(e^u-u)}$.
Other than the $e^{u/2}$ factor, the integral agrees with that discussed in appendix C.2 of \cite{Harlow:2011ny}.
Therefore, we can follow their discussion to choose the integration contour.
Due to the extra $e^{u/2}$ factor in the integrand,  the integral converges as $u\to-\infty$ but is not absolutely convergent as $u\to\infty$.
For negative $E$, this problem can be avoided by taking the $u$ integral for large Re($u$) to approach $\infty+i\eps$ for some small, positive number $\eps$.
\label{fo5}
}
\be
e^{i\phi_0(E)}=\frac{e^{\frac{\pi i}{4}-\frac{\pi}{2}\, E}}{\sqrt{2\pi}}\,\Gamma\(iE+\hf\).
\label{scatphase0}
\ee
The function $\phi_0(E)$ can be recognized as the scattering phase of the fermions off the inverse oscillator potential, which in turn encodes the scattering of tachyons in the  background \eqref{CFTc=1} \cite{Moore:1991zv}.
Thus, the wave functions $\psepm$ can be regarded as describing ingoing and outgoing states, while $\hat S$ is nothing but the S-matrix operator \cite{Alexandrov:2002fh}.

The phase $\phi_0(E)$ \eqref{scatphase0} is not real, but has an imaginary part given by
\be
\Im \phi_0(E)=\frac{1}{2}\, \log\(1+e^{2\pi E}\)=\sum_{n=1}^\infty\frac{(-1)^{n-1}}{2n}\, e^{2\pi n E}.
\label{Im-scatphase}
\ee
This implies that the S-matrix restricted to asymptotic states purely on the right hand side of the potential is not unitary.
Since the occupied energy levels have $E<0$, \eqref{Im-scatphase} is exponentially suppressed, and this imaginary part is a manifestation of the tunneling of fermions to the left hand side of the potential.
Thus, the chiral description of MQM can also be used to extract non-perturbative information about the system.

\subsection{Tachyon perturbations} \label{subsec-tachyon}

The chiral representation is particularly useful for describing backgrounds of 2d string theory with a non-trivial tachyon condensate \cite{Alexandrov:2002fh}.
To get a perturbed background, we should change the state of the free fermion system in such a way that on the dual side it incorporates tachyons in a coherent state.
This can be achieved by changing the asymptotic form of the one-fermion wave functions.
It was found that the perturbed wave functions should take the following form
\be
\Pepm(\xpm)=\frac{ e^{\mp i \vp_{\pm}(\xpm;E)}}{\sqrt{2\pi }}\,  \xpm^{ \pm i E-{1\over 2}},
\label{enPsipm}
\ee
where the phases can be split into three pieces
\be\label{enekk3}
\vp_\pm (\xpm;E)= V_\pm(\xpm)+\hf\, \phi(E)  + v_\pm(\xpm;E)\, .
\ee
These three pieces are such that $V_\pm$ vanishes at $\xpm=0$, $v_\pm$ vanishes at $\xpm\to\infty$,\footnote{
If the perturbation involves, in the string language, vertex operators of momenta larger than 1, the vanishing condition on $v_\pm$ is more intricate because there are branch cuts extending toward infinity.
For the discrete spectrum of perturbations \eqref{edeftpm} considered below, it can be easily formulated in terms of the parametrization \eqref{xmptau} and requires that $v_\pm$ vanish at $\tau\to\pm \infty$, respectively.
}
and $\phi(E)$ is independent of the coordinates.
The potentials $V_\pm$ encode the form of the tachyon condensate and are assumed to be given, whereas $\phi$ and $v_\pm$ are to be determined by the compatibility condition that both functions \eqref{enPsipm} represent the same physical state in conjugate representations
\be
\hat S  [\Pep](\xm)= \Pem(\xm).
\label{SmatrixMQM}
\ee
Here $\hat S$ is the same operator acting by the Fourier transform as in \eqref{ennormcond1} playing the role of S-matrix
so that $\phi(E)$ can be seen as a perturbed scattering phase.

It is important to note that the wave functions \eqref{enPsipm} are {\it not} eigenfunctions of the Hamiltonian \eqref{chiralH} and therefore the corresponding solution to the time-dependent Schr\"odinger equation is obtained not by a simple multiplication by $e^{-iEt}$ as for the unperturbed wave functions \eqref{enekk1}, but by the following replacement
\be
\Pepm(\xpm,t)= e^{\mp t/2}\Pepm(e^{\mp t}\xpm).
\label{solSchr}
\ee
Thus, the perturbed state of the system is time-dependent, while $\Pepm(\xpm)$
describe only the $t=0$ slice.
This is in agreement with the time-dependence of string backgrounds with a non-trivial tachyon condensate like in \eqref{genCFT}.

However, one can in fact define a deformed Hamiltonian for which $\Pepm(\xpm)$ are eigenstates with eigenvalue $E$.
Once $\vp_\pm$ are fixed, it is defined as a solution of either of the two equations \cite{Alexandrov:2002fh}
\be
H = H_0 + x_\pm \, \p_{\xpm} \vp_\pm (x_\pm; H)\, ,
\label{totalH}
\ee
where the ordering ambiguities between $x_\pm$ and $H$ inside $\p_{\xpm}\vp_\pm$ are fixed by demanding that acting on an eigenstate of $H$ of eigenvalue $E$, $\p_{\xpm}\vp_\pm(x_\pm,H)$ gives $\p_{\xpm}\vp_\pm(x_\pm,E)$.
We can see the equality of the two Hamiltonians by noting that $\Psi^E_+(x_+)$ is an eigenstate of $H_0 + x_+ \, \p_{x_+} \vp_+ (x_+; H)$ with eigenvalue $E$, and
$\Psi^E_-(x_-)$ is an eigenstate of $H_0 + x_- \, \p_{x_-} \vp_- (x_-; H)$ with
eigenvalue $E$. Since $\Psi^E_+(x_+)$ and $\Psi^E_-(x_-)$ are different representations of the same state, it follows that $H_0 + x_\pm \, \p_{\xpm} \vp_\pm (x_\pm; H)$ have same eigenvalues and eigenstates, and hence describe the same operator.
Although this Hamiltonian does not seem to have a direct physical meaning in MQM,
it will allow us to define the usual thermodynamic quantities using the stationary picture it represents.
And it is these quantities that will be shown to be equal to their stringy counterparts.

\subsection{Free energy} \label{subsec-free}

In this paper we are interested in a particular class of perturbations that are dual to the backgrounds where the worldsheet action takes the form \eqref{genCFT}.
The vertex operators appearing in this action correspond to the spectrum of tachyons of Euclidean theory compactified on a circle of radius $R$.
In the MQM formalism, such backgrounds are generated by the potentials
\be
\label{edeftpm}
V_\pm(\xpm)= \sum\limits_{k=1}^{\kmax} t_{\pm k} \, \xpm^{k/R},
\ee
where $t_{\pm k}$ are parameters which can differ from those in the CFT action by normalization factors.
We have determined the precise relation between the parameters in appendix \ref{sappnorm}.

This class of perturbations is expected to have a thermodynamic description as a system at finite temperature $1/(2\pi R) $ \cite{Alexandrov:2002pz}.
In particular, one can define the grand canonical free energy\footnote{
Note that we associate the label $E$ with energy.
This implies that we silently passed to the interpretation of the system based on the Hamiltonian \eqref{totalH} where the deformed Fermi sea is stationary and one can apply the standard thermodynamic relations.
}
\be
\cF\(\mu \)=  \int_{-\infty}^\infty
d E\, \rho(E )\log\left(1+e^{-2\pi R(\mu+E)}\right),
\label{grandF}
\ee
where $\rho(E)$ is the density of states.
To compute $\rho(E)$ in our system, let us introduce a cut-off $\Lambda\gg\mu$ and impose the following boundary condition on the wave functions in the chiral representation\footnote{
To justify this boundary condition, let us put a completely reflecting wall at $x=\sqrt{\Lambda/2}$ which requires the wave function in the $x$-representation to vanish at this point.
For large $\Lambda$, the wave function at the wall is a sum of an ingoing and outgoing waves which are captured by $\Psi_+$ and $\Psi_-$, respectively.
Since asymptotically $p\approx \pm x$, we find that the boundary condition equates the above two components evaluated at $\sqrt{\Lambda}$.
Then \eqref{bndcond} follows from \eqref{SmatrixMQM} up to an irrelevant phase.
}
\be
\hat S[\Psi_+](\sqrt{\Lambda}) =\Psi_+(\sqrt{\Lambda}).
\label{bndcond}
\ee
Applying this condition to the deformed wave functions \eqref{enPsipm} and using \eqref{SmatrixMQM}, one obtains that, for large $\Lambda$, it is satisfied for a discrete set of energies $E_n\ (n\in \Zint )$ defined by
\be
\phi(E_n) - E_n\log \Lambda+V(\Lambda)=2\pi n,
\qquad V(\Lambda)=\sum\limits_{k}(t_k+t_{-k})\Lambda^{k/2R}.
\label{enquantcond}
\ee
Therefore, the density of the energy levels is given by
\be
\rho(E)= {\log \Lambda\over 2 \pi}-
{1\over 2\pi } {d\phi(E)  \over d E} .
\label{endensity}
\ee
Substituting this result into \eqref{grandF}, dropping a $\Lambda$-dependent non-universal contribution, and integrating by parts, one obtains
\be
\cF(\mu)
= - R \int_{-\infty}^{\infty}dE \, \frac{\phi(E)}{1+e^{2\pi R(\mu+ E)}} \, .
\label{enepart}
\ee
From this representation it is easy to establish the following relation \cite{Alexandrov:2003qk}
\be
2\sin \frac{\p_\mu }{2R} \cdot \cF(\mu ) = \phi(-\mu) ,
\label{relFphi}
\ee
where the operator on the left hand side can be seen as a finite difference operator: $2\sin\(\frac{\partial_\mu}{2R}\) \cF = \frac{1}{i} \(\cF(\mu+\frac{i}{2R}) -\cF(\mu-\frac{i}{2R})\)$.
If we evaluate this difference using \refb{enepart}, then naively the result vanishes since the integrands in the expressions for $\cF(\mu\pm {i\over 2R})$ take the same values.
However, the integrands have a pole on the real axis at $E=-\mu$ and one needs to carefully examine the presription for evaluating the contribution from this pole.
For this we can take \refb{enepart} as the definition
of $\cF(\mu)$ on the real $\mu$ axis and define it everywhere else in the complex plane via analytic continuation. In that case one can easily check that as we approach the point $\mu+i/(2R)$ starting from the real
axis, a pole approaches the integration contour over $E$ from the upper half plane, while as we approach the point $\mu-i/(2R)$ starting from the real
axis, a pole approaches the integration contour over $E$ from the lower half plane. The difference between the two integrals can now be computed using the residue theorem and gives the result $\phi(-\mu)$.

The relation \refb{relFphi} shows that to get the free energy, it is sufficient to compute $\phi(E)$.

\subsection{Complex curve and integrability}
\label{subsec-integr}

A useful way to visualize the perturbation is to consider the profile of the Fermi sea, i.e. the boundary of the region in phase space filled by the fermions.
This profile emerges in the quasi-classical approximation, which corresponds to the tree level of string theory and is obtained by applying the stationary phase approximation to the integral in \eqref{SmatrixMQM} or its inverse taken at $E=-\mu$.
This results in the following two equations
\be
\xp\xm=\mu+\xpm\p_{\xpm}\vpz_{\pm}(\xpm;-\mu),
\label{Fermiprof}
\ee
where $\vpz_{\pm}$ is the leading part of $\vp_{\pm}$ in the large $\mu$ limit.
In this approach, it is the condition that the two equations describe the same curve that allows to determine the functions $\vpz_{\pm}$.
In the absence of perturbation, they both reduce to the usual hyperbola $\hf(x^2-p^2)=\mu$.
In section \ref{sec-MQMSL} we shall give a different perspective on the same equations.

The profile of the Fermi sea determined by \eqref{Fermiprof} plays an important role in many aspects of the theory as, being continued to complex values of $\xpm$, it defines the MQM complex curve \cite{Alexandrov:2003qk,Alexandrov:2004ks}.
This curve, regarded as a two-dimensional surface embedded in $\IC^2$, turns out to encode various perturbative and non-perturbative information about the system even though \refb{Fermiprof} was derived in the quasi-classical approximation.
In particular, the instanton effects studied below will emerge from certain special points of the complex curve.

The complex curve also appears as a crucial element of the integrable structure governing the perturbations \eqref{edeftpm}.
Namely, in \cite{Alexandrov:2002fh} it was shown that these perturbation are generated by a set of commuting Hamiltonians, which together form the Toda lattice hierarchy.
This furnishes a powerful technique for finding exact solutions.
Among other things, this implies the existence of a $\tau$-function whose logarithm coincides with the grand canonical free energy, $\cF=\log\tau(\mu,\{t_n\})$, and can be shown to satisfy a hierarchy of bilinear differential equations known as Hirota equations \cite{Jimbo:1983if}.
In particular, the first equation in this hierarchy is called Toda equation and is a differential equation on $\cF$ with respect to $\mu$ and $t_{\pm 1}$.

While an exact solution for the free energy and correlation functions can be found by solving the Toda (or more general Hirota) equation supplemented by a boundary condition at $t_{\pm k}=0$, there is an easier way to get it.
It is based on the use of the so called string equation which at large $\mu$ reduces precisely to the equation \eqref{Fermiprof} on the Fermi profile.
The idea is to use the following ansatz for the solution of \eqref{Fermiprof}
\be
\xpm(\tau) =e^{\pm \tau-{\chi\over 2R}}
\(1+ \sum\limits_{k\ge 1} a_{\pm k}\, e^{\mp {k\over R}\, \tau} \),
\label{xmptau}
\ee
where $\chi=\p_\mu^2\cF_{(0)}$ with $\cF_{(0)}$ being the leading term in $\cF$ in the large $\mu$ limit.
Equation \eqref{xmptau} provides a parametrization of the MQM curve where $\tau$ plays the role of a uniformization parameter.
To find the coefficients $a_k$ and the free energy, one should substitute the ansatz \eqref{xmptau} into \eqref{Fermiprof}, expand the result for $\tau\gg 1$ (resp. $\tau\ll -1$), and extract the constant term as well as coefficients in front of the positive (resp. negative) powers of $e^{\tau/R}$ in the equation with $\xp$ (resp. $\xm$) dependent right hand side.
In particular, for the simplest case where $t_1=t_{- 1}$ are the only non-vanishing parameters, which is supposed to describe the SL theory, this gives an algebraic equation for $\chi$ and an explicit expression for $a_1$ \cite{Alexandrov:2002fh}
\be
\mu e^{ {\chi\over R} } -\frac{1}{R^2}\(1-\frac{1}{R}\)t_1^2 \, e^{\(2-\frac{1}{R}\) \frac{\chi}{R}} =1,
\qquad
a_1= \frac{t_1}{R} \,e^{\(1-\frac{1}{2R}\) \frac{\chi}{R}}.
\label{eqchi}
\ee
Substituting \eqref{xmptau} back into \eqref{Fermiprof} and using \refb{enekk3}, it is also easy to get $v^{(0)}_{\pm}$, the leading part of the unknown piece of the perturbing phase.
Denoting by $\Xpm(\xmp)$ the mutually inverse functions defined in the parametric form by \eqref{xmptau}, the general result is given by
\be
v^{(0)}_\pm(\xpm;-\mu)=  \int^{\xpm}_{\xpm^{0}} \Xmp(\ypm)d\ypm -\mu\log\xpm-V_\pm(\xpm),
\label{solv0}
\ee
where $\xpm^{0}$ are chosen so that the right hand side vanishes at large $\xpm$.

It is useful to note that the functions $\xpm(\tau, \mu)$ describe a canonical transformation in the sense that
\be
{\p x_+ \over \p\tau}\, {\p x_-\over \p\mu} - {\p x_- \over \p\tau} \, {\p x_+\over \p\mu}=1\, .
\label{canontr}
\ee
Since $-\mu$ is an eigenvalue of the perturbed Hamiltonian $H$ \eqref{totalH}, this agrees with the fact that $\tau$ is a canonical parameter along trajectories generated by $H$.

For what follows, it is useful to note that if we introduce
\be
\lambda= t_1 \, \mu^{\frac{1}{2R}-1}
\label{deflam}
\ee
and decompose
\be
\chi=-R\log \mu+R \scX,
\ee
then the relations \eqref{eqchi} can be rewritten as
\be
e^\scX -\frac{1}{R^2}\(1-\frac{1}{R}\)\lambda^2 \, e^{\(2-\frac{1}{R}\) \scX}=1,
\qquad
a_1=\frac{\lambda}{R} \,e^{\(1-\frac{1}{2R}\) \scX(\lambda)},
\label{XthroughL}
\ee
which shows that $\scX$ and $a_1$ depend on $\mu$ and $t_1$ only through the combination $\lambda$.

\subsection{Instanton effects} \label{subsec-instantons}

Let us now briefly review what is known about instanton corrections to the free energy
of the theory compactified on a circle of radius $R$.
First, in the unperturbed MQM, they can be extracted either
from the well-known exact formula for $\cF(\mu)$ \cite{Klebanov:1991qa},
or by inverting the difference operator in \eqref{relFphi} and applying it to \eqref{Im-scatphase}.
In both cases, one finds\footnote{The second method does not allow us to fix the coefficients
in the second sum since all terms are annihilated by the operator in the left hand side of \eqref{relFphi}.
Note also that for rational $R$ some terms in both sums become singular.
However, in most of the paper, we assume $R$ to be irrational. Therefore, we are not concerned about these singularities
and mixing of different contributions appearing for rational radius.}
\be
\cF_{\rm np}(\mu)=i\sum_{n=1}^\infty {e^{-2\pi n \mu}
\over 4n(-1)^{n}\sin{\pi n\over R}}+
i\sum_{n=1}^\infty{e^{-2\pi R n \mu} \over 4n(-1)^{n} \sin(\pi R n )}\, .
\label{encFnp}
\ee
In 2d string theory, there are two types of non-perturbative corrections, which have their origin in the two kinds of branes.
They both have a ZZ-boundary condition for the Liouville field and differ by the boundary condition imposed on the matter field: Dirichlet in the first type and Neumann in the second.
In \refb{encFnp}, the first set of terms correspond to D-instantons, while the second set of terms correspond to D0-branes wrapping the compactification circle.
Note that due to degeneracy between $n$ (1,1) ZZ-branes and $(n,1)$ ZZ-brane (as well as their possible fractions), their contributions are mixed in the $n$-th term of each sum in \eqref{encFnp} and cannot be distinguished.

After adding tachyon perturbations, the fate of the two types of non-perturbative corrections is quite different.
Those which come from wrapped D0-branes do not change, whereas the first term in \eqref{encFnp}
acquires a very non-trivial dependence on the parameters $t_n$ \cite{Alexandrov:2003nn}.
This dependence has been computed explicitly for the SL theory in the quasi-classical approximation, i.e.
to the first two orders in the large $\mu$ (or small string coupling) limit.
Namely, by solving a linearization of the Toda equation around its perturbative solution,
one can show that the $n$-th term in the first sum in \eqref{encFnp}
is replaced by \cite{Alexandrov:2003nn,Alexandrov:2003un,Alexandrov:2004cg}
\be
A_n (\mu,\lambda) \, e^{-\bS_n(\mu,\lambda)},
\label{nonpert-n}
\ee
where
\be
\bS_n=2\mu \[\theta_n(\lambda)+\(2-\frac{1}{R}\) \lambda \,e^{-\frac{\scX(\lambda)}{2R}}\sin\( \frac{1}{R}\,\theta_n(\lambda)\)\],
\label{Sn-SL}
\ee
\be
A_n=\frac{ C\, \mu^{-1/2} \,e^{\hf \scX(\lambda)}}{\sin(\theta_n(\lambda))\sin\(\frac{\theta_n(\lambda)}{R}\)}
\[\cot(\theta_n(\lambda))- \(\frac{1}{R}-1\) \cot\(\frac{1-R}{R}\, \theta_n(\lambda)\)\]^{-1/2},
\label{resAnSL}
\ee
and we used the quantities appearing in the perturbative solution of the SL theory:
the scaling variable $\lambda$ \eqref{deflam} and the functions $\scX(\lambda)$ and $a_1(\lambda)$ defined in \eqref{XthroughL}.
Finally, the functions $\theta_n(\lambda)$ are solutions of
\be
\sin \theta_n =  a_1(\lambda)\, \sin \( \frac{1-R}{R}\, \theta_n\),
\qquad \theta_n(0)=\pi n,
\label{defthnSL}
\ee
and $C$ is an undetermined constant. It cannot be fixed by this method since $A_n$ is a solution of a homogeneous equation.

Instead, the constant $C$ was found in \cite{Alexandrov:2004cg} by another method
which is briefly explained in appendix \ref{ap-cutoff}
and is based on the use of the relation \eqref{SmatrixMQM}, combined with orthonormality of the wave functions $\Pepm$,
to evaluate the scattering phase $\phi(E)$. However, the obtained result, given in \eqref{finalC},
was unsatisfactory because of its dependence on the cut-off $\Lambda$.
In fact, as we show in the same appendix, the analysis in \cite{Alexandrov:2004cg} has missed
a series of instanton contributions which can be responsible for disappearance of the cut-off dependence.
However, the argument explaining how this comes about is not rigorous and gives an instanton normalization factor
that differs from the one found later in section \ref{sec-sublead} by a factor of $i$.
Therefore, in the next section we present an alternative derivation from first principles, leading to the results consistent with the subsequent string theory analysis.

\section{Instantons from wave functions: SL theory}
\label{sec-MQMSL}

In this section we shall give a derivation of instanton effects in the SL theory using the relation \eqref{SmatrixMQM} between the two chiral representations.
For the purposes which will become clear in section \ref{sec-sublead}, we consider a slightly generalized version of the SL theory with an additional integer parameter $k$, which is obtained from the general perturbation \eqref{edeftpm} by taking $t_k=t_{-k}$ nonzero and all other parameters vanishing.

Before we start, we would like to make a general comment.
In principle, a theory like MQM, which is defined at the perturbative level, can have several non-perturbative completions \cite{Marino:2012zq}.
This manifests, for example, as an ambiguity in the choice of an integration contour, which in turn determines which saddle points contribute to the integral and which do not.
We resolve this ambiguity by requiring that our specific non-perturbative completion at finite SL parameter is the one that is obtained by analytic continuation from small $t_k$. We assume that the string theory side is also defined using the
same procedure.
This is expected to ensure that if a match with string theory has been achieved in a small domain around $t_k=0$, where matrix model quantities can be compared against string correlation functions, it will continue to hold for finite $t_k$ as well.

\subsection{Determination of the quasi-classical wave functions}
\label{subsec-oneint}

Our starting point is the relation \eqref{SmatrixMQM} which, using \refb{enPsipm} and \refb{enekk3},  can be rewritten as an explicit expression for the scattering phase
\begin{align}
e^{i\phi(-\mu)} &=
\frac{1}{\sqrt{2\pi}}
\int_0^\infty \frac{d\xp}{\xp} \,\cI(\xp,\xm),
\non \\
\cI(\xp,\xm) &= \exp\Bigg[ i\xp\xm -\(i\mu-\hf\) \log (\xp\xm) - it_k\(\xp^{k/R}+\xm^{k/R}\)
\label{enefou1}\\
& \hspace{0.5in} - i v_+(\xp;-\mu)- i v_-(\xm;-\mu)\Bigg] ,
\nonumber
\end{align}
where we set $E=-\mu$, recalling that $\mu$ is a positive parameter.
We are interested in the large $\mu$ expansion and
introduce the scaling variable similar to \eqref{deflam},
\be
\lambda_k= t_k \, \mu^{\frac{k}{2R}-1},
\label{deflamk}
\ee
which is kept constant in this limit. Then
under the scaling
\be
\xpm\sim \mu^{1/2},
\qquad
v_\pm\sim \mu\,,
\ee
the terms in the exponent scale as $\mu$. Let us denote by $v^{(0)}_\pm$ and $v^{(1)}_\pm$ the order $\mu$ and order one contributions to $v_\pm$. Then
the first two orders of the expansion of the integrand $\cI(\xp,\xm)$ can be written as
\be
\cI(\xp,\xm)=B(\xp,\xm)\, e^{iS(\xp,\xm)},
\ee
where
\bea
S(\xp,\xm) &=&\xp\xm -\mu\log (\xp\xm) - t_k\(\xp^{k/R}+\xm^{k/R}\)-v^{(0)}_+(\xp)- v^{(0)}_-(\xm),
\label{Sxpxm}
\\
B(\xp,\xm)&=&(\xp\xm)^{1/2}\, e^{- i \(v^{(1)}_+(\xp)+ v^{(1)}_-(\xm)\)}.
\label{Bxpxm}
\eea
The `effective action' $S(\xp,\xm)$ scales as $\mu$ so that one can apply the saddle point approach to evaluate the integral \eqref{enefou1}.

At leading order, one finds the following saddle point equation obtained by variation of $S(\xp,\xm)$ with respect to $x_+$:
\be
\xp\xm=\mu+\frac{k}{R}\, t_k \xp^{k/R}+\xp\p_{\xp} v^{(0)}_+(\xp;-\mu),
\label{sp-eq}
\ee
which coincides with one of the equations \eqref{Fermiprof} for the profile of the Fermi sea.
In principle, it determines $\xp$ as a function of $\xm$ for an arbitrary function $v^{(0)}_+$.
But we have an additional condition that after substitution of this function into $S(\xp,\xm)$, the result should be independent of $\xm$ since $\phi(-\mu)$ on the left hand side of \eqref{enefou1} is a constant. This is equivalent to the requirement
$\frac{d}{d\xm}\,S(\xp(\xm),\xm)= (\p_{\xm}S)(\xp(\xm),\xm)=0$
and leads to an equation analogous to \eqref{sp-eq}:
\be
\xp\xm=\mu+\frac{k}{R}\, t_k \xm^{k/R}+\xm\p_{\xm} v^{(0)}_-(\xm;-\mu).
\label{sm-eq}
\ee
It is easy to check that the two equations are compatible provided\footnote{A systematic approach to arriving
at the equations below, which also works for the more general case discussed in the next section, is as follows.
We take the change of variables \eqref{xpmSLk} as an ansatz and try to choose the constants $a_k$ and $\scX$
such that $\tau_+
=\tau_-$ is a solution to the saddle point equation \eqref{sp-eq} for all $\tau_-$.
Substituting \eqref{xpmSLk} into \eqref{sp-eq},
setting $\tau_+=\tau_-$ and requiring that the equation holds for large $\tau_-$, we get \eqref{XthroughLk}.
The knowledge of $v^{(0)}_+$ is not needed at this stage since it vanishes for large value of its argument. Next we
can integrate the full equation \eqref{sp-eq} at $\tau_+=\tau_-$ to arrive at the form of $v^{(0)}_+$ given \eqref{solvpmSL}. The constant of integration is fixed by requiring that $v^{(0)}_+$ vanishes for large value
of its argument.
A similar analysis of \eqref{sm-eq} leads to the form of $v^{(0)}_-$ given in \eqref{solvpmSL}.}
\be
\begin{split}
v^{(0)}_\pm(\xpm,-\mu)=&\, \mu\biggl[
\pm \tau_\pm + a_k\, e^{-\scX}\( \frac{R}{k}\, e^{\pm\frac{k}{R}\tau_\pm}- \(\frac{R}{k}-1\) e^{\mp\frac{k}{R}\tau_\pm}\)
\\
&\,
-\log(\xpm/\sqrt{\mu})-\lambda_k(\xpm/\sqrt{\mu})^{k/R}
-\hf\, \scX+a_k^2 \, e^{-\scX}\biggr],
\end{split}
\label{solvpmSL}
\ee
where
$\tau_\pm$ are reparametrizations of $\xpm$
\be
\xpm(\tau_\pm) =\sqrt{\mu}\, e^{\pm \tau_\pm-\hf\scX}\(1+  a_k\, e^{\mp {k\over R}\,\tau_\pm} \),
\label{xpmSLk}
\ee
and $\scX$ and $a_k$ are functions of $\lambda_k$ satisfying
\be
e^\scX -\frac{k^2}{R^2}\(1-\frac{k}{R}\)\lambda_k^2 \, e^{\(2-\frac{k}{R}\) \scX}=1,
\qquad
a_k= \frac{k}{R} \, \lambda_k \,e^{\(1-\frac{k}{2R}\) \scX}.
\label{XthroughLk}
\ee
The last two terms in \eqref{solvpmSL} ensure the condition that $v^{(0)}_\pm$ vanish at $\tau_\pm\to \pm\infty$, respectively.

After substitution of \eqref{solvpmSL} into \eqref{sp-eq} and \eqref{sm-eq}
and using relations \eqref{XthroughLk},
the two equations simplify, respectively, to\footnote{Note that at this point the status of the two equations is different: the first one is the saddle point equation, while the second ensures that at the saddle point the effective action is $\tau_-$-independent.}
\be
\xm(\tau_-)=\xm(\tau_+)
\label{spe-xpxm}
\ee
and
\be
\xp(\tau_+)=\xp(\tau_-)\, .
\label{spe-xpxm2}
\ee
These have an obvious common solution
\be
\tau_+=\tau_-.
\label{sp-pertsol}
\ee
Plugging \eqref{solvpmSL} and \eqref{xpmSLk} into \eqref{Sxpxm} and evaluating the effective action at the saddle point \eqref{sp-pertsol},
it is straightforward to check that all $\tau_-$-dependent terms cancel
and one remains with the following constant contribution to the scattering phase
\be
\phi^{(0)}(-\mu)= S(\xp(\tau),\xm(\tau))
=-\mu\log \mu+\mu\(\scX+e^{-\scX}(1-a_k^2)\).
\label{phi0-SL}
\ee
Using \eqref{relFphi}, this result can then be integrated to reproduce the expression for the genus-0 free energy $\cF_{(0)}$,
first found in the T-dual formulation in \cite{Kazakov:2000pm}.
In particular, by calculating $\p_\mu\scX$ and $\p_\mu a_k$
at fixed $t_k$ from \eqref{deflamk} and \eqref{XthroughLk}, and using \eqref{relFphi} for large $\mu$, we get
\be\label{efkck}
\p_\mu^2 \cF_0(\mu) =R\, \p_\mu \phi^{(0)}(-\mu) = -R\, \log\mu + R\, \scX .
\ee

One can relate these to the results reviewed in section \ref{sec-MQM}
by noting that \eqref{sp-pertsol}, substituted into \eqref{xpmSLk}, gives us back \eqref{xmptau} and eqs. \eqref{XthroughLk} agree with \eqref{eqchi}.
Besides, by integrating \eqref{sp-eq} and \eqref{sm-eq},
the leading order solution \eqref{solvpmSL} for $v_\pm$ may also be expressed as in \eqref{solv0}:
\be
v^{(0)}_\pm(\xpm;-\mu)=  \int^{\xpm}_{\xpm^0} \Xmp(\ypm)d\ypm -\mu\log\xpm-V_\pm(\xpm),
\label{solv03}
\ee
where $\Xpm(\xmp)$ are the mutually inverse functions defined by the identification $\tau_+=\tau_-$ in \eqref{xpmSLk}
and $\xpm^0=\xpm(\pm\tau_{0})$ with $\tau_0$ being a solution of the following equation
\be
\tau_0 + a_k\, e^{-\scX}\( \frac{R}{k}\, e^{\frac{k}{R}\tau_0}- \(\frac{R}{k}-1\) e^{-\frac{k}{R}\tau_0}\)
=-\hf\, \log\mu +\frac12 \,\scX - a_k^2 \, e^{-\scX}\, .
\ee
Due to this, the effective action \eqref{Sxpxm} takes the elegant form
\be
\begin{split}
S(\xp,\xm)=&\, \xp\xm- \int^{\xp}_{\xp^0}\Xm(\yp)\,d\yp
-\int^{\xm}_{\xm^0} \Xp(\ym)\,d\ym
\\
=&\,  x_+x_- -\int_{\tau_0}^{\tau_+} \xm(\tau)\,
\pxp(\tau)\, d\tau
-\int_{-\tau_0}^{\tau_-} \xp(\tau)\, \pxm(\tau)\, d\tau\, ,
\end{split}
\label{Sxpxmpre}
\ee
where $\pxpm=\p_{\tau}\xpm(\tau)$.
Note that this equation is valid for all $(x_+,x_-)$, not just those that lie on the extrema of $S$.

We also need to evaluate the subleading (one-loop) contribution to the integral \eqref{enefou1}.
It comes from two sources: the subleading terms in the integrand evaluated at the saddle point \eqref{sp-pertsol}
which produce
\be
\frac{B(\xp,\xm)}{\sqrt{2\pi}\xp}
=\(\frac{\xm}{2\pi\xp}\)^{1/2} e^{-i \(v^{(1)}_+(\xp)+v^{(1)}_-(\xm)\)},
\label{sublead-int}
\ee
and the result of the Gaussian integral around the saddle point which gives the factor
\be
\(\frac{2\pi \pxp}{i \,\pxm}\)^{1/2} ,
\label{sublead-Gint}
\ee
where all functions in \eqref{sublead-int} and \eqref{sublead-Gint} are supposed
to be evaluated at $\tau_-$.
Requiring that all $\tau_-$-dependence cancels, we can now fix the subleading contribution to $v_\pm(\xpm)$ to be given by
\be
\begin{split}
v^{(1)}_\pm(\xpm;-\mu)
=&\, \pm{i\over 2}  \log\(\pm \frac{\xpm}{\pxpm}\)
\\
=&\, \pm {i\over 2}  \log \[\frac{1 + a_k\, e^{\mp k\tau_\pm/R}}{1 + a_k \(1-{k\over R}\) e^{\mp k\tau_\pm/R}}\].
\end{split}
\label{vpm-sublead}
\ee
Substituting these back into the product of \refb{sublead-int} and \refb{sublead-Gint}, we see that at order $\mu^0$ the only contribution to the scattering phase is the non-universal constant term $\pi/4$ coming from the factor $i$ in \eqref{sublead-Gint}
(the same that appears in \eqref{scatphase0}), consistently with the fact that the free energy has an expansion in even powers of the string coupling. This gives, to this order,
\be
\phi_{\rm pert}(-\mu) =-\mu\log \mu+\mu\(\scX+e^{-\scX}(1-a_k^2)\) +{\pi\over 4}\, .
\ee

Substituting the above results into \eqref{enPsipm}, one finds that in the quasi-classical approximation the perturbed wave functions take the following form
\be
\Psi^{{-\mu}}_\pm(\xpm)\approx \frac{e^{\mp \frac{\pi i}{8}}}{\sqrt{\pm 2\pi \pxpm}}\,
\exp\[\mp i \int^{\xpm}_{\xpm(\pm \tau_0)} \Xmp(\ypm)d\ypm \].
\label{pertPepm}
\ee
This result agrees with the quasi-classical wave functions found in \cite{Alexandrov:2004cg}.

Proceeding in a similar way, one can, in principle, compute further corrections in $1/\mu$ to the perturbed wave functions $\Psi^{{-\mu}}_\pm$.
However, we shall not do this and stop at the subleading order since our main objects of interest in this paper are not perturbative, but instanton corrections.

\subsection{Instanton contributions}
\label{subsec-instcontr}

So far we analyzed only perturbative contributions to the integral \eqref{enefou1}.
But there are also non-perturbative contributions which may have two origins.
First of all, as we will see, the saddle point equation \eqref{sp-eq} (or \eqref{spe-xpxm})
has other solutions besides \eqref{sp-pertsol}.
These solutions generate exponentially suppressed contributions which can be written as
\be
\frac{1}{2\sqrt{2\pi}}\int_{\rm inst} \frac{d\xp}{\xp} \,B(\xp,\xm)\, e^{iS(\xp,\xm)}
\equiv \cJ(\xm),
\label{instcontr}
\ee
where the subscript ``inst" denotes that the integral over $\xp$ is along the steepest descent contour of the instanton saddles. We have included an extra factor of 1/2 in the integral since the full integration contour runs over only half of the steepest descent contour of each instanton.
This factor is present even in the unperturbed theory \cite{Sen:2021qdk}.\footnote{
Due to the relation to the integral in  appendix C.2 of \cite{Harlow:2011ny} mentioned in footnote \ref{fo5}, we can use their result to express the original integration contour as a union of steepest descent contours of various saddle points.
In particular, for large $\mu$, we can follow \cite{Harlow:2011ny} to show that the
steepest descent contour of the perturbative saddle reaches the first instanton saddle.
Therefore, we can express the original integration contour as a sum of the steepest descent contour of the perturbative saddle ending at the first instanton saddle, half of the steepest descent contour of the first instanton saddle and additional contributions that are further suppressed.
}

Second, the functions $v_\pm(\xpm)$ entering the integrand in \eqref{enefou1} may also get instanton corrections needed to cancel the $\xm$-dependence introduced by the contributions \eqref{instcontr}.
Let us call them $\vin_\pm$.
In this paper we are interested only in the one-instanton effects and therefore in \eqref{instcontr} these corrections can be ignored because their effect will have double suppression.
On the other hand, we must take them into account in the contribution from the perturbative saddle.
To leading order in the instanton expansion, these corrections change this contribution to
\be\label{ene126}
\exp\[i\phi_{\rm pert} - i \langle\vin_+(\xp)\rangle- i \vin_-(\xm)\] ,
\ee
where both $\xpm$ are expressed as functions of $\tau_-$.
Here, $\langle\vin_+(\xp)\rangle$ denotes the expectation value of $\vin_+(\xp)$ at the perturbative saddle.
At the leading order this is given by $\vin_+(\xp)$ itself, but there are subleading corrections involving derivatives of $\vin_+(\xp)$. For our analysis we shall not need the explicit form of these terms.

Adding \eqref{instcontr} to \eqref{ene126}, we get the instanton corrected value of $e^{i\phi}$. Taking the
logarithm on both sides, one finds
\ben\label{enemaster}
\phi &=& \phi_{\rm pert} -   \langle\vin_+\rangle -  \vin_-
+\tcJ,
\qquad
\tcJ(\xm)=- i e^{-i\,\phi_{\rm pert} }\,\cJ(\xm),
\een
where as usual we have ignored double suppressed terms.

To proceed further, in principle, we should find $\vin_\pm$ by requiring them to cancel the $\xm$-dependent terms produced by the last term.
In appendix \ref{ap-Smatrix} we do this in the small $t_k$ expansion.
However, we will see now that to find the scattering phase to the desired order, we do not need to determine $\vin_\pm$ explicitly.

A crucial observation is that the integral \eqref{instcontr}, and hence the last term in \eqref{enemaster}, is periodic in $\tau_-$ with period $2\pi i R$, at least for small SL parameter and near $\Re \tau_- \approx 0$.\footnote{
In this case we could also use periodicity under shift by $2\pi i R/k$. But our choice allows to provide a uniform analysis that will also work in the next section where several parameters $t_k$ are turned on.
}
Indeed, after changing variables to $\tau_\pm$ by means of \eqref{xpmSLk}, it is easy to see that all terms depend on them either through $e^{k \tau_\pm/R}$ or via the combination $\tau_+-\tau_-$.
Shifting the integration variable $\tau_+$ by $\tau_-$, one achieves that the only remaining dependence on $\tau_-$ is via $e^{k \tau_-/R}$, and thus periodic.
In principle, the periodicity could be spoiled by the presence of branch cuts.
For example, the saddle point evaluation of \eqref{instcontr} gives rise to the factor
\be
\(\frac{\pxm(\tau^{\rm (in)}_+(\tau_-))}{\pxm(\tau_-)}\)^{1/2},
\ee
where $\tau^{\rm (in)}_+(\tau_-)$ is a solution of the saddle point equation \eqref{spe-xpxm}, different from \eqref{sp-pertsol} and  giving rise to an instanton effect.
However, from \eqref{xpmSLk} it follows that this factor is still periodic for small $t_k$ and small $\Re\tau_-$.
This will be enough for our purposes.

As a consequence of this periodicity, the last term in \eqref{enemaster} near $\Re\tau_- =0$ can be expanded in Fourier series
\be\label{efourier}
\tcJ(\xm)=\sum_{n\in \IZ} \tcJ_n\, e^{n \tau_-/R},
\ee
with bounded $|\tcJ_n|$.
Then $\langle\vin_+\rangle$ and $\vin_-$ can be chosen to be equal to the parts of this series
corresponding to negative and positive modes, respectively:
\be
\langle\vin_+(\xp(\tau))\rangle=\sum_{n=1}^{\infty} \tcJ_{- n}\, e^{- n \tau/R}, \qquad
\vin_-(\xm(\tau))=\sum_{n=1}^{\infty} \tcJ_{ n}\, e^{ n \tau/R}. \label{evpmeq}
\ee
This ensures that, on the one hand, the right hand side of \eqref{enemaster} is independent of $\tau_-$, and on the other hand,
the series for $\langle\vin_+\rangle$ (resp. $\vin_-$) is absolutely convergent, and hence analytic, in the region $\Re\tau>0$ (resp. $\Re\tau<0$)
and vanishes at $\tau\to\infty$ (resp. $\tau\to-\infty$). Since $\langle\vin_+\rangle$ is given by $\vin_+$ plus subleading corrections involving derivatives of $\vin_+$, we can find $\vin_+$ from the expression for $\langle\vin_+\rangle$.

As a consequence of \eqref{efourier} and \eqref{evpmeq}, only the first term and the constant Fourier mode $\tcJ_0$ survive on the right hand side of \eqref{enemaster}.
The latter can be obtained by integrating $\tcJ(\xm)$ over $\tau_-$ in the range $0\le \tau_-<2\pi i R$ and dividing the result by $2\pi i R$.
Thus, we arrive at
\be
\label{ene63a}
\phi=\phi_{\rm pert}
- \frac{e^{-i\,\phi_{\rm pert} }}{4\pi R\sqrt{2\pi}}
\int\limits_0^{2\pi iR} d\tau_-
\int\limits_{\rm inst}d\tau_+ \(-\pxp\pxm\)^{1/2} e^{iS(\xp,\xm)},
\ee
where we changed the integration variable from $\xp$ to $\tau_+$
and substituted
\be
B(\xp,\xm)= \xp\(-\frac{\pxm}{\pxp}\)^{1/2}
\label{Bxpxm-gen}
\ee
following from \eqref{Bxpxm} and \eqref{vpm-sublead}.
Although the reasoning leading to this representation applies only for small SL parameter, as explained in
the beginning of this section, the finite parameter case is obtained by analytic continuation from small $t_k$.
Therefore, if no singularities appear on the way, the representation \eqref{ene63a} should still be true.

Before analyzing the instanton contributions in the deformed theory, it is instructive to reproduce the results \eqref{encFnp} in the undeformed theory from \eqref{ene63a}.
In this case $\lambda_k$ and $a_k$ vanish and a set of solutions to \eqref{spe-xpxm}, \eqref{spe-xpxm2} are given by
\be
\tau_+ = \tau_- - 2\pi i n, \qquad n=1,2,\dots .
\ee
Compared to the perturbative contribution, the only extra effect is that the $\log x_+$ term  in the expression for $\cI(x_+,x_-)$ in \eqref{enefou1} is shifted by $-2\pi i n$, producing an extra multiplicative factor of $(-1)^n e^{-2\pi n\mu}$.
We focus on the case $n=1$.
In this case in \eqref{ene63a} we get
\be
{1\over \sqrt{2\pi}}\int\limits_{\rm inst}d\tau_+ \(-\pxp\pxm\)^{1/2} e^{iS(\xp,\xm)} = - e^{-2\pi \mu}
e^{i\,\phi_{\rm pert}}\, .
\ee
This gives
\be \label{eunperturbed}
\phi=\phi_{\rm pert} + {i\over 2} \,e^{-2\pi \mu}
\qquad \Rightarrow \qquad
\cF^{(1)}_{\rm np} = -{i\, e^{-2\pi \mu}\over 4 \sin(\pi/R)}
\, ,
\ee
where in the last step we have used \eqref{relFphi}.
This gives the $n=1$ term in the first sum on the right hand side of \eqref{encFnp}.

Let us now turn to the deformed theory.
To evaluate the last term in \eqref{ene63a}, first of all, one should understand which other saddle points, besides the perturbative one \eqref{sp-pertsol}, contribute to the integral over $\tau_+$ in \eqref{ene63a}.
It turns out that, instead of solving this problem, it is easier to do this for the double integral in \eqref{ene63a}.
The point is that the original saddles, which solve \eqref{sp-eq} or \eqref{spe-xpxm}, have a very non-trivial dependence on $\tau_-$, whereas in the double integral this dependence is fixed by a second saddle point equation coming from the extremization in the $\tau_-$ variable.
Besides, the double integral is closely related to the one that was already analyzed in \cite{Alexandrov:2004cg} and reviewed in appendix \ref{ap-cutoff}, so that we can borrow some of those results.

\subsubsection{Double points}
\label{subsec-dpMQM}

Since the effective action $S(\xp,\xm)$ \eqref{Sxpxm} is symmetric in $\xp$ and $\xm$,
the two saddle point equations for the double integral in \eqref{ene63a}
must also be symmetric. Due to \eqref{spe-xpxm}, they simply read
as\footnote{Note that even though the second equation is identical to \eqref{spe-xpxm2}, they have slightly different origin. Here it is a consequence of the saddle point equation, whereas in \eqref{spe-xpxm2} it was derived by demanding $\tau_-$ independence of the phase.}
\be
\xp(\tau_+)=\xp(\tau_-),
\qquad
\xm(\tau_+)=\xm(\tau_-).
\label{sp-eq2}
\ee
Solutions of these two equations can be given
a nice geometric meaning as `double points' of the MQM complex curve \cite{Alexandrov:2004ks}
as follows. As discussed in section \ref{subsec-integr},
the perturbative saddle, corresponding to \eqref{xpmSLk}
with $\tau_+=\tau_-\equiv\tau$, can be seen as a complex curve embedded into
$\IC^2$ parametrized by $(\xp,\xm)$ if we regard $\tau$ as a complex variable.
Then a solution to \eqref{sp-eq2} with $\tau_+\ne \tau_-$ implies that
the two values of the uniformization parameter $\tau$
correspond to the same point of this curve.
Mathematically, it represents a pinched cycle where the curve touches itself.

Although one cannot draw objects in four-dimensional space,
the double points can easily be visualized by taking a two-dimensional section by the plane $\xp=\bar x_-$ \cite{Alexandrov:2004ks}.
It results into a one-dimensional curve which has the same parametrization $\xpm(\tau)$ as in \eqref{xpmSLk},
but now with $\tau$ being pure imaginary, $\tau=i\theta$, $\theta\in \IR$.\footnote{This is to be contrasted with the profile of the Fermi sea
which is a section of the MQM curve by the orthogonal plane $\xpm=\bar x_\pm$ and parametrized by $\xpm(\tau)$ with $\tau\in\IR$.}
Then the double points are simply the self-intersection points of this curve.

We illustrated this in Fig. \ref{fig-lotsections} where we showed such sections and the corresponding double points
in the two cases of $R/k=4/3$ and $R/k=3/4$. We have chosen rational values just for the illustrative purpose because otherwise the
curve would never close. But as a result, there is only a finite number of double points. In the case of irrational $R$,
which we are really interested in, their number is always infinite.

\twofig{Section of the MQM complex curve by the plane $\xp=\bar x_-$ and its double points for $R/k=4/3$ and $R/k=3/4$.
The positions of the double points are captured by \eqref{dbpoints-SL}
with the parameters shown in the figure.
Due to rationality of $R$ and relations $\theta_1^\sind=\theta_3^{1-\sind}$, $\theta_2^0=\theta_2^1$ ($R/k=4/3$)
and $\theta_1^\sind=\theta_2^\sind$ ($R/k=3/4$), double points with various parameters coincide and there is only a finite number of them.}
{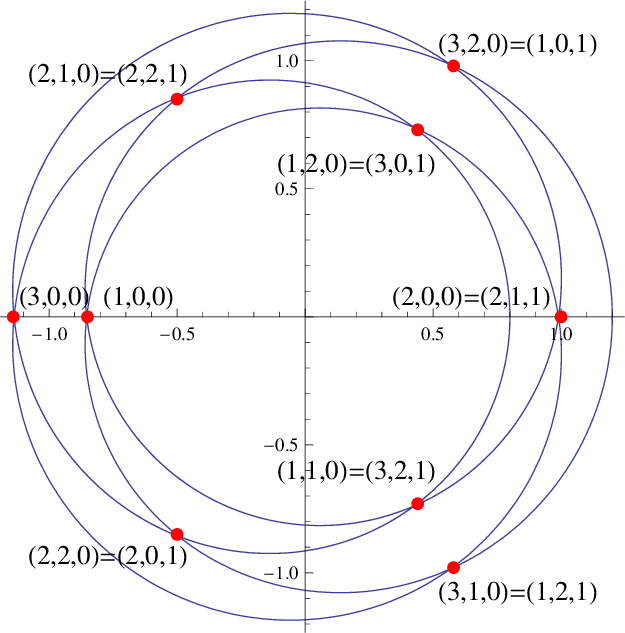}{7cm}{1cm}{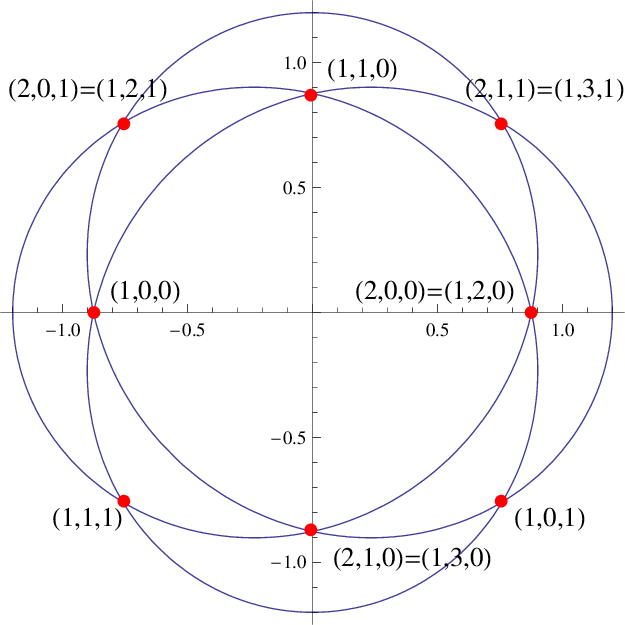}{7cm}{0.2cm}{fig-lotsections}

It is easy to realize that for irrational $R$ there are two infinite 2-parametric sets of double points.
We will label them by $(n,m,\sind)$ with  $n\in\IN$, $m\in\IZ$ and $\sind=0,1$.
The corresponding values of $\tau_\pm$ which solve \eqref{sp-eq2} are given by
\be
\tauipm{n,m}=i\(\pi \, (2m+\sind)\,\frac{R}{k} \mp \thpm_n(\lambda_k)\),
\label{dbpoints-SL}
\ee
where the functions $\thpm_n(\lambda_k)$ are defined to be solutions of the following algebraic equations
\be
\sin \( \thpm_n\) =(-1)^\nu a_k \sin \( \frac{k-R}{ R}\, \thpm_n\),
\qquad
\thpm_n(0)=\pi n,
\label{eq-thetas}
\ee
with $a_k$ given in \eqref{XthroughLk}.
Several comments are in order:
\begin{itemize}
\item
Since $\thpm_n$ are independent of $m$, in each of the two sets the double points with different index $m$ are related
by a shift of the uniformization parameter by a multiple of $2\pi i R/k$.
This is a manifestation of a discrete rotation symmetry of the MQM curve \cite{Alexandrov:2004ks}.
In particular, this symmetry implies that the instanton contributions generated by these double points are also $m$-independent. Furthermore,
due to the restriction on the range of $\tau_-$ integration between 0 and $2\pi i R$, only $k$ different values of $m$ contribute to the integral.

\item
The two series of double points labelled by $\sind=0,1$ correspond to maxima and minima of $iS(\xp,\xm)$.
The precise identification depends on the value of the parameters $n$, $R/k$ and $\lambda_k$.
One can see that at small $\lambda_k$, the integration contour in the undeformed theory is deformed to the union of the steepest descent contours of only the maxima.
Indeed, according to \eqref{dbpoints-SL}, the maxima and minima lie
on the imaginary $\tau_-$ axis along which the original integration contour lies.
Therefore, as we move along this axis, the integrand decays as we move away from the maxima of  $iS(\xp,\xm)$ and increases as we move away from its minima.
This shows that for small $\lambda_k$ the steepest descent contours of the maxima of $iS(\xp,\xm)$ lie along the original integration contour, while the steepest descent contours of its minima lie orthogonal to the original integration contour.
As a result, for each $n$ only one set of double points -- the ones that give the dominant contribution -- contribute to the double integral in \eqref{ene63a}.
Following the comment in the beginning of this section, at finite $\lambda_k$ we prescribe to integrate over the steepest descent contours of those double points that produce dominant contributions at small SL parameter.

\item
In \cite{Alexandrov:2004cg}, only the double points $(n,m,\sind)=(n,0,0)$ (for $k=1$) have been taken into account and shown to generate the instanton contributions \eqref{nonpert-n}.
They are distinguished by the fact that they all lie on the real line in the complex $\xp=\bar x_-$ plane so that the fermion momentum $p=(\xp-\xm)/\sqrt{2}$ vanishes.
In appendix \ref{ap-cutoff} we argue that taking into account the double points $(n,m,0)$ with $m\ne 0$, which were omitted in \cite{Alexandrov:2004cg}, may be responsible for cancellation of the cut-off dependence in the instanton normalization factor found in that paper.

\item
The second set of double points, $(n,m,1)$, can be obtained from the first one by a simple flip of the sign of the parameter $\lambda_k$, which generates a shift in the uniformization parameter by $\pi i R/k$
and exchanges $\theta^0_n$ and $\theta^1_n$.
Therefore, the instanton contributions associated to the two sets should also be related by this simple flip.
Actually, the instanton contributions corresponding to $\sind=1$ in \eqref{eq-thetas} were earlier found in \cite{Alexandrov:2003nn} as solutions of the linearized Toda equation, but they were dismissed because it was observed that for $n=k=1$, $1/2<R<1$ and large SL parameter, the values relevant for the description of (T-dual of) the black hole background, the instanton action becomes negative.
However, this is not true for other values of the parameters
and therefore, in general, we cannot disregard them.

\end{itemize}

Thus, we arrive at the following result for the scattering phase
\be
\label{phi-evalSL}
\phi=\phi_{\rm pert}
- \frac{k}{4\pi R\sqrt{2\pi}} \sum_{n=1}^\infty \Delta_n^{\bind(n)} \, e^{-\bS_n^{\bind(n)}},
\ee
where $\bS_n^\sind$ and $\Delta_n^\sind$ are the leading and subleading parts, respectively, of the contribution generated by the double point $(n,m,\sind)$, and
\be
\bind(n)=\left\{ \begin{array}{ll}
0 &\mbox { if } \bS_n^0<\bS_n^1
\\
1 & \mbox { if } \bS_n^0>\bS_n^1
\end{array}\right.
\quad \mbox{at } |\lambda_k|\ll 1.
\label{choicesaddle}
\ee
Finally, the factor $k$ comes from the fact that within the period $2\pi i R$ which we integrate over there are altogether $k$ saddle points producing dominant contributions, and the integration contour after a suitable deformation runs along the steepest descent contour of all of them.
The factor of $1/2$ appearing in \eqref{instcontr} is associated with the $\tau_+$ integration and accompanies the contribution from each of the $k$ saddles.
We shall now evaluate $\bS_n^\sind$ and $\Delta_n^\sind$ explicitly.

\subsubsection{Instanton action}

The Euclidean instanton action is given by
\be\label{esnfirst}
\bS_n^\sind=i\phi^{(0)}-i S\(\xp\bigl(\tauip{n,m}\bigr),\xm\bigl(\tauim{n,m}\bigr)\).
\ee
Using \eqref{Sxpxm}, \eqref{solvpmSL}, \eqref{phi0-SL} and \eqref{eq-thetas},
it is straightforward to obtain the following explicit expression
\be
\bS^\sind_n(\mu,\lambda_k)=2\mu \[\theta^\sind_n +(-1)^\sind \(\frac{2R}{k}-1\) a_k \,e^{-\scX}\sin\( \frac{k}{R}\,\theta^\sind_n\)\],
\label{Sn-SLk}
\ee
which for $\sind=0$ and $k=1$ reduces to \eqref{Sn-SL} found in \cite{Alexandrov:2003nn,Alexandrov:2003un,Alexandrov:2004cg}.

The quantity ${\bf S}_n^\sind$ can also be written in more geometric terms so that
the resulting expression is valid also in more general cases, as will be shown in the next section.
Using the expression for $\phi^{(0)}$ derived earlier, we can express \eqref{esnfirst} as:
\be\label{esnsecond}
\bS_n^\sind=i S(\xp(\tau_1), \xm(\tau_1))-i S\(\xp\bigl(\tauip{n,m}\bigr),\xm\bigl(\tauim{n,m}\bigr)\),
\ee
where $\tau_1$ is an arbitrary constant. Using \eqref{Sxpxmpre}, we can express this as
\be
\begin{split}
\bS_n^\sind =&\, i\, x_+(\tau_1)x_-(\tau_1) - i \,
\xp\bigl(\tauip{n,m}\bigr)\xm\bigl(\tauim{n,m}\bigr) + i \int_{\tau_1}^{\tauip{n,m}} \xm(\tau) \,\pxp(\tau)\, d\tau
\\
&\, + i \int_{\tau_1}^{\tauim{n,m}} \xp(\tau) \,\pxm(\tau)\, d\tau
=i \int_{\tauim{n,m}}^{\tauip{n,m}} x_-(\tau)\, \pxp(\tau)\, d\tau\, ,
\end{split}
\ee
where in arriving at the last expression we have used the condition on the double points \eqref{sp-eq2}
and integration by parts on the last term of the middle expression. This result can be written as an integral over a closed cycle on the complex curve\footnote{The leading perturbative contribution $\phi^{(0)}$
to the scattering phase
can also be given a similar geometric representation since, up to non-universal terms analytic in $\mu$,
it is equal to a similar integral along the non-compact cycle on the curve that coincides with the Fermi profile \cite{Alexandrov:2003qk}.}
\be
\bS^\sind_n=-i\oint_{\gamma_n^\sind} \xm d\xp\, ,
\label{Sinst-int}
\ee
where $\gamma_n^\sind$ is the image of the interval
$[\tauip{n,0},\tauim{n,0}]$
under the map $\tau\mapsto (\xp(\tau), \xm(\tau))$.
Using Stokes' theorem, one can also obtain another useful representation
\be
\bS_n^\sind=-i\int_{D[\gamma^\sind_n]} d\xm d\xp
=-i\int_0^\mu d\eps
\int_{\tauip{n,0}}^{\tauim{n,0}}
d\tau
=2 \int_0^{\mu} \theta^\sind_n\(\eps^{\frac{k}{2R}-1} t_k \) d\eps,
\label{actSn}
\ee
where $D[\gamma^\sind_n]$ is a disk in $\IC^2$ with the boundary given by $\gamma^\sind_n$
and $\eps$ and $\tau$ are related to $x_\pm$ via the
equations \eqref{XthroughLk}
and \eqref{xpmSLk} with $\mu$ replaced by $\eps$ and $\tau_\pm$
replaced by $\tau$ everywhere.
$\theta_n^\nu$ appearing in the integrand is computed from \refb{eq-thetas}, \refb{XthroughLk} after replacing $\lambda_k$ by $t_k \eps^{\frac{k}{2R}-1}$ everywhere.
In arriving at \eqref{actSn} we have used the fact that
due to \eqref{canontr},
the Jacobian for the change of variables from $(x_+,x_-)$ to $(\eps,\tau)$ is 1.

Now let us analyze which of the
saddle points labelled by $\sind$ for fixed $n$ gives the dominant contribution.
We will restrict ourselves to the leading instanton contribution with $n=1$ and assume that $\lambda_k>0$.
Then one should distinguish two parameter ranges: $1/2<R/k<1$ and $R/k>1$.\footnote{For $R/k<1/2$,
the vertex operator $\cV_{k/R}$ \eqref{genCFT} grows in the weak coupling region and therefore
it is not marginal and is non-renormalizable.
Therefore, we will not consider this range.}
The qualitative behavior in the two ranges is presented in Fig. \ref{fig-instaction}.
In the first case, for small $\lambda_k$, $\bS_1^0<  \bS_1^1$ while for large $\lambda_k$ the relation is inverted.
Furthermore, at some point $\bS_1^1$ becomes even negative  and one may worry about the consistency of our expansion.
However, as specified in \eqref{choicesaddle}, it is the behavior at small $\lambda_k$ that determines which saddle point contributes
to the integral and in this case this is the saddle labeled by $\sind=0$. Thus, the fact that $\bS_1^1<0$ for sufficiently large $\lambda_k$
is irrelevant because this saddle does not actually contribute. Note however that for this argument to be
valid, one needs to ensure that $\Delta_1^0$ remains
non-zero as we increase $t_k$ so that the steepest descent contour of this saddle
varies continuously.
We have checked numerically that this is indeed the case.

\twofig{$\bS_1^\sind$ as functions of the SL parameter in the two ranges $1/2<R/k<1$ and $R/k>1$, for $\lambda_k>0$. The blue curves
correspond to $\bS_1^0$ and the purple ones to $\bS_1^1$.}
{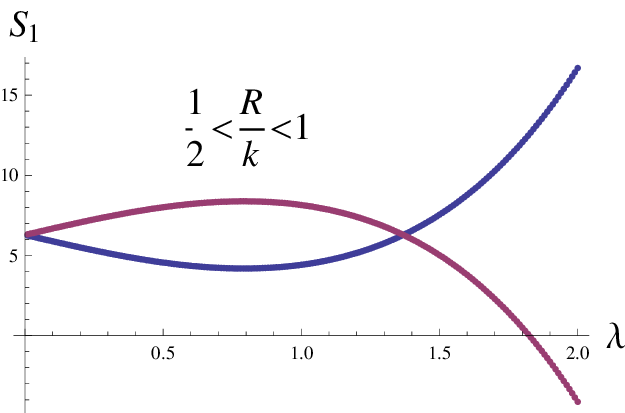}{7cm}{1cm}{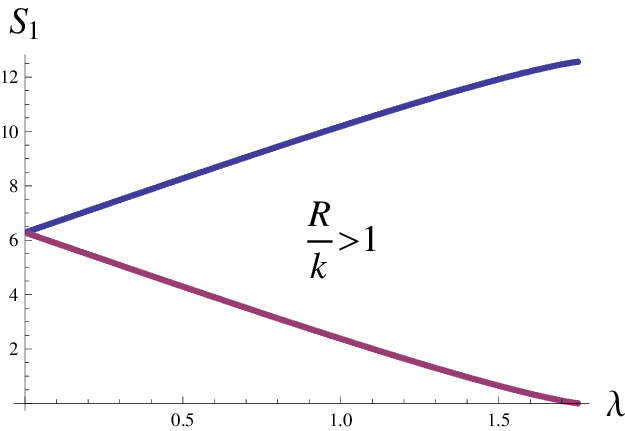}{7cm}{0.2cm}{fig-instaction}

In the range $R/k>1$, $\bS_1^0>  \bS_1^1$ for all $\lambda_k$ and thus it is the saddle point labeled by $1$ that is dominant and contributes to the free energy.
But again there is a point $\lambda_k^{\rm cr}$ where  $\bS_1^1$ vanishes which raises the same worries as in the previous case.
Fortunately, as we show in appendix \ref{ap-crit}, this is precisely the critical point where the system exhibits the $c=0$ critical behavior: at this point the $c =1$ scalar field settles into one of the minima of the SL potential and decouples, leaving behind a $c = 0$ system coupled to gravity \cite{Hsu:1992cm,Kazakov:2000pm}.
As a result, we never reach a point where ${\bf S}_1^1$ can become negative.
We have also checked that $\Delta_1^1$ remains non-zero all the way up to the critical point $\lambda_k^{\rm cr}$ so that the steepest descent contour associated with this saddle point varies continuously as we vary $\lambda_k$.

\subsubsection{Instanton prefactor}

Next, we compute the factor $\Delta_n^\sind$ in \eqref{phi-evalSL}.
As in the case of the perturbative saddle point, it receives two contributions: from the subleading terms in the integrand of \eqref{ene63a} multiplied by $e^{-\pi i/4}$ coming from the perturbative scattering phase, and from the Gaussian integral around the relevant instanton saddle.
As a result, it is given by
\be
\begin{split}
\Delta_n^\sind=&\,  2\pi\,e^{-\frac{\pi i}{4}}\,\sqrt{-\pxp\pxm}
\[(-i)^2\det\(\begin{array}{cc}
\frac{\p^2 S}{\p\tau_+^2} & \frac{\p^2 S}{\p\tau_+\p\tau_-}
\\
\frac{\p^2 S}{\p\tau_+\p\tau_-} & \frac{\p^2 S}{\p\tau_-^2}
\end{array}\)\]^{-1/2}
\\
=&\, 2\pi \Bigl[i\bigl(\pxm(\tau_+)\pxp(\tau_-)-\pxp(\tau_+)\pxm(\tau_-)\bigr) \Bigr]^{-1/2},
\end{split}
\label{Deltan}
\ee
where the right hand side is evaluated at $\tau_+=\tauip{n,0}$ and $\tau_-=\tauim{n,0}$, and
at the second step we have used the representation \eqref{Sxpxmpre}
to compute the derivatives of the effective action.
Substituting \eqref{xpmSLk} and using \refb{dbpoints-SL}, \refb{eq-thetas}, one obtains
\be
\Delta^\sind_n=\pi\mu^{-1/2}\sqrt{R/k}\,\frac{ e^{\scX/2}}{\sin(\theta^\sind_n)}
\[\cot(\theta^\sind_n)- \(\frac{k}{R}-1\) \cot\(\frac{k-R}{R}\, \theta^\sind_n\)\]^{-1/2}.
\label{Deltapm}
\ee

This completes the computation of the instanton corrections to the scattering phase in the first two orders in the large $\mu$ expansion.
However, we are interested in the free energy which is related to the scattering phase by \eqref{relFphi}.
When the differential operator on the left hand side of this relation acts on a non-perturbative term,
in our approximation it is sufficient to consider its action only on $\bS_n^\sind$.
Due to \eqref{actSn}, its effect is to bring down the factor
$-2\sin (\theta_n^\sind/R)$. Thus, the non-perturbative part of the free energy from
the $n$-th saddle point is found to be
\be
\cF^{(n)}_{\rm np}(\mu,\lambda_k)= A_n^{\bind(n)} \, e^{-\bS_n^{\bind(n)}},
\label{cFnp}
\ee
where $\bind(n)$ is defined in \eqref{choicesaddle},
$\bS_n^\sind$ is given in \eqref{Sn-SLk} and
\be
\begin{split}
A_n^\sind = &\, \frac{k}{8\pi R\sqrt{2\pi}} \,  \frac{\Delta_n^\sind}{\sin (\theta_n^\sind/R)}
\\
=&\, \frac{\mu^{-1/2}}{8 \sqrt{2\pi R/k}} \,
\frac{ e^{\scX/2}}{\sin(\theta^\sind_n)\sin (\theta_n^\sind/R)}
\[\cot(\theta^\sind_n)- \(\frac{k}{R}-1\) \cot\(\frac{k-R}{R}\, \theta^\sind_n\)\]^{-1/2}.
\end{split}
\label{res-An-SL}
\ee
Again, for $\sind=0$ and $k=1$, this function perfectly agrees with \eqref{resAnSL}.
But now we have also obtained a concrete finite value for the overall numerical coefficient
\be
C=\frac{1}{8 \sqrt{2\pi R/k}}\, .
\label{newC}
\ee
This is the main new result of this section. The other difference from \eqref{resAnSL} is that the
dominant saddle does not always correspond to $\nu=0$.

Although we have given the results for the $n$-th saddle point, only the $n=1$ contribution
is significant. This is due to the fact that we have only computed the leading order term in
$g_s$ expansion for $n=1$, and all higher order terms in $g_s$ expansion around this instanton which we omitted dominate over
the $n\ge 2$ contributions. Furthermore, in obtaining \eqref{enemaster} we have approximated
$\log(1+e^{-i\,\phi_{\rm pert}} \cJ)$ as $e^{-i\,\phi_{\rm pert}}\cJ$
and thus neglected multi-instanton contributions. In particular, powers of the $n=1$ contribution compete with the terms given above for $n\ge 2$.
In contrast, since $\phi_{\rm pert}$ is real and the $n=1$ contribution is pure imaginary,
they do not compete with each other and
the saddle corresponding to $n=1$ indeed gives the dominant contribution to the imaginary part
of the free energy.

\subsection{Small $\lambda_k$ limit}

To facilitate the comparison of the above results to string amplitudes in sections \ref{sec-instact} and \ref{sec-sublead},
we will expand them for small $\lambda_k$.

First, solving perturbatively \eqref{XthroughLk}, one finds
\be
\scX= \frac{k^2}{R^2}\(1-\frac{k}{R}\) \lambda_k^2 +\cO(\lambda_k^4),
\qquad
a_k=\frac{k}{R}\, \lambda_k +\cO(\lambda_k^3).
\ee
Using this in the equation \eqref{eq-thetas}, one obtains
\be
\theta^\sind_n= \pi n + (-1)^\sind \frac{k}{R}\,\lambda_k\,  \sin\(\frac{\pi k n}{R} \)
-\frac{k^2}{2R^2}\(1-\frac{k}{R}\) \lambda_k^2\, \sin\(\frac{2\pi k n}{R} \)  +\cO(\lambda_k^3).
\ee
Plugging these results into \eqref{Sn-SLk} and \eqref{res-An-SL}, one arrives at the following expansions
\bea
\hspace{-0.9cm}
\bS^\sind_n&=& \mu\biggl[2\pi n+ (-1)^\sind 4\lambda_k \sin\(\frac{\pi k n}{R} \)
+\frac{k^2}{R^2}\,\lambda_k^2\, \sin\(\frac{2\pi k n}{R} \) +\cO(\lambda_k^3)\biggr],
\label{SMQM-SL}
\\
\hspace{-0.9cm}
A^\sind_n
&=& \frac{(-1)^n}{8\sin\(\frac{\pi n}{R} \)}\,  \frac{\mu^{-1/2}[(-1)^\sind\lambda_k]^{-1/2}}{\sqrt{2\pi\sin\(\frac{\pi k n}{R} \)}}
\[1- (-1)^\sind
\frac{k\lambda_k}{R^2}\, \cot\(\frac{\pi n}{R}\) \sin\(\frac{\pi k n}{R} \) +\cO(\lambda_k^2)\] .
\label{AMQM-SL}
\eea
Note several features of these results:
\begin{itemize}
\item
The subleading contribution $A_n$ has a different scaling in $\mu\sim g_\str^{-1}$ compared
to the non-deformed theory \eqref{encFnp}.
\item
$A_n$ diverges for small SL parameter as $\lambda_k^{-1/2}$, and therefore the result is apparently
discontinuous at $\lambda_k=0$ where it is expected to be given by \eqref{eunperturbed}.
This can be traced to the fact that for $\lambda_k=0$, we had a continuous family of saddle points labelled by $\tau_-$ spanning the range
$0\le\tau_-\le 2\pi i R$, while for $\lambda_k\ne 0$ the $\tau_-$ integral becomes Gaussian of width
of order $(\mu\lambda_k)^{-1/2}$ for fixed $n$ and $k/R$.
Clearly, when the width of the Gaussian becomes comparable
to $2\pi R$, the current formula breaks down and we make a smooth transition to the earlier formula
valid for $\lambda_k=0$. The transition occurs at $\lambda_k\sim 1 / (\mu R^2)$.
\item
While $\bS_n$ depends on the radius only through the combination $R/k$, this is not true for $A_n$.
\item
From \eqref{SMQM-SL} we conclude that
\be
\bind(n) = \begin{cases}  0\ \ \hbox{if} \  \ \lambda_k \sin\(\frac{\pi k n}{R}\)<0\cr
1\ \ \hbox{if} \ \  \lambda_k \sin\(\frac{\pi k n}{R}\)>0
\end{cases}\, .
\label{sigman}
\ee
This implies that $A_n^{\bind(n)}$ is pure imaginary.
\end{itemize}

Finally, using \eqref{eunperturbed} and \eqref{AMQM-SL}, we can
give the result for the ratio of the leading non-perturbative corrections to the free energy
in the deformed and non-deformed theories at the leading order at small $\lambda_k$:
\be
r_1\equiv \lim_{\lambda_k\to 0} \frac{\cF_{\rm np}^{(1)}(\mu,\lambda_k)}{\cF_{\rm np}^{(1)}(\mu)}
=4i\sin\(\frac{\pi}{R}\)\lim_{\lambda_k\to 0} A_1^{\bind(1)}
=\frac{1}{2\sqrt{2\pi\mu}}\left|
\lambda_k \sin\(\frac{\pi k }{R}\)\right|^{-1/2}\, ,
\label{ratioLam}
\ee
where it is understood that we pick the contribution from the dominant saddle
specified by $\bind(1)$ in \eqref{sigman}.
Note that the ratio reduces to the ratio of the prefactors of the deformed and undeformed instanton contributions because ${\bf S}_1^{\bar\sind(1)}$ is smooth in the $\lambda_k\to 0$ limit.

\section{Generic perturbation and double SL}
\label{sec-MQMdSL}

In this section we generalize the previous analysis to arbitrary set of parameters $t_k$ appearing in the potentials \eqref{edeftpm}
and then illustrate the results in the case of just two non-vanishing parameters.
Since the derivation follows closely the one for the usual SL perturbation, we will not provide as much details as in the previous section, but just state the main results.

\subsection{Generic perturbation}
\label{subsec-MQMgenV}

Let us consider a perturbation generated by the potentials \eqref{edeftpm} with the only restriction $t_k=t_{-k}$.
Starting from the relation \eqref{SmatrixMQM}
and following exactly the same steps as in the previous section,
one can arrive at the following generalization of the results of section \ref{sec-MQMSL}:
\begin{itemize}

\item
Generalization of \eqref{enefou1} is given by
\be
\label{enefou1gen}
\begin{split}
e^{i\phi(-\mu)}=&\,
\frac{1}{\sqrt{2\pi}}
\int_0^\infty \frac{d\xp}{\xp} \,\cI(\xp,\xm),
\\
\cI(\xp,\xm)=&\, \exp\bigg[i\xp\xm -\(i\mu-\hf\) \log (\xp\xm) - i\sum_{k=1}^{\kmax} t_k\(\xp^{k/R}+\xm^{k/R}\)
\\
&\qquad - i v_+(\xp;-\mu)- i v_-(\xm;-\mu)\bigg] .
\end{split}
\ee

\item
Generalization of \eqref{xpmSLk}, defining the new variables $\tau_\pm$ in terms of
$x_\pm$, takes the form
\be
\xpm(\tau_\pm) =\sqrt{\mu} \, e^{\pm \tau_\pm-\hf\scX}
\(1+ \sum_{k=1}^{\kmax} a_k\, e^{\mp {k\over R}\, \tau_\pm} \)\, .
\label{xmptauX}
\ee

\item
Generalization of \eqref{XthroughLk} determining the coefficients $\scX$ and $a_k$ is
a system of algebraic equations
\bea
a_k+\!\!\sum_{l=k+1}^{k_{\max}}a_l a_{l-k}&=&\frac{k}{R}\, \lambda_k \, e^{\(1-\frac{k}{2R}\)\scX}
+\!\!\sum_{l=k+1}^{k_{\max}}\frac{l}{R}\, \lambda_l \, e^{\(1-\frac{l}{2R}\)\scX}
\!\!\!\!\sum_{\sum_{i=1}^n d_i=l-k}\!\!
\frac{\Gamma\(\frac{l}{R}+1\)}{n!\,\Gamma\(\frac{l}{R}+1-n\)}\prod_{i=1}^n a_{d_i},
\nn
\\
1+\sum_{l=1}^{k_{\max}}a_l^2&=& e^{\scX}+ \sum_{l=1}^{k_{\max}}\frac{l}{R}\, \lambda_l \, e^{\(1-\frac{l}{2R}\)\scX}
\sum_{\sum_{i=1}^n d_i=l}\frac{\Gamma\(\frac{l}{R}+1\)}{n!\,\Gamma\(\frac{l}{R}+1-n\)}\prod_{i=1}^n a_{d_i},
\label{eq-chiallk}
\eea
where the first equality holds for $k=1,\dots, k_{\max}$,
the sums go over ordered decompositions into $d_i\ge 1$,
and we used the parameters $\lambda_k$ defined in \eqref{deflamk}.
As we prove in appendix \ref{ap-proofalt}, with help of the first relation, the second one can be rewritten
in a much simpler form
\be
e^{\scX}=1+\sum_{l=1}^{k_{\max}}\(1-\frac{l}{R}\)a_l^2.
\label{eq-chiallk-simple}
\ee

\item
Generalization of \eqref{solvpmSL} and \eqref{vpm-sublead} take the form:
\ben\label{e1720}
v^{(0)}_\pm(x_\pm) &=&\mu \bigg[
\pm \tau_\pm +e^{-\scX}\sum_{k=1}^\km a_k
\( \frac{R}{k}\, e^{\pm\frac{k}{R}\tau_\pm}- \(\frac{R}{k}-1\) e^{\mp\frac{k}{R}\tau_\pm}\)
\bigg]
\non\\
&&
+e^{-\scX}\sum_{k,\ell=1\atop k\ne \ell}^\km a_k a_\ell \(1-{k\over R}\) {R\over \ell-k}\,
e^{\pm(\ell-k)\tau_\pm/R}
- \log (x_\pm/\sqrt\mu)- \sum_{k=1}^\km  \lambda_k
\,   (x_\pm/\sqrt\mu)^{k/R}
\non\\
&&
-\frac{\scX}{2}+ \sum_{k=1}^\km
 \lambda_k e^{-k\scX/2R}\sum_{n=1}^k {1\over n!}
\!\sum_{\sum_{i=1}^n d_i=k}\!\!
\frac{\Gamma\(\frac{k}{R}+1\)}{\Gamma\(\frac{k}{R}-n+1\)}\prod_{i=1}^n a_{d_i}\, ,
\non\\
v^{(1)}_\pm &=&\mp {i\over 2}  \log \[{1 + \sum_{k=1}^\km a_k \(1-{k\over R}\) e^{\mp k\tau_\pm/R}\over
1 + \sum_{\ell=1}^\km a_\ell\, e^{\mp \ell\tau_\pm/R}}
\]\, .
\een
\item The saddle point equation of the $x_+$ integral in
\eqref{enefou1gen} for fixed $x_-$, obtained by extremizing the leading part of the exponential with respect to $x_+$,
takes the same form as
\eqref{spe-xpxm}:
\be\label{expmgen}
x_-(\tau_+) = x_-(\tau_-)\, ,
\ee
with the perturbative solution being $\tau_+=\tau_-$.
\item
As a consistency check, one can verify that the contribution to the integral in \eqref{enefou1gen}
from
the perturbative saddle point $\tau_+=\tau_-$ is independent of $\tau_-$ and gives
the following result for the leading part of the scattering phase
\be\label{e47phifrel}
\begin{split}
\phi^{(0)}(-\mu)=&\, -\mu\log\mu +\mu\Biggl[ \scX +e^{-\scX} \(1+\sum_{k=1}^{\kmax} a_k^2\)
\\
&\,
-2\sum_{k=1}^{\kmax}
\lambda_k e^{-\frac{k}{2R}\scX}\sum_{n=1}^k
\frac{\Gamma\(\frac{k}{R}+1\)}{n!\,\Gamma\(\frac{k}{R}-n+1\)}\sum_{\sum_{i=1}^n d_i=k}\prod_{i=1}^n a_{d_i}\Biggr]\, .
\end{split}
\ee

\item
For instanton contribution one arrives at the
same expression \eqref{ene63a} for the scattering phase
\be\label{ene63agen}
\phi=\phi_{\rm pert}
- \frac{e^{-i\,\phi_{\rm pert} }}{4\pi R\sqrt{2\pi}}
\int\limits_0^{2\pi iR} d\tau_-
\int\limits_{\rm inst}d\tau_+ \(-\pxp\pxm\)^{1/2} e^{iS(\xp,\xm)},
\ee
where the effective action takes again the same form as in \eqref{Sxpxmpre},
\be
\begin{split}
S(\xp,\xm) =&\,
\xp\xm- \int^{\xp}_{\xp(\tau_0)}\Xm(\yp)\,d\yp
-\int^{\xm}_{\xm(-\tau_0)} \Xp(\ym)\,d\ym
\\
=&\,  \xp\xm- \int^{\tau_+}_{\tau_0} \xm(\tau)\,\pxp(\tau)\, d\tau
-\int^{\tau_-}_{-\tau_0} \xp(\tau)\,\pxm(\tau)\, d\tau\, .
\end{split}
\label{Sxpxmgen}
\ee
We have not written down the value of $\tau_0$ since it will not be needed for calculating
the difference $S(\xp,\xm)-\phi_{\rm pert}$.

\item
The saddle points of the double integral in (4.8) are given by solutions to the equations identical to \eqref{sp-eq2}:
\be
\xp(\tau_+)=\xp(\tau_-),
\qquad
\xm(\tau_+)=\xm(\tau_-),
\label{sp-eq2gen}
\ee
but with $x_\pm$ given by \eqref{xmptauX}.
\end{itemize}

It is obvious from \eqref{sp-eq2gen} that the instanton saddles coincide
with the double points of the complex curve.
To find the double points, let us make the ansatz
\be
\tau_\pm=i\(R\zeta \mp \theta\).
\label{doublepoint-allm}
\ee
Then for real $(\zeta,\theta)$ the two equations \eqref{sp-eq2gen} become
complex conjugates of each other and the resulting complex equation can be conveniently
rewritten  as two real ones
\be
\begin{split}
&
\sum\limits_{k=1}^{\kmax} a_{k}\, \sin (k\zeta)\sin\(\frac{k-R}{R}\,\theta\)=0\, ,
\\
&
\sin\theta =\sum\limits_{k=1}^{\kmax} a_{k}\, \cos (k\zeta)\sin\(\frac{k-R}{R}\,\theta\) \, .
\end{split}
\label{eqthetam}
\ee
One can make a few general comments on their solutions:
\begin{itemize}
\item
Let $k_0=\gcd\{k \ : t_k\ne 0\}$. Then an obvious solution to \eqref{eqthetam} is
\be
\begin{split}
\zeta_m=&\, \pi m/ k_0,
\\
\sin(\theta_{n,m}) =&\, \sum\limits_{k=1}^{\kmax}(-1)^{mk/k_0}\, a_{k}\, \sin\(\frac{k-R}{R}\,\theta_{n,m}\),
\qquad \theta_{n,m}|_{t_k=0}=\pi n,
\end{split}
\label{gendp}
\ee
where $m\in \IZ$, $n\in \IN$ and $\theta_{n,m}$ depends on $m$ only mod 2.
This solution is a direct generalization of the double point \eqref{dbpoints-SL} and exists for any set of parameters $t_k$.

\item
In general, there are additional solutions, whose existence however can depend on the values of parameters.
An explicit example of such kind of double points will be considered in the next subsection.

\item The transformation $\zeta\to\zeta + 2\pi/k_0$ is a symmetry of the equations \eqref{eqthetam}.
Acting on the set \eqref{gendp}, it shifts the value of $m$ by 2. Therefore, it is a general feature that
the instanton contribution from the saddle labelled by $(n,m)$ depends on $m$ only via the combination $m$ mod 2.

\item
The transformation $\zeta\to 2\pi -\zeta$ is also a symmetry of the equations \eqref{eqthetam}.
Acting on the set \eqref{gendp}, it exchanges the double points labeled by $(n,m)$ and $(n,2k_0-m)$
and therefore preserves whether the second index is odd or even.
Since these double points are already related by the shift symmetry mentioned above, in this case
this symmetry does not lead to anything new. However,
for more general solutions it could relate double points that are not related
by any other symmetry.

\end{itemize}

Let us now evaluate a contribution from a solution to \eqref{eqthetam} to the free energy.
As usual, it takes the form
\be
A(\zeta,\theta) \, e^{-\bS_{\rm inst}(\zeta,\theta)}.
\label{nonpert-gen}
\ee
Here $\bS_{\rm inst}$ is given by ($i$ times) the difference of
$S(x_+, x_-)$ given in \eqref{Sxpxmgen}
evaluated at the perturbative saddle $\tau_+=\tau_-$ and the instanton
saddle \eqref{doublepoint-allm}. The final result, generalizing \eqref{Sn-SLk},
is\footnote{Naively, one also obtains an imaginary contribution.
However, rewriting it using only trigonometric functions with arguments $k\zeta$ and $\(\frac{k}{R}-1\)\theta$,
it can be shown to actually vanish.}
\be
\begin{split}
\bS_{\rm inst}(\zeta,\theta)=&\, 2\mu\left[\theta+ e^{-\scX}\left\{\sum_k\(\frac{2R}{k}-1\)a_k\, \cos(k\zeta)\,\sin\(\frac{k}{R}\, \theta\)
\right.\right.
\\
&\, \left.\left.
+\sum_{k\ne l}\frac{2R-(k+l)}{2(k-l)}\, a_k a_l \,\cos\((k-l)\zeta\)\sin\(\frac{k-l}{R}\, \theta\) \right\}\right],
\end{split}
\label{Sinst-all}
\ee
and can be rewritten in an elegant geometric form similarly to \eqref{Sinst-int} and \eqref{actSn}
\be
\bS_{\rm inst}(\zeta,\theta)=-i\oint_{\gamma} \xm d\xp=2 \int_0^{\mu} \theta\(\eps^{\frac{k}{2R}-1} t_k \) d\eps,
\label{Sinst-int-gen}
\ee
where $\gamma$ is a closed cycle on the complex curve obtained as the image of the interval $( i(R\zeta-\theta), i(R\zeta+\theta))$ under the map $\tau\mapsto (\xp(\tau), \xm(\tau))$.
The generalization of \eqref{res-An-SL} for the prefactor is
\be
A(\zeta,\theta)=\(-\frac{1}{2\sin (\theta/R)}\)\(-\frac{\Delta(\zeta,\theta)}{4\pi R\sqrt{2\pi}} \) ,
\ee
where the first factor comes from inverting the relation \eqref{relFphi} with help of \eqref{Sinst-int-gen}
and the second factor is the subleading contribution to the second term in \eqref{ene63agen}.
As in \eqref{Deltan}, the function $\Delta(\zeta,\theta)$ can be found to be
\be
\Delta(\zeta,\theta)= 2\pi \Bigl[i\bigl(\pxm(\tau_+)\pxp(\tau_-)-\pxp(\tau_+)\pxm(\tau_-)\bigr) \Bigr]^{-1/2}\, ,
\label{Delta-gen}
\ee
where $\tau_\pm$ are given by \eqref{doublepoint-allm}.
Collecting all factors together and substituting \eqref{xmptauX}, one obtains
an explicit result
\be
\label{eDeltam}
A(\zeta,\theta)=  \frac{\mu^{-1/2}\, e^{\scX/2} }{4 R\sqrt{2\pi}\, \sin(\theta/R)}\,
\[2\Im\(
e^{2i\theta}\prod_{\sigma=\pm} \( 1 +\sum_{k=1}^{\kmax} a_k \(1 - {k\over R}\) e^{ik \(\sigma \zeta-{\theta\over R}\)}\)
\)\]^{-1/2}\!\!\! .
\ee

Finally, the instanton contributions \eqref{nonpert-gen} should be summed over all (relevant) solutions of \eqref{eqthetam}
with $\zeta\in[0,2\pi)$. In particular, since these contributions are invariant under $\zeta\to \zeta + 2\pi/k_0$,
which is also a symmetry of \eqref{eqthetam}, there are at least $k_0$ identical contributions of each type.
If one takes into account also the symmetry $\zeta\to 2\pi -\zeta$, one gets an additional factor of 2,
which is however absent for the instantons associated to \eqref{gendp}
since in this case the symmetry reshuffles the contributions that have been already counted.

\subsection{Double sine-Liouville}
\label{subsec-dSL}

Let us now illustrate the general results of the previous subsection on a particular example
where there are only two non-vanishing parameters $t_k$ and $t_{2k}$ which we dub ``double sine-Liouville".
In this case the equations \eqref{eq-chiallk} and \eqref{eq-chiallk-simple} reduce to
\be
\begin{split}
&
a_{2k}= \frac{2k}{R}\, \lambda_{2k} \, e^{\(1-\frac{k}{R}\)\scX},
\qquad
a_k=\frac{\frac{k}{R}\, \lambda_k \, e^{\(1-\frac{k}{2R}\)\scX}}
{1+\frac{2k}{R}\(1-\frac{2k}{R}\) \lambda_{2k} \, e^{\(1-\frac{k}{R}\)\scX}}\, ,
\\
&\qquad\qquad
e^{\scX}=1+\(1-\frac{k}{R}\)a_k^2+\(1-\frac{2k}{R}\)a_{2k}^2,
\end{split}
\label{eqchi-2SL}
\ee
while the first equation in \eqref{eqthetam} takes a factorized form
\be
\sin(k\zeta)\[ a_k \sin\(\frac{k-R}{R}\,\theta\)+2a_{2k}
\cos(k\zeta)\, \sin\(\frac{2k-R}{R}\,\theta\)\]=0.
\label{eqzeta-2SL}
\ee
This implies that the MQM complex curve has two different sets of double points and hence in this theory
there are two types of instanton effects. We will consider them now one by one.

\subsubsection{First set}

The first set is the one described in \eqref{gendp} which exists for any values of the
SL parameters. It has $\zeta_m=\pi m/k$, $m\in \IZ$,
and since the corresponding solution for $\theta$ and the instanton corrections depend only on whether $m$ is even or odd, we
will label them by $\sind= m \mod 2$, similarly to the usual SL case. In particular, we have
\be
\sin(\theta^\sind_n) =(-1)^\sind  a_k \sin\(\frac{k-R}{R}\,\theta^\sind_n\)+a_{2k}\sin\(\frac{2k-R}{R}\,\theta^\sind_n\),
\label{eqtheta-2SL}
\ee
where as usual different solutions are distinguished by the
`initial condition' at vanishing values of the parameters:
$\theta^\sind_n|_{\lambda_k=\lambda_{2k}=0}=\pi n$.
The instanton corrections to the free energy take the same form as in \eqref{cFnp} where $\bS_n^\sind$
is obtained from \eqref{Sinst-all} as
\be
\bS^\sind_n =  2\mu\Biggl\{\theta^\sind_n\,
+e^{-\scX}\biggl[\(\frac{R}{k}-1\)a_{2k}\sin\(\frac{2k}{R}\,\theta^\sind_n\)
+(-1)^\sind\(\frac{2R}{k}-1+\(\frac{2R}{k}-3\)a_{2k}\)a_k\sin\(\frac{k}{R}\,\theta^\sind_n\)\biggr]\Biggr\}\, .
\label{Smn-2SL-1}
\ee
The prefactor is found as a special case of \eqref{eDeltam}
with an additional factor $k$ counting the number of the double points within the period $2\pi i R$,
\bea
A^\sind_n&=&  \frac{\mu^{-1/2}\sqrt{k/R}\, e^{\scX/2} }{8\sqrt{2\pi}\, \sin(\theta^\sind_n/R)}
\Biggl\{\[2a_{2k} \sin\(\(\frac{2k}{R}-1\) \theta^\sind_n\)+(-1)^\sind a_k \sin\(\frac{k-R}{R}\, \theta^\sind_n\)\]
\label{Delta-2SL-1}\\
&\times & \biggl[\cos(\theta^\sind_n)
-\(\frac{2k}{R}-1\) a_{2k} \cos\(\frac{2k-R}{R}\, \theta^\sind_n\)
-(-1)^\sind\(\frac{k}{R}-1\) a_k \cos\(\frac{k-R}{R}\, \theta^\sind_n\)
\biggr] \Biggr\}^{-1/2}.
\non
\eea

In the limit of small $\lambda_k$ and $\lambda_{2k}$ (with their ratio fixed), one finds
\be
\begin{split}
a_{2k}=&\,\frac{2k}{R}\, \lambda_{2k} +\cO(\lambda^3)
\qquad
a_k=\frac{k}{R}\, \lambda_k-\frac{2k^2}{R^2}\(1-\frac{2k}{R}\) \lambda_k\,\lambda_{2k} +\cO(\lambda^3),
\\
&\quad
e^{\scX}= 1+\frac{k^2}{R^2}\(1-\frac{k}{R}\)\lambda_{k}^2
+\frac{4k^2}{R^2}\(1-\frac{2k}{R}\) \lambda_{2k}^2 +\cO(\lambda^3).
\end{split}
\ee
Using this in the equation \eqref{eqtheta-2SL}, one obtains
\be
\begin{split}
\theta^\sind_n= &\,\pi n +(-1)^\sind\frac{k}{R}\,\lambda_k\,  \sin\(\frac{\pi k n}{R} \)
+\frac{2k}{R}\, \lambda_{2k} \sin\(\frac{2\pi k n}{R} \)
\\
&\,
-\frac{k^2}{2R^2}\(1-\frac{k}{R}\) \lambda_k^2 \sin\(\frac{2\pi k n}{R} \)
-\frac{2k^2}{R^2}\(1-\frac{2k}{R}\) \lambda_{2k}^2 \sin\(\frac{4\pi k n}{R} \)
\\
&\,
-(-1)^\sind \frac{2k^2}{R^2}\(2-\frac{3k}{R}\) \lambda_k \,\lambda_{2k}\cos\(\frac{\pi k n}{R} \)\sin\(\frac{2\pi k n}{R} \)
+\cO(\lambda^3).
\end{split}
\ee
Plugging these results into \eqref{Smn-2SL-1} and \eqref{Delta-2SL-1}, one arrives at the following expansions
\be
\begin{split}
\bS^\sind_n=&\, \mu \biggl[ 2\pi n +(-1)^\sind 4 \lambda_k \sin\(\frac{\pi nk}{R}\)
+4\lambda_{2k} \sin\(\frac{2\pi k n}{R} \)
+\frac{k^2}{R^2}\,\lambda_k^2 \sin\(\frac{2\pi n k}{R}\)
\label{Smn-2SL-1-exp}\\
&\,
+\frac{4k^2}{R^2}\,\lambda_{2k}^2 \sin\(\frac{4\pi k n}{R} \)
+(-1)^\sind 16\frac{k^2}{R^2}\, \lambda_k \lambda_{2k}\cos^2\(\frac{\pi k n}{R} \)\sin\(\frac{\pi k n}{R} \)
+\cO(\lambda^3)\biggr],
\end{split}
\ee
\bea
A^\sind_n
&=& \frac{(-1)^n\mu^{-1/2}}{8\sin\(\frac{\pi n}{R} \)}\(2\pi\sin\(\frac{\pi k n}{R} \)\)^{-1/2}
\((-1)^\sind \lambda_k +8\lambda_{2k} \cos\(\frac{\pi kn}{R}\) \)^{-1/2}
\nn\\
&\times&
\Biggl[1-\frac{k}{R^2}\, \cot\(\frac{\pi n}{R}\) \sin\(\frac{\pi k n}{R} \)
\((-1)^\sind\lambda_k +4\lambda_{2k}\cos\(\frac{\pi kn}{R}\)\)
\label{resAn-2SL-1-exp}\\
&&\qquad
-\frac{\frac{2k^2}{R^2}\, \lambda_k \lambda_{2k} \cos^2\(\frac{\pi kn}{R}\)}
{ \lambda_k +(-1)^\sind 8\lambda_{2k}\cos\(\frac{\pi kn}{R}\)}+\cO(\lambda^2)
\Biggr].
\nn
\eea

\subsubsection{Second set}

The second set of double points corresponds to vanishing of the second factor in \eqref{eqzeta-2SL}.
Using the resulting relation in the second equation in \eqref{eqthetam}, one finds that there is a two-parameter
set of such double points characterized by $(\zeta_{n,m},\theta_n)$ which satisfy
\be
\cos(k\zeta_{n,m})=-\frac{a_k}{2a_{2k}}\, \frac{\sin\(\frac{R-k}{R}\,\theta_{n}\)}{\sin\(\frac{R-2k}{R}\,\theta_{n}\)}\, ,
\qquad
\sin(\theta_n)=a_{2k} \sin\(\frac{R-2k}{R}\,\theta_n\).
\label{2SL-2set}
\ee
Note that the equation on $\theta$ does not depend on $\zeta$. This is why $\theta_n$ carries only one index
distinguishing the initial conditions at vanishing values of the parameters: $\theta_n|_{\lambda_k=\lambda_{2k}=0}=\pi n$.
The index $m$ on $\zeta_{n,m}$ labels solutions related by the symmetries $\zeta\to \zeta+2\pi/k$ and $\zeta\to 2\pi-\zeta$,
so that on an interval of length $2\pi$ there are $2k$ such solutions. Since the different
saddle points labelled by $m$ are related by the symmetry transformations, the contribution
to the integral from these saddle points will be independent of $m$.

\begin{figure}[t]
\centerline{\includegraphics[width=5.5cm]{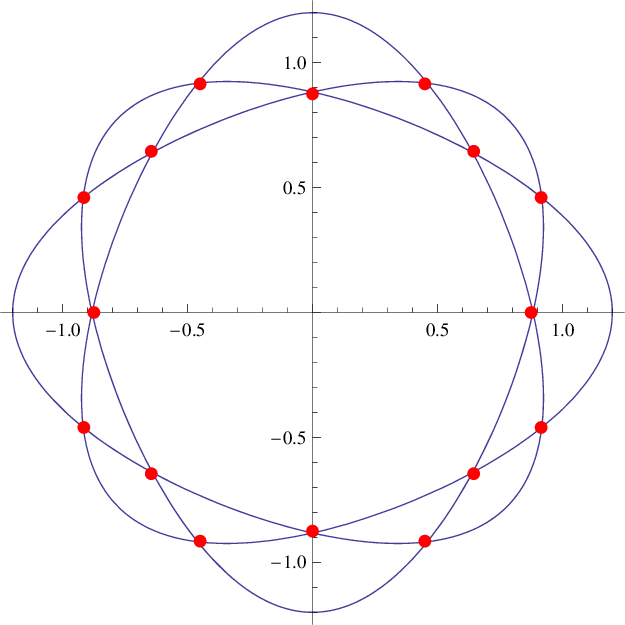}\hspace{0.3cm}\includegraphics[width=5.5cm]{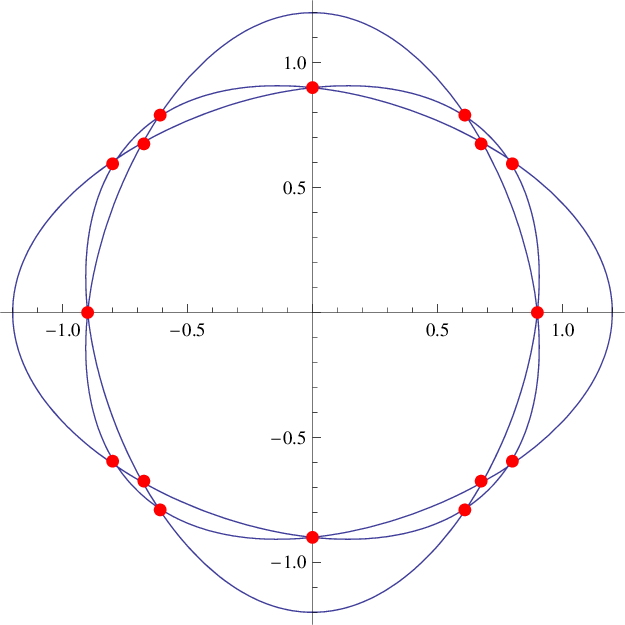}
\hspace{0.3cm}\includegraphics[width=5.5cm]{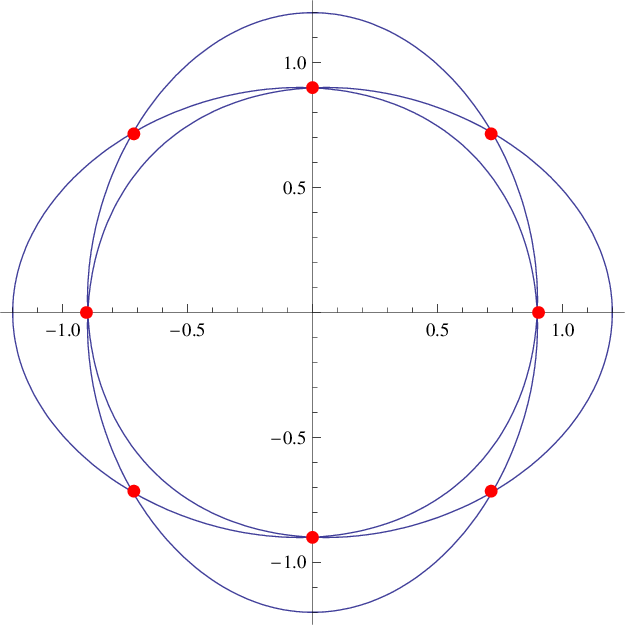}}
\vspace{0.2cm}
\caption{Section of the MQM complex curve of dSL for $R/k=3/4$ and various values of the parameters $\lambda_k$ and $\lambda_{2k}$
shown in the order of decreasing ratio $\lambda_{2k}/\lambda_k$.
In the last picture one has passed a critical point where three double points merge together and leave behind only one.
\label{fig-double}}
\end{figure}

Note that decreasing $|\lambda_{2k}|$, one eventually reaches a critical point $\lambda_{2k}^{\rm cr}$
where the right hand side of the first equation becomes larger than 1 and the solution
ceases to exist.
We illustrate this situation in Fig. \ref{fig-double}.
One can see how three double points, two of the second set and one of the first, approach and then merge together
turning into a single double point. In fact, this is a generic situation as the condition for the critical point $\cos(k\zeta)=\pm1$
coincides with the equation on $\zeta$ from the first set. Physically, at this point nothing dramatic happens to the system
except that, exactly at $|\lambda_{2k}|=\lambda_{2k}^{\rm cr}$, the one-loop determinant vanishes signaling
the emergence of a zero mode of the effective action. As a result, in the neighborhood of the critical point
the above analysis becomes inapplicable.

Let us now assume that $|\lambda_{2k}|>\lambda_{2k}^{\rm cr}$ and evaluate the instanton effects generated by
the solutions of \eqref{2SL-2set}. The contribution of the $n$-th set of saddles
to the free energy can be written as
\be
A_n \, e^{-\bS_n}\, ,
\ee
where $\bS_n$ and $A_n$ are obtained from \eqref{Sinst-all} and \eqref{eDeltam},
respectively, upon specializing to our case and using \eqref{2SL-2set}.
Besides, $A_n$ should include the factor $2k$, the number of the double points within the period $2\pi i R$.
This results in
\be
\bS_{n}=  2\mu\Biggl\{\theta_{n}
- e^{-\scX}\Biggl[\(\frac{R}{k}-1\)a_{2k}\sin\(\frac{2k}{R}\,\theta_{n}\)
+\(\frac{R}{k}+\(\frac{R}{k}-2\)a_{2k}\)\frac{a_k^2}{a_{2k}}\,
\frac{\sin\(\frac{k}{R}\,\theta_{n}\)\sin\(\frac{R-k}{R}\,\theta_{n}\)}{2\sin\(\frac{R-2k}{R}\,\theta_{n}\)}\Biggr]\Biggr\},
\label{Smn-2SL-2}
\ee
\be
\begin{split}
A_n=&\,  \frac{\mu^{-1/2}\sqrt{k/R}\, e^{\scX/2} }{4\sqrt{2\pi}\, \sin(\theta/R)}
\Biggl\{
\[\frac{a_k^2}{2a_{2k}}\, \frac{\sin^2\(\frac{k-R}{R}\, \theta_n\)}{\sin\(\frac{2k-R}{R}\,\theta_n\)}
-2a_{2k} \sin\(\frac{2k-R}{R}\,\theta_n\)\]
\\
&\,\quad \times\[\cos(\theta_n)
+\(\frac{2k}{R}-1\) a_{2k} \cos\(\frac{2k-R}{R}\, \theta_n\)\]\Biggr\}^{-1/2}.
\end{split}
\label{Delta-2SL-2}
\ee

In the limit of small $\lambda_k$ and $\lambda_{2k}$ (with their ratio fixed), one finds
\be
\theta_{n}=
\pi n-\frac{2k}{R}\, \lambda_{2k}\sin\(\frac{2\pi k n}{R} \)
-\frac{2 k^2}{R^2}\(1-\frac{2k}{R}\) \lambda_{2k}^2 \sin\(\frac{4\pi k n}{R} \)
+\cO(\lambda^3),
\ee
\be
\cos(k\zeta_{n,m})=
-\frac{\lambda_k}{8\lambda_{2k}\cos\(\frac{\pi k n}{R}\)}
-\frac{k^2 }{2R^2} \,\lambda_k \, \cos \(\frac{\pi k n}{R} \)
+\cO(\lambda^2).
\ee
Substituting these results into \eqref{Smn-2SL-2} and \eqref{Delta-2SL-2}, one gets
\be
\bS_n= \mu\[ 2\pi n
-4\lambda_{2k}\sin\(\frac{2\pi k n}{R} \)
-\frac{\lambda_k^2}{4\lambda_{2k}}\, \tan\(\frac{\pi k n}{R}\)
+\frac{4k^2}{R^2}\,\lambda_{2k}^2 \sin\(\frac{4\pi k n}{R} \)
+\cO(\lambda^3)\],
\label{Smn-2SL-2-exp}
\ee
\be
\begin{split}
A_n= &\,
\frac{(-1)^n\mu^{-1/2}}{4\sqrt{2\pi}\sin\(\frac{\pi n}{R} \)}
\(\frac{\lambda_k^2}{8\lambda_{2k}}\, \tan \(\frac{\pi k n}{R}\) -4\lambda_{2k} \sin\(\frac{2\pi k n}{R}\) \)^{-1/2}
\label{resAn-2SL-2-exp}\\
&\,\times
\Biggl[1+\frac{2k}{R^2}\, \lambda_{2k}\cot\(\frac{\pi n}{R}\) \sin\(\frac{2\pi k n}{R} \)
-\frac{\frac{4k^2}{R^2}\,  \lambda_{2k}\cos^2\(\frac{\pi kn}{R}\)}
{1 -\frac{64\lambda_{2k}^2}{\lambda_k^2} \cos^2\(\frac{\pi kn}{R}\)}+\cO(\lambda^2)
\Biggr].
\end{split}
\ee

\bigskip

Comparing the first order terms in \eqref{Smn-2SL-1-exp} and \eqref{Smn-2SL-2-exp},
we can find which of the two sets of double points generates the dominant contributions.
One obtains
\be
\bS^{\sind}_n-\bS_n=8\lambda_{2k} \sin\(\frac{2\pi k n}{R} \)
\[1+\frac{(-1)^\sind \lambda_k}{8\lambda_{2k} \cos\(\frac{\pi nk}{R}\)}\]^2
+\cO(\lambda^2),
\ee
which implies that if $\lambda_{2k} \sin\(\frac{2\pi k n}{R} \)$ is positive
then the dominant contribution is due to the second set. Otherwise
it is the first set that is dominant.
Furthermore, it is straightforward to check that the condition
\be
|\lambda_{2k}|>\lambda_{2k}^{\rm cr}
=\left|\frac{\lambda_k}{8\cos\(\frac{\pi nk}{R}\)}\right|
+\cO(\lambda^2)
\ee
ensuring the existence of the two sets of double points, also ensures
that the prefactor of the dominant instanton contribution, \eqref{resAn-2SL-1-exp}
or \eqref{resAn-2SL-2-exp}, is pure imaginary.
If it is spoiled, the second set of double points does not exist anymore
and we return to the situation described in section \ref{sec-MQMSL}
where a half of double points from the first set competes with the other half.

\section{D-instanton action in string theory}
\label{sec-instact}

Our goal in this section is to compute the D-instanton action in the sine-Liouville
theory using CFT technique and compare it with the results of the matrix model analysis described in
section \ref{sec-MQMSL}. We will restrict ourselves to the case of (1,1) ZZ boundary conditions for the Liouville field,
although a generalization to $(1,n)$ ZZ branes is straightforward.\footnote{Up to first order in the SL parameter
it has already been done in \cite{Alexandrov:2003un}.}

\subsection{String field theory conventions}
\label{sec-sftconventions}

We begin with a brief recollection of the string field theory results  from section 3.3 of \cite{Sen:2021tpp}
that we shall be using. Let $\psi_c=c\bar c V_c$ be an on-shell closed string field with $V_c$ being a dimension (1,1) primary in the matter plus Liouville sector. Then the string field theory one-point function of  $\psi_c$ is given by
\be\label{erev1}
A_{\rm disk}(\psi_c)= {T\over 2}\, \langle c_0^- \psi_c\rangle\, ,
\qquad
c_0^-\equiv {1\over 2} (c_0-\bar c_0)\, ,
\ee
where $\langle\, \cdot \,\rangle$ denotes upper half plane correlation function and $T$ is the Euclidean action of the D-instanton.
The string field theory two-point function of $\psi_c$ and an on-shell open string field $\psi_o$ is given by
\be\label{erev2}
A_{\rm disk}(\psi_c \psi_o)=
i\pi T \langle
\psi_c \psi_o
\rangle\, .
\ee
The factor of $i$ was found in appendix A of \cite{Alexandrov:2021shf}. The string field theory three-point function of $\psi_c$ and a pair of on-shell open string states $\psi_o^{(1)}=c \cV_o^{(1)}$ and $\psi_o^{(2)}=c \cV_o^{(2)}$ is given by
\be \label{eCOO}
A_{\rm disk}(\psi_c \psi_o^{(1)} \psi_o^{(2)})=
i\pi T \int du\left\langle
\psi_c(i) \psi_o
^{(1)}(0) V^{(2)}_o(u)\right\rangle ,
\ee
where the integral over $u$ runs along the boundary of the upper half plane, i.e.\ along the real axis.
Inserting more on-shell open string vertex operators will lead to insertion of more factors of $\int du V_o(u)$ into the correlation function.
Finally, the effect of inserting a $\psi_c$ into an existing amplitude with required number of integrated and unintegrated vertex operators is to insert an integrated vertex operator
\be\label{erev3}
\int {dx dy\over \pi}\, V_c
\ee
into the correlation function.

We shall also need the correct normalization of the string field theory two-point function of two on-shell closed string vertex operators $c\bar c V_c^{(1)}$ and $c\bar c V_c^{(2)}$. The result for this, analyzed in appendix A of \cite{Eniceicu:2022xvk}, is
\be\label{erev4}
A_{\rm disk}(c\bar c V_c^{(1)}\, c\bar c V_c^{(2)}) =
{i\, T\over 2} \int_0^1 dy\, \left\langle c\bar c V_c^{(1)}(i)\, (c(z)+\bar c(\bar z))
V_c^{(2)}(z,\bar z)\right\rangle\, , \qquad z\equiv iy\, .
\ee

Finally, the results for the sphere amplitude can be read out from section 3.1 of \cite{Sen:2021tpp}.
The on-shell three point function takes the form:
\be
4\, \kappa^{-2} \left\langle c\bar c V_c^{(1)}\, c\bar c V_c^{(2)}\, c\bar c V_c^{(3)}\right\rangle_{\rm sphere},
\label{3ptspheresft}
\ee
where $\kappa$ is the closed string coupling constant.
Its relation to the D-instanton tension $T$ needs
to be determined by comparing the decomposition of the annulus partition function into the closed
and open string channels, and depends on the specific D-instanton we are considering. Every additional
external closed string state requires insertion of integrated vertex operator given in \eqref{erev3}.

\subsection{Normalization of the coupling constants}

Let us now fix some normalizations.
In the undeformed theory the instanton action is given by
\be
T = {1\over g_\str} = 2\pi\mu\, ,
\label{relmugs}
\ee
which defines $g_s$ and establishes its relation to the matrix model parameter $\mu$.
Note that this relation could receive corrections. However, since it is fixed by comparing perturbative amplitudes
and since perturbation theory gives a power series expansion in $g_\str^2$,
this relation does not get modified at order $g_\str$.

Next, we describe the sine-Liouville deformation
as the result of adding the following operator to the world-sheet action
\be\label{edeform}
\repla_k \int d^2 z \, \cV_{k/R} (z,\bar z),
\qquad
d^2 z = dx dy \quad \mbox{and $z=x+iy$}\, ,
\ee
where the precise normalization of the vertex operator $\cV_\omega$ introduced in  \eqref{genCFT}
is fixed by specifying the one-point disk amplitude
\be
\label{ecorrln}
\langle \cV_\omega(i)\rangle =\sin(\pi \omega) \, \cos(\omega x),
\ee
with $x$ being the D-instanton location along the time direction.
Various numerical factors and a power of $\mu$ that usually appear in this amplitude have been absorbed into
the overall normalization of $\cV_\omega$.
The ghost part of the disk amplitude is normalized as
\be\label{eghostcor}
\langle c(z_1) c(z_2) c(z_3)\rangle_{\rm ghost}=-(z_1-z_2)(z_2-z_3)(z_1-z_3)\, .
\ee

Using \eqref{erev1}, \eqref{ecorrln} and \eqref{eghostcor}, we get
\be\label{eadiskV}
\begin{split}
A_{\rm disk}(c\bar c\cV_\omega) = &\, {T\over 2}\,  \langle c_0^- c\bar c \cV_\omega(i)\rangle
=\frac{ T}{4}\left\langle (\p c(i)-\bar \p \bar c(i)) c\bar c \cV_\omega(i)\right\rangle
\\
=&\, 2\,  T\sin(\pi\omega) \, \cos(\omega x)\, .
\end{split}
\ee
This amplitude can be used to establish a relation between the string theory parameter $\repla_k$ and
the matrix model parameter $\lambda_k$. In appendix \ref{sappnorm} it is shown that
in the normalization of the vertex operator in which \eqref{eadiskV} holds,
$\repla_k$ and  $\lambda_k$ are related by
\be
\lambda_k = \pi^2 \, \repla_k\, .
\label{identlam}
\ee

\subsection{First order correction} \label{sfirst-order}

Deforming the world-sheet action by \eqref{edeform} corresponds to inserting into the world-sheet correlation function the following factor
\be
\exp\left[-\repla_k \int d^2 z \, \cV_{k/R} (z,\bar z)\right]
= \sum_n { (-\repla)^n\over n!}
\left( \int d^2 z \, \cV_{k/R} (z,\bar z)\right)^n\, .
\ee
Expectation value of this operator on the disk, regarded as a function of the instanton position $x$ along the Euclidean time direction, may be viewed as a contribution to $-V_{\rm eff}(x)$ where $V_{\rm eff}(x)$ is the effective potential for the D-instanton position. Using \eqref{erev3}, we see that every insertion of $-\repla_k\int d^2 z \cV_{k/R}(z,\bar z)$ corresponds to insertion of a string field $-\pi\repla_k c\bar c \cV_{k/R}$. Hence we can express $V_{\rm eff}(x)$ as
\be\label{e513}
V_{\rm eff}(x)=V^{(1)}(x)+V^{(2)}(x) + \cdots\, ,
\ee
where $V^{(n)}$ denotes the correction of order $(\repla_k)^n$:
\be
V^{(n)}=(-1)^{n-1} \frac{(\pi\repla_k)^{\, n}}{n!}\, A_{\rm disk}\((c\bar c\cV_{k/R})^n\).
\ee
In particular, due to \eqref{eadiskV}, the order $\repla_k$ contribution is given by
\be\label{ev1result}
V^{(1)}(x)=2 \pi  \repla_k\, T\sin{\pi k\over R} \cos{kx\over R}\, .
\ee

The potential has $k$ minima and $k$ maxima at
\be
x^\sind_m=\pi(2m+\sind)R/k,
\qquad
m=0,\dots,k-1, \quad \sind=0,1.
\label{maxmin}
\ee
Whether $\sind=0$ or $\sind=1$ corresponds to the minima
depends on the sign of $\repla_k\,\sin{\pi k\over R}$.
Adding the value of the potential at the extrema to the action $T$ of
the instanton in the undeformed theory, we see that in the presence of deformation we
have two different types of instantons with action $T+ \Delta T_\sind^{(1)}$, where
\be\label{rdetltat1}
{\Delta T_\sind^{(1)}\over T}=(-1)^\sind\, 2 \pi \repla_k \sin{\pi k\over R}\, .
\ee
This result perfectly agrees with \eqref{SMQM-SL} taken for $n=1$ provided we use the relation
\eqref{identlam}. Furthermore, this shows that the critical points $x^\sind_m$ correspond to the double points
of the MQM complex curve labelled by $(1,m,\sind)$. Since the dominant saddle corresponds
to negative $\Delta T_\sind^{(1)}$, we have at the dominant saddle
\be \label{edominant}
(-1)^\sind\,  \repla_k \sin{\pi k\over R} <0\, .
\ee
This is the same condition as \eqref{sigman}.

As will be discussed in section \ref{sec-sublead}, eventually we need to integrate over the instanton position $x$.
One can give an argument similar to that in MQM that the  integration contour along the real $x$ axis can be expressed as the union of the steepest descent contours of the dominant saddle points. Indeed, since the dominant saddle points describe local maxima of the integrand $e^{-V_{\rm eff}}$, the integrand falls off along the real $x$ axis away from these saddle points and therefore the steepest descent contours of the dominant saddles are directed along the integration contour of $x$. In contrast, the steepest descent contours of the subdominant saddles are along the imaginary $x$ axis and are orthogonal to the $x$ integration contour. The reader would have noticed that this argument is identical to that in MQM, with the role of $x$ being played by $i\tau_-$ in MQM. This feature continues to hold in more general situations, and the general result is that the $x$ integration contour lies along the steepest descent contours of the local minima of $V_{\rm eff}$ .

\subsection{Second order correction}

We can also calculate the correction to the action to the order $\repla_k^2$ in terms
of the disk two-point function.  It is given by
\be\label{e3.6a}
V^{(2)} = -{\pi^2\over 2} \, \repla_k^2 \, A_{\rm disk}(c\bar c\cV_{k/R}\, c\bar c\cV_{k/R})\, ,
\ee
where, using \eqref{erev4},
\be\label{e3.7a}
A_{\rm disk}(c\bar c\cV_{\omega_1}\, c\bar c\cV_{\omega_2}) =
{i\over 2 g_\str} \int_0^1 dy\, \left\langle c\bar c \cV_{\omega_1}(i)\, (c(z)+\bar c(\bar z))
\cV_{\omega_2}(z,\bar z)\right\rangle ,
\qquad z\equiv iy\, .
\ee
This integral however has a divergence as $y\to 0$ which needs to be regulated using string field theory.
This has been done in \cite{Balthazar:2019rnh,Sen:2020eck}. Let us define
\be
f_E(\omega_1,\omega_2)={g_\str^{-1}\, A_{\rm disk}(c\bar c\cU_{\omega_1} \, c\bar c\cU_{\omega_2})\over
A_{\rm disk}(c\bar c\cU_{\omega_1})A_{\rm disk}(c\bar c\cU_{\omega_2}) }\, ,
\label{deffE}
\ee
where $\cU_{\omega}$ is obtained from $\cV_\omega$ by replacing
the $c=1$ matter part $\cos(\omega X)$ by $e^{i\omega X}$.
Then we have
\be\label{eratiodisk}
{ A_{\rm disk}(c\bar c\cV_\omega \, c\bar c\cV_\omega)\over (A_{\rm disk}(c\bar c\cV_\omega))^2 }
= {g_\str\over 4\cos^2{\omega x}}\, \Bigl[ f_E(\omega, \omega)\, e^{2i\omega x}
+ f_E(-\omega, -\omega)\, e^{-2i\omega x} + 2 f_E(\omega, -\omega)\Bigr]\, .
\ee
An explicit finite integral representation for the function $f_E(\omega_1,\omega_2)$ is given in
appendix \ref{ap-2pf-disk}.
As is discussed there, an important feature of this representation is that it involves the additive contribution
$-\omega_1 \omega_2 \log\beta^2$ depending on an arbitrary constant parameter $\beta$,
representing a field redefinition ambiguity of string field theory. Physical results however should be $\beta$-independent.

Using \eqref{eadiskV}, \eqref{e3.6a} and \eqref{eratiodisk}, as well as $T=1/g_\str$ and the
evenness of $f_E(\omega_1,\omega_2)$ under $\omega_i\to -\omega_i$, we get
\be\label{et2inter}
{V^{(2)}}(x) =-\pi^2 \, T\, \repla_k^2\,\sin^2\({\pi k\over R}\) \, \[ f_E(\omega,\omega)\, \cos{2k x\over R}
+f_E(\omega,-\omega)\] ,
\qquad
\omega=k/R.
\ee
Note that this correction does not destroy the extrema \eqref{maxmin} of the effective action.
Therefore, the order $\repla_k^2$ shift in the D-instanton action is obtained by evaluating $V^{(2)}(x^\sind_m)$.
This gives,
\be\label{et2fin}
{\Delta T_\sind^{(2)}\over T} =-  \pi^2 \repla_k^2\,\sin^2\({\pi k\over R}\)
\Bigl[ f_E(\omega,\omega)+f_E(\omega,-\omega)\Bigr].
\ee
Importantly, this combination is independent of the parameter $\beta$, as is expected for physical quantities.
Although an analytic expression of $f_E(\omega_1,\omega_2)$ is not known in string theory, its numerical evaluation shows that it satisfies a set of remarkably simple relations \eqref{relfE}.
In particular, the first relation \eqref{epred} ensures that
\be
\label{resDT2}
{\Delta T_\sind^{(2)}\over T} =  \frac{\pi^3 k^2}{2R^2}\,  \repla_k^2\,\sin\({2\pi k\over R}\),
\ee
which is perfectly consistent with the matrix model prediction found in \eqref{SMQM-SL} after using \eqref{relmugs} and \eqref{identlam}.
Thus, we found a perfect match between instanton actions in MQM and string theory up to second order in the SL deformation parameter.

\section{Normalization of the instanton contribution in string theory}
\label{sec-sublead}

In this section we analyze the overall normalization of the  D-instanton contribution to the amplitudes in string theory, given by the exponential of the annulus partition function.
First, we evaluate it in the undeformed background.
Next, we find how it is modified by the SL perturbation at the leading order in the perturbation in the parameter $\repla_k$.
Then we extract the second order correction.

\subsection{Unperturbed theory}

Let us first consider the subleading instanton contribution in the unperturbed theory.
It coincides with the partition function of the ZZ instanton which is given by
\be\label{eff0}
A_0={1\over 2} \, \exp\left[ \int_0^\infty{dt\over 2t}\, Z(t)\right],
\qquad
Z(t) = (e^{2\pi t}-1) \sum_{n=-\infty}^\infty
e^{-2\pi t n^2 R^2}\, .
\ee
We have included a multiplicative factor of 1/2 since the full integration contour contains only half of the steepest descent contour passing through the D-instanton saddle point \cite{Sen:2021qdk}.
The $n$-th term in the sum in \eqref{eff0} is the contribution from the open strings on the D-instanton that wind $n$ times along the compact time direction.
Particular attention must be paid to the $n=0$ term because the integral over $t$ diverges as $t\to\infty$.
This divergence can be treated using open string field theory and yields \cite[Eq.(3.16)]{Sen:2021qdk}
\be\label{eff1}
\exp\left[\int {dt\over 2t} \(e^{2\pi t}-1\)\right] = {i\over 4\pi^2}\int dx = {i R\over 2\pi}\, .
\ee
The contribution from the $n\ne 0$ terms gives a finite integral, at least for $R>1$, which is evaluated as follows \cite[Eq.(3.16)]{Eniceicu:2022nay}
\be
\begin{split}
\label{eff2}
\exp\left[\sum_{n\ne 0}\int {dt\over 2t} \(e^{2\pi t(1 - n^2 R^2)}-e^{-2\pi t n^2 R^2}\)\right]
=&\,  \prod_{n\ne 0} \left({n^2 R^2 \over n^2R^2-1} \right)^{1/2}
\\
=&\, \prod_{n=1}^\infty{1\over 1 - {1\over n^2 R^2}}
= {\pi/R\over \sin(\pi/R)}\, .
\end{split}
\ee
Substituting \eqref{eff1} and \eqref{eff2} into \eqref{eff0},  we get
\be \label{eF0a}
A_0 ={i\over 4\sin(\pi/R)}\, ,
\ee
which agrees with the $n=1$ term in the first sum in \eqref{encFnp} up to a sign.
In fact, the sign of \eqref{eff1} is ambiguous and depends on the particular choice of the integration contour.

\subsection{First non-trivial order perturbation}

We shall now compute the correction to the exponential of the annulus partition function to the first non-trivial order in $\repla_k$.
A crucial observation is that the zero mode associated with the D-instanton
position $x$ is lifted in the deformed theory.
Let us therefore carefully analyze the effect of this lifting.

In the undeformed theory, let $\xi$ be the zero mode of the open string field theory
that describes the freedom of translating  the D-instanton along the Euclidean time direction.
In order to evaluate the $\xi$ integral, we need a relation between $\xi$ and the position $x$ of the D-instanton
along the Euclidean time direction. It is given by \cite{Sen:2021qdk,Sen:2021tpp}
\be\label{e26}
d\, \xi = {1\over g_o \pi\sqrt 2} \, d x \, ,
\ee
where the open string coupling constant $g_o$ is related to the tension $T$ of the instanton
via the relation
\be\label{etexp}
T = {1\over 2\pi^2 g_o^2}\, .
\ee
Therefore, the contribution from the integration over $\xi$ is given by
\be \label{efinfactor}
\int {d\xi\over \sqrt{2\pi}} =
\sqrt{\frac{T}{2\pi}}\int dx
= ({2\pi T})^{1/2} R\, .
\ee
This goes in as a factor in the final expression given on the right hand side
of \eqref{eff1}.

In the deformed theory, the mode $\xi$ acquires a mass $h$ given by the $L_0$ eigenvalue of the
corresponding vertex operator. As a result, the contribution to the partition
function from this mode is now given by
\be
\int {d\xi\over \sqrt{2\pi}} \,e^{-{1\over 2} h \xi^2} = h^{-1/2}\, .
\label{contrib0}
\ee
Comparing \eqref{efinfactor} and \eqref{contrib0},
and taking into account that in the deformed theory there are $k$ critical points on the circle that give identical contributions,
we find that
the ratio of the zero mode contribution in the deformed theory to that in the
undeformed theory is equal to
\be
r_1=\frac{k}{R}\, (2\, \pi\, h\, T)^{-1/2} \, ,
\label{rstr}
\ee
where the index 1 indicates that this result corresponds to the $(1,1)$ ZZ boundary condition.

It remains to evaluate the induced mass $h$ of the zero mode.
The direct string theory calculation involves computing
the quadratic term in $\xi$ that appears in the effective action in the
presence of the deformation \eqref{edeform}. It is proportional to the disk amplitude
with one closed string vertex operator $c\bar c V$ and a pair of open string vertex
operators $ i\sqrt 2 c\p X$, and its calculation is described in appendix \ref{sa}.
Here instead we shall calculate $h$ by an indirect method based on the
tachyon potential obtained up to second order in the SL parameter in \eqref{ev1result} and \eqref{et2inter}.
For this we use the fact that the string field theory action,
expanded to quadratic order in $\xi$, gives the result $h\xi^2/2$.
This can be compared with the potential $V_{\rm eff}(x)$ expanded to quadratic order in $(x-x_c)$ where
$x_c$ describes an extremum of the potential. Thus, we have
\be\label{e610a}
{1\over 2} V_{\rm eff}''(x_c) (x-x_c)^2 = {1\over 2} \, h\, \xi^2
= \frac{h}{4\pi^2 g_o^2}\, (x-x_c)^2,
\ee
where in the last step we have used \eqref{e26}. Using \eqref{etexp}, \eqref{ev1result} and
\eqref{et2inter} and substituting for the critical point $x^\sind_m$ \eqref{maxmin}, we now get
\be\label{ehfin}
h^\sind = {V_{\rm eff}''(x^\sind_m)\over T} =(-1)^{\sind+1} \, 2\, \pi\,  \repla_k\,\frac{k^2}{ R^2}\, \sin\({\pi k\over R} \)
+4\pi^2\repla_k^2\, \frac{k^2}{R^2}\, \sin^2\({\pi k\over R}\)
f_E(\omega,\omega)+{\cal O}(\repla_k^3)\, ,
\ee
where $\omega=k/R$.

Substituting the order $\repla_k$ contribution to \eqref{ehfin} into \eqref{rstr},
one finds
\be
r_1^\sind=\frac{k}{R}\, (2\, \pi\, h^\sind\, T)^{-1/2}
=(2\pi)^{-3/2}\[(-1)^{\sind+1}\repla_k\sin\(\frac{\pi k}{R}\)\]^{-1/2}\mu^{-1/2},
\label{SFTr1}
\ee
where in the last equality we used \eqref{relmugs}. Using \eqref{edominant} we
see that at the dominant saddle,
\be
r_1^\sind=(2\pi)^{-3/2}\left|\repla_k\sin\(\frac{\pi k}{R}\)\right|^{-1/2}\mu^{-1/2}\, .
\ee
This result agrees with \eqref{ratioLam},  provided we use the relation \eqref{identlam}.

The presented derivation sheds light on the apparent singularity of the instanton contributions in the SL theory
at small $\lambda_k$ noticed in \cite{Alexandrov:2004cg}: it arises due to the lifting of the zero mode of the instanton
with the mass parameter proportional to $\lambda_k$. Furthermore, the divergence in the limit $\lambda_k\to 0$
is an artifact of the approximation used to get the above results. We see from \eqref{e610a} that the $x$ integral
is Gaussian of width of order $g_o/\sqrt h\sim (\mu\repla_k)^{-1/2}$ for fixed $k/R$. When this becomes
of order $2\pi R$, which is the range of $x$ integration, the Gaussian approximation breaks down.
This occurs at $\repla_k\sim 1/(\mu R^2)$, in agreement with the MQM results. 
Once $\lambda_k$ becomes sufficiently small, the expansion must be rearranged and instead of first expanding in powers of $g_\str$
and then in $\lambda_k$, the procedure should be reversed. This is precisely what is done in the MQM framework in appendix \ref{ap-Smatrix}
and, as is expected, the results show no sign of singularity.

\subsection{Second order contribution}
\label{ssecond}

Note that in the computation of the previous subsection we have used only the order $\repla_k$ term in the
expression for $h$. The inclusion of the quadratic order term is meaningful only when we also
take into account the effect of order $\repla_k$ shift in the conformal weights of the
non-zero modes in the sum in \eqref{eff0}. We shall now describe a procedure for doing this.

First, let us note that the expression for the quadratic correction to $h$, as given
in \eqref{ehfin}, is not quite correct since the relation \eqref{e26} between $\xi$ and $x$ can also get
corrections of order $\repla_k$. We shall deal with this by expressing the integral over
$\xi$ as an integral over $x$, picking up a Jacobian factor. In that case
$\exp(-h\xi^2/2)$ will be replaced by $\exp(-V_{\rm eff}''(x_c) (x-x_c)^2 /2)$ and we can perform the Gaussian
integral over $x$, picking up a factor proportional to $\left(V_{\rm eff}''(x_c)\right)^{-1/2}$, while the Jacobian from the
change of variable will give an additional contribution that can be regarded as a part of the
annulus amplitude as described in \cite{Sen:2020eck}. A similar Jacobian factor
is involved in relating the string field theory gauge transformation parameter and the $U(1)$
transformation parameter on the D-instanton, but it is also included in the annulus amplitude.

To facilitate the analysis, it will be useful to first consider a toy model where we have two open
string modes -- a zero mode $x$ and a non-zero mode $y$ in the undeformed background. Let us suppose
that in the presence of a background closed string field $\lambda$, the potential takes the form:
\be
V = {1\over 2} \(C_1 y^2 + C_2 \lambda x^2 + C_3 \lambda^2 x^2 + C_4 \lambda y^2\) .
\ee
Then after changing variables from
string field theory degrees of freedom to $x,y$ and suppressing the corresponding Jacobian factors, the partition function in the deformed theory is given by
\be
\begin{split}
\int {dx\over \sqrt{2\pi}} \, {dy\over \sqrt{2\pi}}\, e^{-V}
=&\,  (C_2\lambda + C_3 \lambda^2)^{-1/2} (C_1+C_4\lambda)^{-1/2}
\\
\approx &\, (C_1 C_2 \lambda)^{-1/2} \left(1 -{C_3\over 2C_2}\,\lambda\right)
\left( 1 - {C_4\over 2 C_1}\,\lambda\right) .
\end{split}
\ee
The first factor on the right hand side of
this expression has the interpretation of the first non-trivial order
result discussed in the last subsection. The second factor can be identified as the result of order
$\lambda^2$ correction to the effective potential for $x$ in the presence of the deformation.
In particular, the quantity $C_3/C_2$ can be interpreted as the ratio of the order $\lambda^2$ and order $\lambda$ correction
to $V''(x)$ at the extremum.
The third factor is the contribution due to the correction to the mass of the non-zero mode.
We can interpret the quantity $C_4/2C_1$  as tadpole diagram in the undeformed theory with
an external closed string field $\lambda$ and the $y$
field propagating in the loop. The factor of $C_4$ comes from the $\lambda$-$y$-$y$
interaction vertex, the factor of $1/C_1$ comes from the $y$ propagator
and the factor of 1/2 comes from the standard symmetry factor of the tadpole
diagram. In full string theory the factor $C_4/2C_1$ will be replaced by the annulus one-point function
in the undeformed theory,
with the contribution from the massless internal modes removed.

Thus, in full string theory, we need to evaluate two factors.
One is the effect of the order $\repla_k^2$ correction in \eqref{ehfin} on  the $x$ integral, which
produces the following factor
\be\label{efirst}
\left(1 +(-1)^{\sind+1} 2\pi \repla_k \sin\(\frac{\pi k}{R}\) f_E(\omega,\omega)\right)^{-1/2}
\simeq 1 +(-1)^\sind
\pi \repla_k \sin\(\frac{\pi k}{R}\) f_E(\omega,\omega),
\ee
where as usual $\omega=k/R$.
The second is obtained from
the annulus one-point function of the external closed string after removing the
zero mode contribution.
Denoting the annulus one-point function of $c\bar c \cV_\omega$  by $\cA(c\bar c\cV_\omega)$,
we can express the second factor as
\be\label{esecond}
1-\pi\repla_k  \, \cA(c\bar c\cV_\omega) .
\ee
Following \cite{Sen:2020eck}, we shall include in this the contribution from the
Jacobian factors from the change of integration
variables from $\xi$ to $x$ and from the string field theory gauge transformation parameter
to the U(1) transformation parameter,  both computed to order $\repla_k$.
As a result, the net extra contribution to the normalization of the partition function is now obtained by
taking the product of \eqref{efirst} and \eqref{esecond}, which can be written as
\be\label{eannulusexpect}
1- \pi\repla_k \, \cA(c\bar c\cV_\omega) +(-1)^\sind \pi \repla_k\,\sin \(\frac{\pi k}{R}\) f_E(\omega,\omega)
\, .
\ee

Let $g_\str\,g_R(\omega)$ be the ratio
\be\label{e3.17}
g_\str\, g_R(\omega) = \frac{\cA(c\bar c\cU_{\omega})}{A_{\rm disk}(c\bar c\cU_{\omega})} \, .
\ee
Then using the symmetry $\omega\to -\omega$ of $g_R(\omega)$,
\eqref{eadiskV} and \refb{relmugs}, one gets
\be
\cA(c\bar c \cV_\omega)
= 2\, g_R(\omega)\, \sin(\pi\omega)\, \cos(\omega\, x)\, .
\label{onepA}
\ee
Evaluating this result at $x^\sind_m$ given in \eqref{maxmin} and substituting into \eqref{eannulusexpect}, one obtains
\be\label{ecomb}
1+(-1)^{\sind+1}\pi\repla_k \, \sin\(\frac{\pi k}{R}\)  \Bigl(2 g_R(\omega)-  f_E(\omega,\omega) \Bigr)\, .
\ee
Due to \eqref{identlam}, this factor agrees with the one appearing in \eqref{AMQM-SL} to order $\repla_k$,
provided the following relation is satisfied
\be\label{e322a}
2 \,g_R(\omega) - f_E(\omega,\omega) =
\frac{\pi\omega}{R}\, \cot\frac{\pi}{R}\, .
\ee

Let us decompose the function $g_R(\omega)$ into two pieces
\be\label{egsplit}
g_R(\omega)=g_E(\omega)+\Delta g_E(\omega),
\ee
where $g_E(\omega)$ is the result for the ratio \eqref{e3.17} at $R=\infty$ keeping $\omega$ fixed
and $\Delta g_E(\omega)$ is the correction due to finite $R$.
Thus, at $R= \infty$, the relation \eqref{e322a} reduces to
\be
2g_E(\omega) - f_E(\omega,\omega)= \omega.
\label{2gfa}
\ee
It follows from the analysis of \cite{Balthazar:2019rnh,Sen:2020eck} that $g_E(\omega)$ is given by the integral in \eqref{egfiniteeuclid}.
Note that like $f_E(\omega_1,\omega_2)$, the function $g_E(\omega)$ involves a $\beta$-ambiguity in the form of an additive term $-{1\over 2}\omega^2 \log\beta^2$, but this ambiguity is cancelled in the combination of $f_E$ and $g_E$ appearing in \eqref{2gfa},
as is expected for all physical quantities.
In appendix \ref{ap-1an-disk}, eq.\refb{2gf}, we show that the relation \eqref{2gfa} actually follows
from the results of \cite{Sen:2020eck} and the relation \eqref{epred}.
We have also checked that it holds numerically.

Substituting \eqref{2gfa} into \eqref{e322a}, we find that
the finite $R$ correction to the annulus one-point function should take the form
\be
\Delta g_E(\omega)= \hf\,\omega\(\frac{\pi}{R}\, \cot\frac{\pi}{R}-1\).
\label{Delga}
\ee
We have derived its analytic integral expression in appendix \ref{ap-anR} and discussed numerical and analytical tests of \eqref{Delga} in appendices
\ref{ap-anR} and \ref{sdergexp}. This confirms agreement between
instanton effects in MQM and in string theory up to subleading order in $g_\str$
and quadratic order in the SL deformation parameter.

Note that if we did not introduce the additional integer parameter $k$ labelling the SL perturbation,
we would not be able to take the limit $R\to \infty$ keeping $\omega$ fixed. In that case it would be impossible to distinguish between $R$ factors
responsible for the $\omega$-dependence  and for finite radius effects.

\section{Double sine-Liouville}
\label{sec-dSL}

In this section we repeat the analysis of the previous sections for the double SL perturbation,
considered in section \ref{subsec-dSL} in the MQM formalism and corresponding to
the presence of two non-vanishing couplings
\be
\lam\int d^2 z \, \cV_\omega (z,\bar z)+\lamm \int d^2 z \, \cV_{2\omega} (z,\bar z),
\ee
where $\omega=k/R$. In this case, to the second order in the parameters $\lam$ and $\lamm$,
the effective action is given by a simple generalization of \eqref{ev1result} and \eqref{et2inter}
\be
\begin{split}
V_{\rm eff}(x)=&\, 2\pi T\Bigl[\lam \sin(\pi\omega)\cos(\omega x)+\lamm \sin(2\pi\omega)\cos(2\omega x)
\\
&\,
-\frac{\pi}{2}\, \lam^2\sin^2(\pi\omega)\Bigl(f_E(\omega,\omega)\cos(2\omega x)+f_E(\omega,-\omega)\Bigr)
\\
&\,
-\frac{\pi}{2}\, \lamm^2\sin^2(2\pi\omega)\Bigl(f_E(2\omega,2\omega)\cos(4\omega x)+ f_E(2\omega,-2\omega)\Bigr)
\\
&\,
-\pi \lam\lamm \sin(\pi\omega)\sin(2\pi\omega)\Bigl(f_E(\omega,2\omega)\cos(3\omega x)+ f_E(\omega,-2\omega)\cos(\omega x)\Bigr)
\Bigr].
\end{split}
\label{V2-2SL}
\ee
Its derivative is
\be
\begin{split}
V_{\rm eff}'(x)=&\,2\pi T \omega \sin(\pi\omega)\sin(\omega x)
\Bigl[-\lam -8\lamm \cos(\pi\omega)\cos(\omega x)
\\
&\,
+2\pi \lam^2\sin(\pi\omega)f_E(\omega,\omega)\cos(\omega x)
\\
&\,
+16\pi \lamm^2\cos(\pi\omega)\sin(2\pi\omega)f_E(2\omega,2\omega)\cos(\omega x)\cos(2\omega x)
\\
&\,
+\pi \lam\lamm \sin(2\pi\omega)\Bigl(f_E(\omega,-2\omega)+3 f_E(\omega,2\omega)(4\cos^2(\omega x)-1)\Bigr)
\Bigr].
\end{split}
\label{dV2-2SL}
\ee
Thus, there are two types of critical points.

\subsection{First type of critical points}

The simplest solution to vanishing of \eqref{dV2-2SL} is given by
\be
x_m=\pi m/\omega, \qquad m=0,\dots, 2k-1.
\ee
At the critical point, the potential takes the following value
\be
\begin{split}
V_{\rm eff}(x_m) =&\, 2\pi T\Bigl[
(-1)^m \lam \sin(\pi\omega)+\lamm\sin(2\pi\omega)
\\
&\,
-\frac{\pi}{2}\, \lam^2 \sin^2(\pi\omega)\Bigl( f_E(\omega,\omega)+f_E(\omega,-\omega)\Bigr)
\\
&\,
-\frac{\pi}{2}\, \lamm^2 \sin^2(2\pi\omega)\Bigl( f_E(2\omega,2\omega)+f_E(2\omega,-2\omega)\Bigr)
\\
&\,
-(-1)^m \pi \lam\lamm \sin(\pi\omega)\sin(2\pi\omega)\Bigl( f_E(\omega,2\omega)+f_E(\omega,-2\omega)\Bigr)
\Bigr]
\\
=&\,
4\pi^2\mu\Bigl[
(-1)^m \lam \sin(\pi\omega)+\lamm\sin(2\pi\omega)\Bigr]
\\
&\,
+\pi^4 \omega^2\mu\Bigl[ \lam^2 \sin(2\pi\omega)
+4\lamm^2 \sin(4\pi\omega)
+16(-1)^m \lam \lamm \cos^2(\pi\omega)\sin(\pi\omega)\Bigr],
\end{split}
\ee
where we used the relations \eqref{epred} and \eqref{identf12} satisfied by the two-point disk amplitude,
as has been confirmed by its numerical evaluation.
This result reproduces the expansion \eqref{Smn-2SL-1-exp} of the instanton action
under the following identifications
\be
\sind =m\mod 2, \qquad
\lam=\pi^{-2} \lambda_k,
\qquad
\lamm=\pi^{-2} \lambda_{2k},
\label{ident-tlam}
\ee
which are perfectly consistent with \eqref{identlam}
and with the identification of critical and double points in the previous section.

Next, we compute
\be
\begin{split}
h_m=&\, \frac{V_{\rm eff}''(x_m)}{T}=2\pi \omega^2\Bigl[(-1)^{m+1}\lam \sin(\pi\omega)-4\lamm\sin(2\pi\omega)
\\
&\,
+2\pi\lam^2\sin^2(\pi\omega)f_E(\omega,\omega)
+8\pi \lamm^2\sin^2(2\pi\omega)f_E(2\omega,2\omega)
\\
&\,
+(-1)^m\pi\lam \lamm\sin(\pi\omega)\sin(2\pi\omega)\Bigl(9f_E(\omega,2\omega)+f_E(\omega,-2\omega)\Bigr)
\Bigr].
\end{split}
\ee
Thus, the ratio of the zero mode contributions in the perturbed and non-perturbed theory is given by
\be
r_{1,m}=\frac{k}{R}\(2\pi T h_m\)^{-1/2}
=\frac{(2\pi)^{-3/2}}{\sqrt{\mu \sin(\pi\omega)}}\,\Bigl((-1)^{m+1}\lam-8\lamm\cos(\pi\omega)+\cO(\repla^2)\Bigr)^{-1/2},
\label{rm-2SL1}
\ee
where the factor $k$ comes from the fact that the critical points with even/odd $m$ generate the same contributions.
Again it is easy to check that, with the identifications \eqref{ident-tlam},
this ratio agrees with \eqref{resAn-2SL-1-exp}.

Following the same analysis that led to \eqref{ecomb}, we can also compute the
first order correction to $r_{1,m}$. It is given by a multiplicative factor
$(1-K_m+\cO(\repla^2))$ where
\bea
K_m&=&
(-1)^m 2\pi \lam \sin(\pi\omega)g_R(\omega)+2\pi \lamm \sin(2\pi\omega)g_R(2\omega)
\non\\
&&
-\frac{\pi\sin(\pi\omega)}{(-1)^m\lam+8\lamm\cos(\pi\omega)}\,\Bigl[
\lam^2 f_E(\omega,\omega)+16\lamm^2 \cos^2(\pi\omega)f_E(2\omega,2\omega)
\non\\
&& \qquad
+(-1)^m\lam\lamm \cos(\pi\omega)\Bigl(9f_E(\omega,2\omega)+f_E(\omega,-2\omega)\Bigr)
\Bigr]
\\
&=& \frac{\pi^2\omega}{R}\, \cot\(\frac{\pi}{R}\) \sin(\pi\omega)\Bigl( (-1)^m\lam+4\lamm \cos(\pi\omega)\Bigr)
+\frac{2\pi^2\omega^2\lam\lamm\cos^2(\pi\omega)}{\lam+(-1)^m 8\lamm\cos(\pi\omega)}\, ,
\non
\eea
where we used \eqref{e322a}, \eqref{identf12} and \eqref{identf0}.
This result perfectly agrees with  \eqref{resAn-2SL-1-exp}.

\subsection{Second type of critical points}

Another solution to vanishing of \eqref{dV2-2SL} is found to be
\be
\begin{split}
\cos(\omega x_m)=&\, -\frac{\lam}{8\lamm\cos(\pi\omega)}
+\frac{\pi \lam^2}{4\lamm}\, \tan(\pi\omega)f_E(\omega,\omega)\cos(\omega x_m)
\\
&\,
+2\pi \lamm\sin(2\pi\omega) f_E(2\omega,2\omega)\cos(2\omega x_m)\cos(\omega x_m)
\\
&\,
+\frac{\pi}{4}\, \lam\sin(\pi\omega)\Bigl(3 f_E(\omega,2\omega)(4\cos^2(\omega x_m)-1)+f(\omega,-2\omega)\Bigr).
\end{split}
\ee
This can be solved iteratively as a power series expansion in $\tilde\lambda$ as
\bea
\cos(\omega x_m)&= & -\frac{\lam}{8\lamm\cos(\pi\omega)}
-\frac{\pi}{4}\, \lam \sin(\pi\omega)\Bigl( 3f_E(\omega,2\omega)-f_E(\omega,-2\omega)-2f_E(2\omega,2\omega)\Bigr)
\non\\
&&
-\frac{\pi\lam^3}{64\lamm^2}\, \frac{\tan(\pi\omega)}{\cos(\pi\omega)}
\Bigl( 2f_E(\omega,\omega)-3f_E(\omega,2\omega)+f_E(2\omega,2\omega)\Bigr)
 +\cO(\repla^2).
\eea
Here the index $m$ is supposed to numerate different solutions and due to the symmetries $x\to x + 2\pi R/k$ and $x\to 2\pi R-x$, we expect to have $2k$ of them.
Note that due to \eqref{identf0}, the last bracket can be simplified to $f_E(\omega,2\omega)-2f_E(\omega,\omega)$.
At the critical point, the potential takes the following value
\bea
V_{\rm eff}(x_m) &=& 2\pi T \biggl[ -\lamm\sin(2\pi\omega)-\frac{\lam^2}{16\lamm}\, \tan(\pi\omega)
\non\\
&&
+\frac{\pi}{2}\, \lam^2 \sin^2(\pi\omega)\Bigl( f_E(\omega,\omega)-f_E(\omega,-\omega)+\hf\,f_E(2\omega,2\omega)
-\frac32\,f_E(\omega,2\omega)+\hf\,f_E(\omega,-2\omega)\Bigr)
\non\\
&&
-\frac{\pi}{2}\, \lamm^2 \sin^2(2\pi\omega)\Bigl( f_E(2\omega,2\omega)+f_E(2\omega,-2\omega)\Bigr)
\non\\
&&
-\frac{\pi \lam^4}{256\lamm^2}\, \tan^2(\pi\omega)\Bigl(4f_E(\omega,\omega)-4 f_E(\omega,2\omega)+f_E(2\omega,2\omega)\Bigr)
+\cO(\repla^3)\biggr]
\non\\
&=&
\pi^2\mu\[
-4 \lamm \sin(2\pi\omega)-\frac{\lam^2}{4\lamm}\,\tan(\pi\omega)
+4\pi^2 \omega^2\lamm^2 \sin(4\pi\omega)+\cO(\repla^3)\],
\eea
where we used \eqref{epred}, \eqref{identf0} and \eqref{identfff}. This result reproduces the instanton action \eqref{Smn-2SL-2-exp}
under the same identifications \eqref{ident-tlam} for $\lam$ and $\lamm$, while it does not depend on index $m$.

Next, we compute
\be
\begin{split}
h=&\, \frac{V_{\rm eff}''(x_m)}{T}=
2\pi\omega^2\biggl[4\lamm\sin(2\pi\omega)-\frac{\lam^2}{8\lamm}\, \tan(\pi\omega)
\\
&\,
+\frac{\pi}{2}\,\lam^2\sin^2(\pi\omega)\Bigl(4f_E(\omega,\omega)-3 f_E(2\omega,2\omega)
-2\pi\omega^2\cot(\pi\omega)\Bigr)
\\
&\,
+8\pi \lamm^2\sin^2(2\pi\omega)f_E(2\omega,2\omega)
-\frac{\pi\lam^4}{64\lamm^2}\, \tan^2(\pi\omega)\Bigl(2f_E(\omega,\omega)-f_E(2\omega,2\omega)\Bigr)
\biggr],
\end{split}
\ee
where we used \eqref{identf12}-\eqref{identfff}.
Thus, the ratio of the zero mode contributions in the perturbed and non-perturbed theory is
\be
r_1=\frac{2k}{R}\(2\pi T h\)^{-1/2}
=2(2\pi)^{-3/2}\mu^{-1/2}\(4\lamm\sin(2\pi\omega)-\frac{\lam^2}{8\lamm}\,\tan(\pi\omega)+\cO(\repla^2)\)^{-1/2},
\label{rm-2SL2}
\ee
where the factor $2k$ comes from the fact that all critical points generate the same contribution.
Again the identifications \eqref{ident-tlam} ensure the agreement of this ratio with \eqref{resAn-2SL-2-exp}.

Finally, we compute the first order correction to $r_1$.
As in the previous case. It is given by a multiplicative factor
$(1-K+\cO(\repla^2))$ where
\bea
K&=&
- \frac{\pi \lam^2}{4\lamm}\, \tan(\pi\omega)g_R(\omega)
+2\pi \lamm \sin(2\pi\omega)g_R(2\omega)\(\frac{\lam^2}{32\lamm^2\cos^2(\pi\omega)}-1\)
\non\\
&&
-\frac{\pi\sin(2\pi\omega)}{\frac{\lam^2}{\lamm}-64\lamm\cos^2(\pi\omega)}\,\Bigl[
\lam^2 \Bigl(4f_E(\omega,\omega)-3 f_E(2\omega,2\omega)-2\pi\omega^2\cot(\pi\omega)\Bigr)
\non\\
&& \qquad
+64\lamm^2 \cos^2(\pi\omega)f_E(2\omega,2\omega)
-\frac{\lam^4}{32\lamm^2}\,\cos^{-2}(\pi\omega)\Bigl(2f_E(\omega,\omega)-f_E(2\omega,2\omega)\Bigr)
\Bigr]
\non\\
&=& -\frac{2\pi^2\omega}{R}\,\lamm \cot\(\frac{\pi}{R}\) \sin(2\pi\omega)
+\frac{4\pi^2\omega^2\lamm\cos^2(\pi\omega)}{1-\frac{64\lamm^2}{\lam^2}\,\cos^2(\pi\omega)}\, ,
\eea
where we used \eqref{e322a} and \refb{relfE}.
This result perfectly agrees with
\eqref{resAn-2SL-2-exp}.

By the general argument given at the end of section \ref{sfirst-order}, the integration contour along real $x$ axis is given by the union of the steepest descent contours along the local minima of $V_{\rm eff}$. Furthermore, these are the same saddles that also contribute to the integral over $\tau_-$ in the MQM.

\section{Discussion}
\label{sec-cocl}

In this paper we revised the evaluation of instanton corrections in the sine-Liouville theory and its generalizations
describing tachyon backgrounds of 2d string theory, which was initiated in \cite{Alexandrov:2004cg}.
First, we found the instanton contributions in the framework of Matrix Quantum Mechanics,
where the results can be obtained non-perturbatively in the SL parameter(s) $\lambda_k$,
and then reproduced their expansion up to second order in $\lambda_k$ from string theory disk and annulus amplitudes.
This has allowed us to solve several issues raised by the old results of \cite{Alexandrov:2004cg} and confirmed once more
the perfect match between matrix model and CFT descriptions of 2d string theory, now extended to non-trivial
time-dependent backgrounds.
By T-duality and FZZ conjecture, our results should also describe instanton effects
in the black hole background of cigar geometry.

It is worth noting that the results of the MQM approach for the instanton action \eqref{Sn-SLk}, or more generally \eqref{Sinst-all},
and the subleading instanton contribution \eqref{res-An-SL}, or \eqref{eDeltam}, provide predictions for an infinite set of
disk and annulus correlation functions of tachyon vertex operators with $(1,n)$ ZZ boundary conditions.
In this paper we extracted only a few of them for the disk two-point and annulus one-point amplitudes at $n=1$,
and checked them against numerical evaluation of their explicit integral representations.
However, many more predictions can be obtained by either taking $n>1$, or going to higher orders in $\lambda_k$,
or adding more perturbation parameters.

An interesting feature of these predictions is that all correlation functions of vertex operators with (Euclidean) momenta
$k_i/R$, $k_i\in \IZ$, appear to be given by elementary functions, which should be contrasted with very complicated
integral expressions derived for them from the CFT formulation (see appendix \ref{ap-CFTampl}).
This suggests that there should exist a way to evaluate these integrals explicitly due to some hidden integrable structure.

Another important feature of these amplitudes is that their original CFT expressions are typically divergent and
one needs to apply string field theory technique to convert them to finite quantities.
While this has been done for the disk two-point and annulus one-point amplitudes,
there are no explicit finite expressions for the higher point correlation functions yet, although the general approach of
how this should be done is understood.

It would also be interesting to go beyond the subleading instanton contribution and compute higher $g_\str$-corrections.
This should be possible using the Toda integrable structure of the MQM formulation, whereas in CFT the complications
increase drastically with each order in perturbation.
Finally, an exciting research direction is to apply resurgence technique to this system and eventually find its complete non-perturbative partition function.

\paragraph{Acknowledgements.}
We are grateful to Ivan Kostov and Spenta Wadia for discussions and to Victor Rodriguez for sharing with us his code for computing numerically string amplitudes.
A.S. is supported by ICTS-Infosys Madhava Chair Professorship and the J. C. Bose fellowship of the Department of Science and Technology, India.

\appendix

\section{Normalization factor and the cut-off}
\label{ap-cutoff}

The first attempt to fix the constant $C$ in \eqref{resAnSL} was made in \cite{Alexandrov:2004cg}. In this appendix we review and extend this approach but
find that this does not quite lead to the correct result. We discuss possible
shortcomings in this approach that eventually force us to the first principle
analysis described in section \ref{sec-MQMSL}.

The approach of \cite{Alexandrov:2004cg} was based on a combination of the relation \eqref{SmatrixMQM} with the orthonormality of the perturbed wave functions $\Pem$,
which allows to rewrite this relation in the presence of the cut-off $\Lambda$ as
\be
e^{i\phi(E)}=\frac{\sqrt{2\pi}}{\log\Lambda }\int_0^{^{\sqrt{\Lambda}}} \!\!\!\! d\xp\,
\int_0^{\sqrt{^{\Lambda}}} \!\!\!\! d\xm\,
\overline{\Pem(\xm)}\, e^{i\xp\xm}\, \Pep(\xp)\,,
\label{doubleint}
\ee
where we took into account that for finite cut-off, $\delta(E-E')$ in the normalization condition
should be replaced by $\rho(E)\delta_{E,E'}$ and then used \eqref{endensity}.
In the quasi-classical approximation, the double integral can be evaluated by saddle point.
Taking into account \eqref{solv0}, one obtains two saddle point equations
\be
\xp=\Xp(\xm),
\qquad
\xm=\Xm(\xp),
\label{saddleeq}
\ee
where $X_\pm$ are functions obtained from the two equations \eqref{xmptau}
by eliminating $\tau$ in one of them using the second.
The main contribution to the integral comes from the trivial solution \eqref{xmptau}.
In the $(\xp,\xm)$-plane it represents a one-dimensional curve which is nothing but the deformed classical fermion trajectory
in the phase space.
As a result, the contribution from this `saddle contour' is proportional to $\log\Lambda$, which is the length of the trajectory
in the presence of the cut-off in terms of the uniformization parameter $\tau$, and cancels the same factor in \eqref{doubleint}.

However, as explained in section \ref{subsec-dpMQM},
there are also other solutions to \eqref{saddleeq} which correspond to the so-called double points of the complex curve.
In \cite{Alexandrov:2004cg}, the double points
corresponding to setting $\tau=\mp i\theta_n$ in the expression for $x_\pm$,
where $\theta_n$ are defined in \eqref{defthnSL}, have been taken into account.
The contributions to the double integral \eqref{doubleint} generated by these double points
are exponentially suppressed compared to the leading contribution and, after being integrated to get
the free energy using \eqref{relFphi}, can be shown to be precisely of the form \eqref{nonpert-n}.
Moreover, in this way one finds that the constant factor $C$ in \eqref{resAnSL} is given by
\be
C=\frac{i\sqrt{2\pi R}}{8\log\Lambda}\, ,
\label{finalC}
\ee
where $\log\Lambda$ in the denominator is the same factor as in \eqref{doubleint}
because the contributions of double points to the integral are finite and do not cancel it.

Of course, such result cannot be correct and it is natural to look for a mistake responsible for the cut-off dependence.
The analysis in section \ref{subsec-dpMQM} suggests a natural candidate for such a mistake. It reveals that for a given $n$
there is actually a discrete parameter family of double points, denoted there by $(n,m,0)$ with $m\in\IZ$,
which generate the same instanton contributions as the ones considered in \cite{Alexandrov:2004cg}.
The latter correspond in this notation to $m=0$ and the other double points are all obtained
by an imaginary shift of the uniformization parameter $\tau\to \tau+2\pi i  m R$.
Therefore, one may expect that taking these additional double points
into account can resolve the issue.

Indeed, due to these additional contributions each instanton correction \eqref{nonpert-n} gets an additional factor
equal to the number of double points with fixed $n$ contributing to the integral \eqref{doubleint}.
Naively, this number is infinite and we arrive at even worse problem than before.
However, we will now provide a tentative argument as to why for a finite cut-off, this number is also finite and actually cancels
the $\Lambda$-dependence in \eqref{finalC}.

The idea is to apply the Bohr-Sommerfeld quantization rule which states that one state of a quantum system
occupies a volume in the phase space equal to $2\pi \hbar$.
As a first step, note that according to \eqref{canontr}
the variable canonically conjugate to the energy is the uniformization parameter $\tau$.
Therefore, if $\Delta E$ is the distance between energy levels,
the maximal distance that a fermion can travel along the uniformization parameter should satisfy
$|\Delta\tau| = 2\pi \hbar /\Delta E$.
On the other hand, from \eqref{enquantcond}, after restoring $\hbar$, one finds
$\Delta E=2\pi \hbar/\log\Lambda$, which finally gives
$|\Delta\tau| = \log\Lambda$.
Assumung that this result is independent of whether the fermion travels in real or imaginary direction, one obtains that the number of double points with fixed $n$
contributing to the integral \eqref{doubleint} is given by
\be
M(\Lambda)=\frac{\log\Lambda}{2\pi R}.
\label{ML}
\ee
Multiplying \eqref{finalC} by this factor, one arrives at a cut-off independent result
\be
C=\frac{i}{8\sqrt{2\pi R}}\, .
\label{finalCnew}
\ee

Remarkably, the new normalization factor differs from \eqref{newC} taken for $k=1$
just by a factor of $i$, which can be traced back to the factor of $i$
appearing in $1/(2\pi i R)\int d\tau_-$ used to extract the constant Fourier mode in \eqref{ene63a}.
On the other hand, it is \eqref{newC} that is shown to reproduce the string amplitude computation.
Thus, the approach we followed here is still unable to produce a fully consistent result.
This is probably related to one or several problems listed below:
\begin{itemize}
\item
First of all, the argument based on the Bohr-Sommerfeld quantization rule is definitely non-rigorous.
Besides, it requires an extension to a complexified domain where the uniformization parameter changes
in the imaginary direction. It is tempting to think that this is precisely
the place where the factor of $i$ is lost, but it is difficult to justify its insertion mathematically.

\item
Second, the representation \eqref{doubleint} is based on the use of the orthonormality of the perturbed wave functions.
It is well justified at the perturbative level, but may fail non-perturbatively.

\item
Finally, this derivation completely ignored instanton corrections to the perturbed wave functions
which do arise as is demonstrated by an explicit calculation in the limit of small SL parameter
in appendix \ref{ap-Smatrix}.

\end{itemize}
Given these problems, in the main text we follow a more solid approach which starts directly from
\eqref{SmatrixMQM} without any additional assumptions.

\section{S-matrix}
\label{ap-Smatrix}

In this appendix we show how the formalism of sections \ref{sec-MQM} and \ref{sec-MQMSL} reproduces the non-perturbative on-shell
amplitudes in the 2d string theory in the  background \eqref{CFTc=1} with a compact time direction.
For this we need to reverse the order of expansion in the two small parameters.
In the computation of the contribution to the free energy in the deformed theory from a given instanton sector,
we first expanded in powers of $g_s\sim 1/\mu$ for fixed $\lambda_k$ and then expanded each term in power series in $\lambda_k$.
However, to compute the S-matrix, we first need to expand in powers of the deformation parameters,
so that the $n$-point S-matrix is given by the coefficient of the $n$-th power of $\lambda_k$'s, and only after that
expand in powers of $g_s$.

We shall work up to second order in the perturbation, which gives us access to two-point amplitudes in the undeformed theory.
Since the energy conservation forces the energies of incoming and outgoing states to be equal,
one can restrict the analysis to a single SL deformation parametrized with fixed $k$.
Our starting point is the expression for the scattering phase \eqref{enefou1}.
Let us change the integration variable from $x_\pm$ to $y_\pm=\log (x_\pm/\sqrt{\mu})$ and
evaluate the integral on the right hand side using the saddle point method.
As we know, there is an infinite number of saddles labelled by $n\in \IN$, with $n=0$ corresponding to the perturbative saddle.
Since we are interested only in the one-instanton contribution, it will be sufficient to consider
the saddles labelled by $n=0$ and 1. Furthermore, for each of these saddle points
we restrict ourselves to the first subleading order in the expansion in the string coupling constant.

Let us express the integrand $\cI\(\sqrt{\mu}\, e^{y_+},\sqrt{\mu} \,e^{y_-}\)$ given in \eqref{enefou1} as
\be\label{eintexp}
\exp\[f(y_+) + \lambda_k g(y_+) +\lambda_k^2 h(y_+) +\cdots\],
\ee
for some functions $f,g,h$, etc. In these rescaled variables the functions $f,g,h$ are of order $\mu$ and for large $\mu$ we can evaluate the integral using the saddle point method. Let the saddle point be at
\be
y_\lambda=y_0 + \lambda_k y_1 + \lambda_k^2 y_2+\cdots\, .
\ee
Then after expanding in $\lambda_k$, the saddle point equation implies the following relations
\be\label{ey1}
f'(y_0)=0,
\qquad y_1 = -{g'(y_0)\over f''(y_0)},
\qquad
y_2=-\frac{h'(y_0)}{f''(y_0)}+\frac{g'(y_0)g''(y_0)}{(f''(y_0))^2}-\frac{(g'(y_0))^2f'''(y_0)}{2(f''(y_0))^3}\, .
\ee
Substituting this into \eqref{eintexp}, keeping terms up to quadratic order in $\lambda_k$ and
$(y_+-y_\lambda)$, and carrying out the Gaussian integration over $y_+$, we get
\be\label{e58a}
\sqrt{2\pi} D^{-1/2}
\exp\[f(y_0) +\lambda_k g(y_0) + \lambda_k^2 \(h(y_0)- \frac{1}{2}\, {g'(y_0)^2\over f''(y_0)}\)\] ,
\ee
where
\be
\begin{split}
D=&\, -f''+ \lambda_k\( \frac{ f'''}{f''}\, g'- g''\)
\\
&\,  +\lambda_k^2\(\frac{f'''}{f''}\, h'-h''
+\frac{g'}{f''}\(g'''-g''\frac{f'''}{f''} +\frac{g'(f''')^2}{2(f'')^2}-\frac{g' f^{(4)}}{2f''}\)\)\, ,
\end{split}
\ee
and all functions are evaluated at $y_0$.

To apply this to our integral, let us expand $v_\pm$ as
\be
v_\pm\(\sqrt{\mu} \, e^{y_\pm}\)=\mu\lambda_k \,u_\pm(y_\pm) +\mu \lambda_k^2\, w_\pm(y_\pm) +\cdots\, .
\ee
At this stage we do not know the functions $u_\pm$ and $w_\pm$ except that they must decay at large values of their arguments.
Then introducing a convenient notation $\hmu=\mu+i/2$, we have the following identifications
\be
\begin{split}
f(y_+)=&\, i\mu \,e^{y_+ +y_-} -i\hmu\, (\log\mu+y_+ +y_-),
\\
g(y_+) =&\, - i \mu\( e^{k y_+/R}+e^{k y_-/R}+ u_+(y_+)+ u_-(y_-)\) ,
\\
h(y_+) =&\, - i \mu\(w_+(y_+)+ w_-(y_-)\) .
\end{split}
\ee
The $n$-th saddle point is given by
\be
y_{0,n}= -\hym- 2\pi i n +\cO(\mu^{-2})\, ,
\qquad
\hym=y_- -\frac{i}{2\mu}\, .
\ee
Substituting these identifications into \eqref{e58a}, taking into account the factor $1/\sqrt{2\pi}$ in \eqref{enefou1},
and multiplying instanton contributions by an extra factor of $1/2$ to account for the fact that the integration
contour runs over only half of the steepest descent contour of the instanton,
we can write the scattering phase in our approximation as
\be
\phi(-\mu)\approx  -i\(\log I_0 + {1\over 2} {I_1\over I_0}\),
\label{phiII}
\ee
where $I_n$ is the contribution from the $n$-th saddle given by
\be
\begin{split}
I_n=&\,  \frac{e^{-2\pi n\hmu }}{\sqrt{D_n}} \exp\Biggl[ i \hmu -i \hmu  \log \hmu
-i\mu \lambda_k \( e^{\frac{k}{R}\,y_{0,n}} +e^{\frac{k}{R}\,y_-}
+ u_+(y_{0,n})+u_-(y_-)\)
\\
&\, -
i \mu\lambda_k^2\Biggl\{\frac{\mu}{2\hmu} \(  {k\over R}\,e^{\frac{k}{R}\,y_{0,n}}
+ u_+'(y_{0,n})\)^2
\! + w_+(y_{0,n})+w_-(y_-)\Biggr\}
\Biggr] ,
\end{split}
\label{efinsaddle}
\ee
where
\be
D_n = -i \mu\Bigg[1-\lambda_k\(\frac{k}{R}\(\frac{k}{R}-1\) e^{\frac{k}{R}\,y_{0,n}}
+u_+''(y_{0,n})-u_+'(y_{0,n}) \)
-\lambda_k^2 D_n^{(2)}\Bigg] ,
\label{calculDn}
\ee
\be
\begin{split}
D_n^{(2)}=&\,
w_+''(y_{0,n})-w_+'(y_{0,n})
\\
&\,
-\({ k\over R}\,e^{\frac{k}{R}\,y_{0,n}} +u_+'(y_{0,n})\)
\(\frac{k^2}{R^2}\(1-\frac{k}{R}\)e^{\frac{k}{R}\,y_{0,n}}
-u_+'''(y_{0,n})+ u_+''(y_{0,n})\),
\end{split}
\ee
and we dropped $\cO(\mu^0)$ terms in $D_n$ because they go beyond our approximation.
The main difference of this analysis from the one in section \ref{sec-MQMSL} will be that here
we shall expand the integrand in a power series in $\lambda_k$,
while in section \ref{sec-MQMSL} the exponential factors containing $\lambda_k$ had not been expanded
as they are accompanied by a factor of $\mu=1/g_s$.

\subsection{Perturbative contribution}

First, let us study the contribution $\log I_0$ from the zero instanton sector.
Taking the logarithm of \eqref{efinsaddle} and neglecting the terms that have negative powers of $\mu$,
one finds
\bea
\log I_0 &\approx &   i \hmu -i \mu  \log \hmu+\frac{\pi i}{4}
\non\\
&&
-i\mu \lambda_k \biggl[ \(1+ \frac{ik^2}{2R^2\mu}\)e^{-\frac{k}{R}\,y_-} +e^{\frac{k}{R}\,y_-}
+ u_+(-y_-)+u_-(y_-)+\frac{i}{2\mu}\,u_+''(-y_-)\biggr]
\label{einexp}\\
&& -
i \mu\lambda_k^2\(\hf \(1-\frac{i}{2\mu}\)\(  {k\over R}\,e^{-\frac{k}{R}\,\hym}
+ u_+'(-\hym)\)^2 +w_+(-\hym)+w_-(y_-)\)+\frac{\lambda_k^2}{2}\, \cD_0\, ,
\non
\eea
where
\be
\cD_0 = D_0^{(2)}+\frac{1}{2}\(\frac{k}{R}\(\frac{k}{R}-1\) e^{-\frac{k}{R}\,y_-}
+u_+''(-y_-)-u_+'(-y_-) \)^2.
\label{cD0}
\ee
Extracting the terms of order $\lambda_k$, requiring that they should be independent of $y_-$ and
taking into account that $u_\pm$ must be decreasing functions, one gets two conditions which fix these
functions at the perturbative level up to second order in $1/\mu$ expansion
\be
\begin{split}
u_+(-y_-)+\frac{i}{2\mu}\,u_+''(-y_- )=
-e^{\frac{k}{R}\,y_-} \Rightarrow
\upert_+(y_+)=&\,  -\(1- \frac{ik^2}{2R^2\mu}\)e^{-\frac{k}{R}\, y_+},
\\
\upert_-(y_-)=&\, -\(1+ \frac{ik^2}{2R^2\mu}\)e^{-\frac{k}{R}\, y_-}.
\end{split}
\label{sol-pertsp}
\ee

Substituting these results into terms quadratic in $\lambda_k^2$ in \eqref{einexp}, one finds
\bea
&&
-i \mu\lambda_k^2\[\frac{k^2}{2R^2}\(1-\frac{i}{2\mu}\)\(  e^{-\frac{k}{R}\,\hym}
+ \(1- \frac{ik^2}{2R^2\mu}\)e^{\frac{k}{R}\, \hym }\)^2
+w_+(-\hym)+w_-(y_-)\]
\non\\
&&
+\frac{\lambda_k^2}{2}\Biggl[
w_+''(-y_- )-w_+'(-y_- )
-\frac{k^3}{R^3}\(e^{-\frac{k}{R}\,y_-}+e^{\frac{k}{R}\, y_-}\)
\(\(1-\frac{k}{R}\)e^{-\frac{k}{R}\,y_-}-\(1+\frac{k}{R}\)e^{\frac{k}{R}\, y_-}\)
\non\\
&&\qquad
+\frac{k^2}{2R^2}\(\(1-\frac{k}{R}\) e^{-\frac{k}{R}\,y_-}+\(1+\frac{k}{R}\)e^{\frac{k}{R}\, y_-} \)^2\,
\Biggr].
\label{pert-lam2-contr}
\eea
As above, requiring that this expression is independent of $y_-$ and
taking into account that $w_\pm$ are decreasing functions, one obtains two conditions fixing them as
\be
\wpert_\pm(y_\pm)=-\frac{k^2}{2R^2} \( 1\pm\frac{ik}{R\mu}\(1-\frac{3k}{2R}\)\)e^{-\frac{2k}{R}\,y_\pm}.
\label{sol-pertsp-w}
\ee

These results are perfectly consistent with the ones obtained for finite $\lambda_k$ in \eqref{solvpmSL} and \eqref{vpm-sublead}.
Indeed, inverting \eqref{xpmSLk}, one finds
\be
\tau_\pm=\pm \[\hf\, \scX +y_\pm - a_k\, e^{-\frac{k}{2R}\,\scX}e^{-\frac{k}{R}\,y_\pm}
- \frac{2k-R}{2R}\, a_k^2 \,e^{-\frac{k}{R}\, \scX} e^{-\frac{2k}{R}\,y_\pm}+\cdots\].
\label{largexpm}
\ee
Substituting this expansion into \eqref{solvpmSL} and \eqref{vpm-sublead} and expanding in power series in
$\lambda_k$, we recover \eqref{sol-pertsp} and \eqref{sol-pertsp-w}, respectively.

Finally, extracting the constant terms from \eqref{pert-lam2-contr}, adding them to the first line in \eqref{einexp},
multiplying by $-i$, and keeping only non-negative powers of $\mu$, one obtains
\be
\begin{split}
\phi_{\rm pert} = &\,
\mu \(1- \log \mu- \frac{k^2}{R^2}\,\lambda_k^2\)+\frac{\pi}{4}+\cO(\mu^{-1},\lambda_k^3) .
\end{split}
\label{pertphi}
\ee
Thus, there are no corrections to the scattering phase at the subleading
order in $1/\mu$ expansion, except the constant term $\pi/4$, as it should be.
Using \eqref{relFphi}, one gets that the free energy, up to non-universal terms, is given by\footnote{Recall that
the derivative in  \eqref{relFphi} is taken keeping constant $t_k$, not $\lambda_k$.}
\be
\cF_{\rm pert}\approx -\frac{R}{2}\, \mu^2 \log \mu-k\mu^2 \lambda_k^2.
\ee
The coefficient of the $\lambda_k^2$ term should be related to the leading perturbative
contribution to the two-point S-matrix after we identify $k/R=\omega$ and $2\pi R=\delta(\omega-\omega')$.
Thus, we find the S-matrix
\be
\frac{\omega}{2\pi}\, \delta(\omega-\omega')\, ,
\ee
which agrees with the explicit computation on the string theory side.

\subsection{Instanton contribution}

Next, we study the leading instanton contribution to the scattering phase which, as in section \ref{subsec-instcontr},
has two sources: the second term in \eqref{phiII} and an instanton correction to the first term appearing due to
possible instanton corrections to $u_\pm$ and $w_\pm$. To write them explicitly, we represent
\be
u_\pm=\upert_\pm+e^{-2\pi \mu} \uinst_\pm,
\qquad
w_\pm=\wpert_\pm+
e^{-2\pi \mu} \winst_\pm,
\ee
where $\upert_\pm$ and $\wpert_\pm$ are given by \eqref{sol-pertsp} and \eqref{sol-pertsp-w},
while $\uinst_\pm$ and $\winst_\pm$ encode instanton corrections and are still to be found.

Thus, on one hand, using \eqref{einexp}, one obtains
\bea
&&- i\(\log I_0\)_{\rm inst}\approx
-\mu\, e^{-2\pi\mu}\biggl[\lambda_k\(\uinst_+(-y_-)+\uinst_-(y_-)+\frac{i}{2\mu}\, \uinstd{+}''(-y_-)\)
\label{logI0inst}\\
&&
+\lambda_k^2\,\biggl( \frac{k}{R}\(  e^{-\frac{k}{R}\,y_-}+e^{\frac{k}{R}\,y_-}\)\uinstd{+}'(-y_-)
+ \(\winst_+(-y_-)+\winst_-(y_-)\)
+\frac{i}{2\mu}\,\winstd{+}''(-y_-)\biggr)\biggr],
\non
\eea
where we neglected $\cO(\mu^0\lambda_k^2 )$ terms multiplying $\uinst_\pm$ because, as we will see,
they contribute to the order beyond our approximation. However, we have kept similar terms involving
$\winst_\pm$ since $\winst_\pm$ will turn out to be of order $\mu$.
On the other hand, from \eqref{efinsaddle} we have
\be
\begin{split}
-\frac{i}{2}\,\frac{I_1}{I_0}\approx &\, \frac{i}{2}\, e^{-2\pi \mu}
\[\frac{1+\frac{k}{R}\,\lambda_k\(\(1-\frac{k}{R}\) e^{-\frac{k}{R}\,y_-}+\(1+\frac{k}{R}\) e^{\frac{k}{R}\,y_-} \)}
{1+\frac{k}{R}\,\lambda_k\(\(1-\frac{k}{R}\) e^{-\frac{k}{R}\(y_- +2\pi i \)}
+\(1+\frac{k}{R}\) e^{\frac{k}{R}\(y_- +2\pi i \)} \)}\]^{1/2}
\\
&\,
\times\exp\biggl[
i\mu \lambda_k\(\(1-e^{-\frac{2\pi i k}{R}} \)e^{-\frac{k}{R}\,\hym}
-\(1- \frac{ik^2}{2R^2\mu}\)\(1-e^{\frac{2\pi i k}{R}} \)e^{\frac{k}{R}\,\hym}\)
\\
&\, \qquad
+\frac{i\mu k^2}{2R^2}\,\lambda_k^2\(1-e^{-\frac{4\pi i k}{R}} \) e^{-\frac{2k}{R}\,y_-}
\biggr],
\end{split}
\label{ratII}
\ee
where we again neglected  $\cO(\mu^0\lambda_k^2 )$ terms.
Expanding \eqref{ratII} in powers of $\lambda_k$, adding the resulting expression to \eqref{logI0inst} and requiring
cancellation of $y_-$-dependent contributions, one can fix all the unknown functions:
\be
\begin{split}
\uinst_\pm(y_\pm) =&\, \,  \pm \hf \(1-e^{\pm\frac{2\pi i k}{R}}\)\(1\mp \frac{ik^2}{2R^2\mu}\) e^{-\frac{k}{R}\,y_\pm},
\\
\winst_\pm(y_\pm)=&\, -\frac{i }{4}\, \mu \(\(1-e^{\pm\frac{2\pi i k}{R}}\)^2
\mp \frac{2i k^2}{R^2 \mu}\(e^{\pm\frac{4\pi i k}{R}}-e^{\pm\frac{2\pi i k}{R}}\)\)e^{-\frac{2k}{R}\,y_\pm}.
\end{split}
\label{inst-u}
\ee
Then the remaining constant terms in \eqref{logI0inst} and \eqref{ratII}
give the instanton contribution to the scattering phase
\be
\phi_{\rm inst}=\frac{i}{2}\, e^{-2\pi \mu}\[1+4\mu^2 \lambda_k^2  \(\sin^2\(\frac{\pi k}{R}\)
-\frac{k^2 }{4R^2 \mu}\, \sin\(\frac{2\pi k}{R}\)\) \] .
\label{phiinst2}
\ee
Finally, from \eqref{relFphi} we obtain the leading instanton correction to the free energy
\be\label{egenR}
\cF_{\rm inst}=-\frac{i\, e^{-2\pi\mu}}{4\sin(\pi/R)}\[1+4\mu^2 \lambda_k^2\, \sin^2\(\frac{\pi k}{R}\)
\biggl\{1 +\frac{k}{2R^2\mu}\biggl( \cot\(\frac{\pi }{R}\) -k\cot\(\frac{\pi k}{R}\)\biggr)\biggr\} \]  .
\ee

The $\lambda_k$-independent part of this result agrees with \eqref{encFnp}, while the term
proportional to $\lambda_k^2$ provides instanton contribution to the two-point amplitude in the undeformed theory.
To get the precise normalization, we use the relation \eqref{identlam}. The two-point function is
given by the second derivative of $\cF_{\rm inst}$ with respect to $\tilde\lambda_{k}$ which gives after substituting $k/R=\omega$
\be
2i\pi^4 \mu^{2}\, \frac{ e^{-2\pi\mu}}{\sin(\pi/R)}\, \sin^2(\pi\omega)
\left\{1 +\frac{1}{2\mu}\( \frac{\omega}{R}\,\cot\(\frac{\pi }{R}\)
-\omega^2\cot(\pi\omega)\)\right\}  .
\label{escatter}
\ee

This needs to be compared to the instanton induced two-point amplitude computed from string theory.
In a power series expansion in $g_s=1/(2\pi\mu)$,
the leading term of this contribution is the product of $e^{-2\pi\mu}$,  two disk
one-point functions and the exponential of the annulus partition function,
while the subleading term is the sum of the disk two-point function with
the product of the disk one-point function and the annulus one-point function,
each multiplied by $e^{-2\pi\mu}$ and the exponential of the annulus partition function.\footnote{We assume that the constant contribution,
representing the partition function on surfaces of Euler number $-1$, continues to vanish even for finite $R$.}
The latter is given in \eqref{eF0a}, but reversing the steps in \eqref{eff1}, it needs to be rewritten as
\be
\label{ezeromodeint}
{i\over 8\pi R \sin(\pi/R)}\, \int_0^{2\pi R} dx
\ee
with the understanding that the integration over the instanton position $x$ needs to be carried
out at the end. From the discussion above \eqref{e513} it follows that the relevant disk one-point
function is that of $\pi c\bar c \cV_\omega$ which can be computed from
\eqref{eadiskV}.
Taking its square and multiplying by \eqref{ezeromodeint},
one obtains the leading contribution
\be
\frac{i\, (4\pi^2\mu \, \sin(\pi\omega))^2}{8\pi  R\sin(\pi/R)}\,  \int_0^{2\pi R} dx \cos^2 (\omega x)
= {2 i \pi^4 \mu^2 \over \sin(\pi/R)}\, \sin^2(\pi\omega).
\label{leadS2}
\ee

Next we consider the subleading contribution.
Using \eqref{ezeromodeint}, \eqref{eadiskV},  \eqref{eratiodisk} and the fact that the relevant
string field accompanying the deformation parameter $\repla_k$ is $\pi\repla_k\, c\bar c\, \cV_\omega$,
we see that the net contribution to the instanton induced two-point amplitude from the disk two-point function is given by
\be
\begin{split}
&\,
\frac{g_\str}{4}\, \frac{i\, (4\pi^2\mu \, \sin(\pi\omega))^2}{8\pi R\sin(\pi/R)}
\int_0^{2\pi R} dx\,
\Bigl( f_E(\omega, \omega) e^{2i\omega x}+ f_E(-\omega, -\omega)
e^{-2i\omega x} + 2\, f_E(\omega, -\omega)\Bigr)
\\
 =&\, i \pi^3  \mu\,  \frac{\sin^2(\pi\omega)}{\sin(\pi/R)} \, f_E(\omega, -\omega)\, .
\end{split}
\label{ediskcont}
\ee
On the other hand, using \eqref{ezeromodeint}, \eqref{eadiskV}  and \eqref{onepA},
the contribution from the annulus one-point function reads
\be
2{g_\str} \,\frac{i\,(4\pi^2\mu \, \sin(\pi\omega))^2}{8\pi R\sin(\pi/R)}\,
g_R(\omega)\,
\int_0^{2\pi R} dx\, \cos^2 (\omega x)
= 2 i \pi^3 \mu\,\frac{\sin^2(\pi\omega)}{\sin(\pi/R)} \, g_R(\omega)\, .
\label{eanncont}
\ee
The factor of 2 on the left hand side accounts for the possibility
of exchanging the two external lines.
Summing up \eqref{leadS2}, \eqref{ediskcont}, \eqref{eanncont}
and multiplying by $e^{-2\pi\mu}$, we obtain the string theory result
for the instanton contribution to the two-point amplitude to the first subleading order in the string coupling:
\be\label{egenRexp}
{2i \pi^4 \mu^2 \, e^{-2\pi\mu}\over \sin(\pi/R)}\, \sin^2(\pi\omega) \[
1+  \frac{1}{2\pi\mu}\, \Bigl(
f_E(\omega,-\omega) + 2 \, g_R(\omega) \Bigr)\] .
\ee
The agreement with \eqref{escatter} then follows immediately from
\eqref{epred} and \eqref{e322a}.

\section{Critical behavior and pure two-dimensional gravity}
\label{ap-crit}

In this appendix we demonstrate that for $R/k>1$, the instanton action, $\bS_1^{\bind(1)}$ vanishes exactly at a critical point
where the system reduces to a theory with vanishing central charge. As in section \ref{sec-MQMSL}, $\bind(n)$ denotes the value of $\sind$ for which $\bS_n^\sind$ has lower value, which according to \eqref{sigman} is 1 for positive $\lambda_k$ and 0 for negative $\lambda_k$.
The existence of such critical point in the SL theory is well known \cite{Hsu:1992cm},
but it was analyzed in relation to the instanton effects only in the region $1/2<R/k<1$ \cite{Alexandrov:2003nn}.
Here we redo the analysis for $R/k>1$ and provide additional consistency checks.

Let us consider the equation \eqref{XthroughLk} for the function $\scX(\lambda_k)$ which is related to
the second derivative of the free energy.
Taking derivative of the l.h.s. with respect to $\scX$ and equating it to zero,
we find that it reaches a maximum value at
\be
e^{-\(1-\frac{k}{R}\)\scX_{\rm cr}}=\frac{k^2 \, \lambda_k^2}{R^2}\(1-\frac{k}{R}\)\(2-\frac{k}{R}\).
\label{Xcr}
\ee
Substituting this value back into \eqref{XthroughLk}, we get the left hand side to
be equal to
\be
{1-{k\over R}\over 2-{k\over R}} \[ {k^2\over R^2} \left( 1-{k\over R}\right) \left( 2-{k\over R}\right)
\lambda_k^2\]^{-{1\over 1-{k\over R}}} \, .
\ee
Equating this to 1,
one finds a critical value for $\lambda_k$
\be
\lamcr_k=\frac{R}{k} \(1-\frac{k}{R}\)^{-\frac{k}{2R}}\(2-\frac{k}{R}\)^{-\(1-\frac{k}{2R}\)}\, .
\ee
The significance of this critical value lies in the fact that for $|\lambda_k|>\lamcr_k$, the left hand side of equation \eqref{XthroughLk} is less than 1 even when $\scX$ is chosen so as to maximize
the left hand side of \eqref{XthroughLk}. Therefore, \eqref{XthroughLk}
does not have solutions anymore.

Solving \eqref{XthroughLk} near the critical point, one finds
\be
\scX= \log \frac{2-k/R}{1-k/R}-2\,\delta^{1/2}
+\frac{2}{3}\(3-\frac{2k}{R}\)\delta+\cO(\delta^{3/2})\, ,
\label{crit-cX}
\ee
where
\be
\delta
=\frac{1-|t_k|\mu^{{k\over 2R}-1}/\lamcr_k}{(1-k/R)(2-k/R)}\, ,
\label{def-critparam}
\ee
and we have chosen the smaller of the two solutions for $\scX$, since as we
increase $|\lambda_k|$ from 0 to $\lamcr_k$, $\scX$ should approach $\scX_{\rm cr}$ from
below.
This implies
\be
a_k\approx \frac{(-1)^{\bind(1)+1}}{1-k/R}\[ 1-\(2-\frac{k}{R}\)\delta^{1/2}
+\frac{1}{6}\(2-\frac{k}{R}\) \(6-\frac{k}{R}\)\delta+\cO(\delta^{3/2})\],
\ee
where we used the relation \eqref{sigman} between $\bind(1)$ and the sign of $\lambda_k$.
Substituting this result into the equation \eqref{eq-thetas}, one observes that at the leading order $\theta_1^{\bind(1)}$ vanishes,
while at the next orders it becomes
\be
\theta^{\bind(1)}_1=\sqrt{6R/k}\,\delta^{1/4}-\frac{\sqrt{6}}{15(k/R)^{3/2}}\(3 - \frac{3k}{R}  + \frac{4 k^2}{R^2}\)\delta^{3/4}+\cO(\delta^{5/4}).
\label{defthnSL-cr}
\ee
Finally, plugging the above expansions into \eqref{Sn-SLk}, we obtain an expansion of the instanton action near
the critical point
\be
\bS^{\bind(1)}_1=2\mu\[\frac{8\sqrt{6}}{5\sqrt{k/R}}\(1-\frac{k}{R}\) \delta^{5/4}+\cO(\delta^{7/4})\].
\ee
In particular, $\bS_1^{\bind(1)}\to 0^+$ and the instanton action never becomes negative.

Now let us verify that at the critical point we indeed reach the $c=0$ theory.
To this end, we define
\be\label{edefbfmu}
\mucz=\mu^{4/5}\delta\, ,
\ee
and take the limit $\delta\to 0$, $\mu\to\infty$ with $\mucz$ fixed.
In terms of the new parameter we get a finite instanton action and the subleading instanton contribution \eqref{res-An-SL}
\be
\begin{split}
\bS_1^{\bind(1)}=&\,\frac{16\sqrt{6}}{5\sqrt{k/R}}\(1-\frac{k}{R}\) \mucz^{5/4},
\\
A_1^{\bind(1)}=&\, \frac{ik\,(k/R)^{1/4}}{16\sqrt{\pi}\, 6^{3/4}} \(1-\frac{k}{R}\)^{-1/2} \mucz^{-5/8}.
\end{split}
\label{SAc=0}
\ee
Both these quantities have the same scaling in $\mucz$ as the instanton effects
in the $c=0$ theory with respect to the cosmological constant \cite{David:1992za,Eynard:1992sg}.

Moreover, we can also check that the numerical coefficients also perfectly match those in the $c=0$ theory.
For this purpose, we compute and compare dimensionless combinations of physical quantities
that do not depend on the normalization of the parameters.
As such combinations, one can take, for example,
\be
\rcz_0=\frac{\p_{\mucz} \Scz_{\rm inst}}{\(-\p_{\mucz}^2\Fcz_0\)^{1/2}}\, ,
\qquad
\rcz_1=\Scz_{\rm inst}^{1/2} \Acz\, ,
\label{rc=0}
\ee
where $\Fcz_0$, $\Scz_{\rm inst}$ and $\Acz$ are the genus zero free energy, instanton action and instanton prefactor
of the $c=0$ theory, respectively.
Their values can be found in the literature, see e.g. \cite{Eynard:1992sg,Alexandrov:2003nn,Eniceicu:2022nay},
\be
\rcz_0=2\sqrt{3},
\qquad
\rcz_1=\frac{i}{4\sqrt{30\pi}}\, .
\label{rrc=0}
\ee

To reproduce \eqref{rrc=0} from \eqref{SAc=0}, first, one must take into account that
the relation between the free energy of the SL theory near the critical point and the one of the $c=0$ theory
is expected to contain the factor $k$ because in the case of perturbation by the operator $\cV_{k/R}$ \eqref{genCFT}
there are $k$ vacua where the scalar field can settle down and hence we get $k$ contributions of the same $c=0$ theory.
This leads to the relations
\be
\cF'_0=k \Fcz_0,
\qquad
\bS_1^{\bind(1)} =\Scz_{\rm inst},
\qquad
A_1^{\bind(1)} =k\Acz,
\label{relFFc=0}
\ee
where $\bS_1^{\bind(1)}$ and $A_1^{\bind(1)}$ are given in \eqref{SAc=0}, while $\cF_0'$ is obtained from $\cF_0$ by i) expressing it in terms of $\mu$ and $\mucz$,
ii) dropping all terms analytic in $\mucz$ since they are not universal,
iii) taking the limit $\mu\to\infty$ keeping $\mucz$ fixed.
This prescription implies
\be
\cF_0'(\mucz) =\lim_{\mu\to\infty}\cF_0^{\rm reg}(\mu,\mucz),
\qquad
\cF_0^{\rm reg}= \cF_0 +{R\over 2}\, \mu^2 \(\ln\mu -{3\over 2}  - \scX^{\rm cr}\).
\ee
The easiest way to find $\cF'_0$ is to note that, re-expressing $\cF_0^{\rm reg}$ back in terms of $\mu$ and $t_k$
and using \eqref{efkck} and \eqref{crit-cX}, one has
\be \label{efpcond}
\p_\mu^2 \cF_0^{\rm reg}\Big|_{t_k} =  R (\scX - \scX^{\rm cr}) = - 2 \, R\, \delta^{1/2} +\cO(\delta)\, ,
\ee
where $|_{t_k}$ indicates that the derivative is taken at constant $t_k$.
On the other hand, the same result is obtained by taking the second $\mu$-derivative of
\be\label{eC16}
\cF'_0=-\frac{32}{15}\, R\(1-\frac{k}{R}\)^{2}\mucz^{5/2}
\ee
with help of \eqref{edefbfmu} and \eqref{def-critparam}.
It now follows from \eqref{eC16} that
\be
\p_{\mucz}^2\cF'_0
=-8R\(1-\frac{k}{R}\)^{2}\mucz^{1/2}.
\label{derFc=0}
\ee
Then substituting \eqref{relFFc=0}, \eqref{derFc=0} and \eqref{SAc=0} into \eqref{rc=0}, one easily checks that one indeed
recovers \eqref{rrc=0}.

\section{Simplified expression for $\scX$}
\label{ap-proofalt}

In this appendix we prove the relation \eqref{eq-chiallk-simple}, i.e. that the equations
\bea
a_k+\!\!\sum_{l=k+1}^{\kmax}a_l a_{l-k}&=&\eta_k
+\sum_{l=k+1}^{\kmax}\eta_l
\sum_{\sum_{i=1}^n d_i=l-k\atop n\ge 1}
\frac{\Gamma\(\frac{l}{R}+1\)}{n!\,\Gamma\(\frac{l}{R}+1-n\)}\prod_{i=1}^n a_{d_i},
\label{eq-ak-ap}
\\
1+\sum_{l=1}^{\kmax}a_l^2&=& e^{\scX}+ \sum_{l=1}^{\kmax}\eta_l
\sum_{\sum_{i=1}^n d_i=l\atop n\ge 1}\frac{\Gamma\(\frac{l}{R}+1\)}{n!\,\Gamma\(\frac{l}{R}+1-n\)}\prod_{i=1}^n a_{d_i},
\label{eq-chiallk-ap}
\eea
where we introduced the notation $\eta_k=\frac{k}{R}\, \lambda_k \, e^{\(1-\frac{k}{2R}\)\scX}$,
imply
\be
e^{\scX}=1+\sum_{l=1}^{\kmax}\(1-\frac{l}{R}\)a_l^2.
\label{eq-chi-ap}
\ee

To prove this fact, let us multiply \eqref{eq-ak-ap} by $ \frac{k}{mR}\sum_{\sum_{i=1}^m d_i=k}\prod_{i=1}^m a_{d_i}$
and sum over $k$ from $m$ to $\kmax$. The resulting equation can be written as
\be
\begin{split}
&
\frac{1}{mR}\sum_{k=m}^{\kmax}k\, a_k\sum_{\sum_{i=1}^m d_i=k}\prod_{i=1}^m a_{d_i}
+\frac{1}{mR}\sum_{l=m+1}^{\kmax} a_l\sum_{\sum_{i=1}^{m+1} d_i=l}\sum_{i=1}^{m}d_i \prod_{i=1}^{m+1} a_{d_i}
\\
=&\, \frac{1}{mR}\sum_{l=m}^{\kmax} \eta_l\sum_{\sum_{i=1}^{n} d_i=l\atop n\ge m}
\frac{\Gamma\(\frac{l}{R}+1\)\sum_{i=1}^{m}d_i}{(n-m)!\,\Gamma\(\frac{l}{R}+1-n+m\)}\prod_{i=1}^n a_{d_i}\, .
\end{split}
\label{sumeqak}
\ee
Next, we symmetrize the factor $\sum_{i=1}^{m} d_i$ over all permutations of $d_i$.
If $i$ runs over $n$ values, the symmetrization amounts to the replacement
\be
\sum_{i=1}^{m} d_i\ \mapsto\ {{n-1\choose m-1}\over {n\choose m}}\, l =\frac{m l}{n}\, ,
\ee
where ${n\choose m}=\frac{n!}{m!(n-m)!}$ are binomial coefficients.
In the second term in \eqref{sumeqak}, this gives rise to the factor $\frac{l}{(m+1)R}$, while in the term on the right hand side
one obtains
\be
\frac{\Gamma\(\frac{l}{R}+1\) \frac{l}{R}}{n(n-m)!\,\Gamma\(\frac{l}{R}+1-n+m\)}
=\frac{\Gamma\(\frac{l}{R}+1\)}{n(n-m)!\,\Gamma\(\frac{l}{R}-n+m\)}
+\frac{\Gamma\(\frac{l}{R}+1\)}{n(n-m-1)!\,\Gamma\(\frac{l}{R}-n+m+1\)}\, .
\ee
Thus, \eqref{sumeqak} takes the form
\bea
&&
\frac{1}{mR}\sum_{k=m}^{\kmax}k\, a_k\sum_{\sum_{i=1}^m d_i=k}\prod_{i=1}^m a_{d_i}
+\frac{1}{(m+1)R}\sum_{l=m+1}^{\kmax} l\, a_l\sum_{\sum_{i=1}^{m+1} d_i=l} \prod_{i=1}^{m+1} a_{d_i}
\label{sumeqaknew}\\
&=& \sum_{l=m}^{\kmax} \eta_l\sum_{\sum_{i=1}^{n} d_i=l\atop n\ge m} \[
\frac{\Gamma\(\frac{l}{R}+1\)}{n(n-m)!\,\Gamma\(\frac{l}{R}-n+m\)}
+
\frac{\Gamma\(\frac{l}{R}+1\)}{n(n-1-m)!\,\Gamma\(\frac{l}{R}-n+1+m\)}
\]\prod_{i=1}^n a_{d_i}\, .
\non
\eea
Noting that in the second term on the right hand side $n$ is forced to be $\ge m+1$ due to the presence of the $(n-1-m)!$ in the denominator, and hence $l$ is forced to be $\ge m+1$ due to the constraint $\sum_{i=1}^n d_i=l$, \eqref{sumeqaknew} can be written as:
\be\label{esmequal}
\cS_m=-\cS_{m+1},
\ee
where
\be
\cS_m=\frac{1}{mR}\sum_{k=m}^{\kmax}k\, a_k\sum_{\sum_{i=1}^m d_i=k}\prod_{i=1}^m a_{d_i}
-\sum_{l=m}^{\kmax} \eta_l\sum_{\sum_{i=1}^{n} d_i=l\atop n\ge m}
\frac{\Gamma\(\frac{l}{R}+1\)}{n(n-m)!\,\Gamma\(\frac{l}{R}-n+m\)}\prod_{i=1}^n a_{d_i}\, .
\label{def-cSm}
\ee
Since the sums over $k$ and $l$ become empty for $m>\kmax$, one concludes from \eqref{esmequal} that all the $\cS_m$'s actually vanish.
Moreover, for $m=1$ the second term in \eqref{def-cSm} coincides with the last term \eqref{eq-chiallk-ap}.
Therefore, using $\cS_1=0$ in that equation, one immediately gets \eqref{eq-chi-ap}, which proves the desired result.

\section{Normalization of the SL parameter} \label{sappnorm}

Our goal in this appendix will be to derive the relation \eqref{identlam} between the deformation parameter $\repla_k$ in the sine-Liouville string worldsheet theory and the MQM parameter $\lambda_k$.
For this we shall exploit the fact that by taking the derivative of the free energy with respect to the deformation parameters we can generate the correlation functions of the theory.
Therefore, comparison of the correlation functions in different formalisms
can be used to find the relation between the deformation parameters.
To this end, we introduce another set of deformation parameters $\hat\lambda_k$ such that by taking the derivative of the free energy with respect to the $\hat\lambda_k$'s we get the perturbative and the instanton amplitudes computed in the convention of \cite{Balthazar:2017mxh,Balthazar:2019rnh}.
We then proceed in three steps.
\begin{enumerate}
\item First, we compare the sphere amplitudes in MQM computed in section \ref{sec-MQMSL} with the sphere
amplitudes computed in \cite{Balthazar:2017mxh} to find the relation between $\lambda_k$ and $\hat\lambda_k$
\item Next, we compare the disk amplitude in string theory computed in section \ref{sec-instact} with the disk
amplitude in string theory computed in \cite{Balthazar:2019rnh} to find the relation between $\repla_k$
and $\hat\lambda_k$.
\item Finally, we combine the two relations to determine the relation between $\repla_k$ and $\lambda_k$.
\end{enumerate}
Note that during this analysis we never make use of the disk amplitudes computed from MQM, so that
the agreement between \eqref{rdetltat1} and \eqref{SMQM-SL} remains a non-trivial test of the equivalence
between the MQM and string theory descriptions.

We begin with step 1.
Our starting point are the results for the scattering amplitudes
in the non-compact $c=1$ string theory for a state of energy $\omega$ to go into $n$ states
of energies $\omega_1,\cdots, \omega_n$, as given in Eqs. (1.3), (1.4) of \cite{Balthazar:2017mxh}.
We shall use a different normalization of the external states by dividing by the energy
$\omega$ carried by the states, so that the amplitudes have smooth $\omega\to 0$
limit.\footnote{As will become clear later, the overall normalization
of the external states is not important, {\it e.g.} we could have divided the amplitudes by $2\omega$
instead of $\omega$ for each external state.}
In this normalization the amplitudes take the form:
\be\begin{split}
    \hbox{3-point amplitude}
    &= i \, \mu^{-1} \, \delta(\omega-\omega_1-\omega_2) \,,
\\
    \hbox{4-point amplitude}
    &=  i \, \mu^{-2} \, \delta(\omega-\omega_1-\omega_2-\omega_3)
    (1 + i \omega) \,,
\qquad \omega, \omega_i>0\, .
\end{split}
\label{3pt-bry2eucllorentz}
\ee
These formulas  are in Lorentzian signature and for non-compact time direction, and the
parameter $g$ in \cite{Balthazar:2017mxh} is identified as $1/\mu$ in our conventions.
To adapt these formulas for Euclidean time, we should
make the analytic continuation $\omega\to i \omega_E$
with $\omega$ on positive (negative) real axis continued to positive (negative) imaginary axis.
We should also replace $i\delta\(\omega-\sum_i\omega_i\)$ by $\delta\(\omega_E-\sum_i \omega_{E,i}\)$.
Finally, in order to account for the fact that
the Euclidean time circle is periodic with period $2\pi R$, we need to take the $\omega_E$'s to be
quantized in units of $1/R$ and replace the Dirac delta function by $R$ times the Kronecker delta function.
Dropping the subscript $E$ to avoid cluttering, we can write the Euclidean version of the amplitudes as
\be\begin{split}
    \hbox{3-point amplitude}
    &= R\, \mu^{-1} \,  \delta_{\omega, \omega_1 + \omega_2} \,,
\\
    \hbox{4-point amplitude}
    &=  R\, \mu^{-2}
    (1 - \omega) \, \delta_{\omega, \omega_1 + \omega_2 + \omega_3} \,,
\qquad \omega, \omega_i>0\, .
\end{split}
\label{3pt-bry2eucl}
\ee

Next, let us introduce couplings $\hat\lambda_k$ and write an expression for the free energy $\cF_0$
as a function of the $\hat\lambda_k$'s so that upon taking  derivative of $\cF_0$ with respect to the
$-\hat\lambda_k$'s we get back \eqref{3pt-bry2eucl} for $\omega_i=k_i/R$.
It is now easy to check that
the following form of $\cF_0$ reproduces the amplitudes given above:
\be\label{eproposedF}
\cF_0 = -{R\over 3!} \,  \mu^{-1}  \hlm^3 +{R\over 4!} \, \mu^{-2}
\hlm^4 - R\mu^{-1}\hlm \sum_{n>0}  \,\hat\lambda_n \hat\lambda_{-n}
+{R\over 2!} \, \mu^{-2} \hlm^2 \sum_{n>0}
\left(1-{n\over R}\right) \hat\lambda_n \hat\lambda_{-n}+\cdots.
\ee
For example, the three-point function of three zero momentum states is $R\mu^{-1}$ according to \eqref{3pt-bry2eucl}. The third derivative of $\cF_0$ with respect to
$-\hlm$ reproduces this result after we set $\hat\lambda_n$ to zero. Similarly, \eqref{eproposedF} correctly
reproduces the amplitude of four zero momentum states, one zero momentum and two non-zero momentum states, and
two zero momentum and two non-zero momentum states.
After restricting to the case where $\hat\lambda_{-k}=\hat\lambda_k$ are the only non-vanishing couplings,
the expansion \eqref{eproposedF} reduces to
\be \label{eproposedFfin}
R^{-1}\cF_0 = -{1\over 3!} \, \mu^{-1} \hlm^3 + {1\over 4!} \, \mu^{-2}\hlm^4
- \mu^{-1}\hlm  \hat\lambda_k^2
+{1\over 2!}\,
\left(1-{k\over R}\right)  \mu^{-2} \hlm^2\hat\lambda_k^2+\cdots\, .
\ee

We shall now compare \eqref{eproposedFfin} with the Taylor series expansion of the MQM free energy
in powers of small fluctuation $\delta\mu$ around a background value of $\mu$ and
in the SL deformation parameter $t_k$.
First, since $\p_\mu\cF_0=R\phi^{(0)}(-\mu)$,
by expanding \eqref{phi0-SL} to quadratic order in $t_k$, we obtain
\be
R^{-1}\p_\mu \cF_0 = -\mu\ln\mu + \mu - \mu  a_k^2 + \cO(a_k^4) =
-\mu\ln\mu + \mu - {k^2\over R^2} \,\mu^{-1+{k\over R}} t_k^2 + \cO(a_k^4) .
\ee
Using this, we can calculate all the derivatives of $\cF_0$ required to expand it around
some background value of $\mu$ in powers of $t_k$ and fluctuation $\delta\mu$.
Keeping only the cubic and higher order terms in the expansion, we get
\be
\label{ecF0MQM}
\begin{split}
R^{-1}\cF_0 =&\, -{1\over 3!} \,\mu^{-1} (\delta\mu)^3 + {1\over 4!}\,\mu^{-2}  (\delta\mu)^4 -   {k^2\over R^2}\,(\delta\mu)\mu^{-1+{k\over R}} t_k^2
\\
&\,
+ {1\over 2!}  \, {k^2\over R^2}\left(1-{k\over R}\right)
(\delta\mu)^2 \mu^{-2+{k\over R}}  t_k^2 +\cdots\, .
\end{split}
\ee
Comparing \eqref{eproposedFfin} and \eqref{ecF0MQM}, one finds the relations
\be\label{ehatnon}
\hlm = \delta\mu, \qquad \hat\lambda_k = \pm {k\over R} \,\mu^{k\over 2R}\,  t_k  =
\pm {k\over R}\,\mu\,
\lambda_k  \, .
\ee
Note that in this derivation we only needed to compare the cubic terms.
The agreement between the quartic terms is an added test of the correspondence between the sphere amplitudes in string theory and MQM.

We now turn to step 2, i.e.\ the determination of the relation between $\hat\lambda_k$ and the parameter $\repla_k$ used in section \ref{sec-instact}.
We shall do this by an analysis similar to the one we did above,
but using disk amplitudes instead of sphere amplitudes.
To this end, we introduce the notion of $\cF_{\rm disk}$ that plays the role of $\cF_0$ for sphere amplitudes, i.e. it is the generating function of disk amplitudes.
We can also identify this with $-V_{\rm eff}$ introduced in section \ref{sec-instact}, but we shall work with a fixed D-instanton location by setting $x=0$.

First, note that according to \eqref{eadiskV}, and the fact that the deformation parameter $\repla_k$ corresponds to switching on a string field background $\pi \repla_k\, c\bar c \cV_\omega$, $\cF_{\rm disk}$ is given by
\be\label{efff1}
- 2\pi \repla_k\, T\, \sin{\pi k\over R}\, ,
\ee
to linear order in $\repla_k$. The minus sign takes into account that the derivative of
$\cF_{\rm disk}$ with respect to $-\repla_k$ should generate the disk one-point function.
We now compare this with the disk one-point function computed in \cite{Balthazar:2019rnh},
{\it with the same normalization of the vertex operators that was used in computing the sphere amplitudes
\eqref{3pt-bry2eucl}}. It is given by $2\sinh(\pi\omega)$ for real positive
$\omega$ in the Lorentzian theory. We need to divide this by $\omega$ in our normalization of the
external states.
After replacing $\omega$ by $i\omega$ and including the minus sign to account for the fact that the
derivative of $\cF_{\rm disk}$ with respect to $-\hat\lambda_k$
should generate the one-point function,
leads to a contribution to
$\cF_{\rm disk}$ of the form
\be
- 4 \,\hat\lambda_k\, {R\over k}\, \sin{\pi k\over R}\, ,
\label{efff2}
\ee
where the additional factor of 2 comes from the two contributions corresponding to $\omega=\pm k/R$.
Comparing \eqref{efff1} and \eqref{efff2} and using $T=2\pi\mu$,
we get
\be\label{efinappe}
\repla_k = {R\over k}\, {\hat\lambda_k\over \pi^2 \mu} = \pm {\lambda_k\over \pi^2}\, ,
\ee
where in the last step we used \eqref{ehatnon}.
This reproduces \eqref{identlam} up to a sign.

The above analysis does not determine the sign of the relation between $\lambda_k$ and
$\repla_k$ since $\lambda_k$ (equivalently $t_k$) appears quadratically in all the formulae. We shall now
show that if we assume that the relative sign is $k$-independent then the double SL theory can
determine the sign. For this we consider the case of a four-point amplitude with outgoing momenta
$0$, $k/R$, $k/R$ and incoming momentum
$2k/R$. According to \refb{3pt-bry2eucl}, the corresponding amplitude is given by
\be
R\, \mu^{-2} \(1-{2k\over R}\) .
\ee
This corresponds to a term in $\cF_0$
\be
{1\over 2}\, R\, \mu^{-2} \(1-{2k\over R}\) \hlm \, \hat\lambda_{-k}^2\,\hat \lambda_{2k}\,  ,
\ee
where the factor of $1/2$ compensates for the fact that the derivative with respect to $\hat\lambda_{-k}$ produces
a factor of 2.
There is a similar term with the signs of the momenta reversed. Therefore, when we set $\hat\lambda_{-k}=\hat\lambda_k$ and
$\hat\lambda_{-2k}=\hat\lambda_{2k}$, we get the following contribution in $\cF_0$
\be\label{ecca1}
R\, \mu^{-2}  \(1-{2k\over R}\) \hlm \,\hat\lambda_{k}^2 \,\hat \lambda_{2k} \, .
\ee
On the other hand, using \refb{e47phifrel} for $\p_\mu \cF_0=R \phi^{(0)}$, and keeping only the terms proportional to
$\delta\mu \,\lambda_{k}^2  \lambda_{2k}$, we get
\be\label{ecca2}
\cF_0 = R\,\mu \(1 - {2k\over R}\) {2\, k^3\over R^3} \,
\delta\mu \, \lambda_{k}^2 \, \lambda_{2k} \, .
\ee
Using \refb{ehatnon} with the same sign for all $k$ we
see that \refb{ecca1} and \refb{ecca2} agree, provided we choose the relative sign between $\hat\lambda_k$ and
$\lambda_k$ to be positive:\footnote{Since in any term in $\cF_0$ involving product of $\lambda_k$'s the
sum of all the $k$'s must be even, it is always possible to switch the signs of $\lambda_k$'s with odd
$k$'s without affecting any of our results. On the string theory side this sign ambiguity
corresponds to a translation
by $\pi$ along the Euclidean time circle.}
\be
\hat\lambda_k = {k\over R}\,  \mu\lambda_k\, ,
\qquad
\hat\lambda_{2k} = {2k\over R}\, \mu\, \lambda_{2k}\, .
\ee
It then follows from the earlier discussion that \refb{efinappe} has positive sign
and hence reproduces \eqref{identlam} exactly.

\section{Direct computation of the conformal weight of the zero mode}  \label{sa}

In this appendix we shall compute the correction to the conformal weight of the zero mode
due to the presence of the SL perturbation by directly analyzing the world-sheet
diagram. For this we need to compute the disk amplitude of a pair of open string vertex operators
corresponding to the zero modes and a closed string vertex operator corresponding to the
SL deformation.

The string field associated with the SL deformation has the
form $\pi \repla_k c\bar c\, \cV_\omega$ with $\cV_\omega$ normalized as in \eqref{ecorrln}.
On the other hand, the normalized vertex operator of the open string zero mode
$\xi$ has the form $i g_o\sqrt 2 c\p X$ where the factor of $i\sqrt 2$ compensates for the factor of $-1/2$ in the operator product expansion
\be \label{exxope}
\p X(z) \p X(w)=-{1\over 2(z-w)^2}+\cdots\, ,
\ee
and the factor of open string coupling $g_o$ arises due to the fact that we work in a convention in which the string field theory kinetic operator for the open string field does not have any factor of $1/g_o^2$. According to
\eqref{eCOO} and the discussion above \refb{e513}, the two-point function of $\xi$ in the presence of the SL deformation is given by
\be\label{equadraticnew}
2 i \pi  T
\int_0^\infty d u \left\langle (-\pi\repla_k) \, c\bar c \cV_\omega (i)\,
 i\sqrt 2 g_o c \p X(0) \, i\sqrt 2 g_o\p X(u)\right\rangle .
\ee
We have restricted the range of $u$ from 0 to $\infty$ since the two vertex operators are identical and the negative $u$ region can be related to the positive $u$ region by an SL(2,R) transformation together with an exchange of two vertex operators. This has been compensated by the explicit factor of 2 multiplying the expression. Now, if $h$ denotes the conformal weight of the zero mode $\xi$ in the deformed theory, then the effective potential will have a term $h\xi^2/2$ and will produce a $\xi$-$\xi$ two-point function $-h$. Equating this to \eqref{equadraticnew}, we get
\be\label{equadraticnewer}
h=-4 i \pi^2  \, T  \, g_o^2\, \repla_k\,
\int_0^\infty d u \left\langle c\bar c \cV_\omega (i)\,
 c \p X(0) \,  \p X(u)\right\rangle .
\ee
We can evaluate the correlation function using \eqref{ecorrln}, \eqref{eghostcor}, \eqref{etexp} and \eqref{exxope}, and find
\be \label{ehufin}
h=-2\, \repla_k\, \sin(\pi\omega) \cos(\omega x) \, \int_0^\infty du\left[-{1\over u^2} +{2\omega^2\over 1+u^2}\right] .
\ee
This integral diverges from the $u=0$ end, but one can extract finite answer from this using string field theory. Instead of developing the whole formalism from the beginning, we shall use the existing results of \cite{Sen:2020eck} by mapping the world-sheet configuration to a more symmetric form where the closed string vertex operator is inserted at $i$ as usual, but the open string vertex operators are inserted at $\pm\beta$ on the real line. Under the SL(2,R) transformation $z\to (z+\beta)/(1-\beta\, z)$, the vertex operators at $i,\pm\beta$  are mapped to the positions:
\be \label{emap}
i\to i, \qquad -\beta\to 0, \qquad \beta\to u=2\beta / (1-\beta^2)\, .
\ee
Using the relation between $u$ and $\beta$ given above, we can express \eqref{ehufin} as,
\be\label{e08}
h=-4\repla_k\sin(\pi\omega)\,\cos(\omega x)
\int_0^1 d\beta\,  \left[{2\omega^2\over (1+\beta^2)^2} - {1\over 4\beta^2} \right]\,
(\beta^2+1)\, .
\ee

The integral \eqref{e08} diverges for small $\beta$. We regulate the divergence
using string field theory whose effect is to regard this contribution as the one coming from the
sum of two Feynman diagrams shown in Fig.2 of \cite{Sen:2020eck}. Of these Fig.2(a)
represents the region $\beta<\eps$ for some small number $\eps$ and Fig.2(b) represents
the contribution from the region $\eps\le\beta\le 1$. The contribution from Fig.2(b) gives
\be
\begin{split}
&\, -4 \repla_k \cos(\omega x)\,\sin(\pi\omega)\,
\int_\eps^1 d\beta\,  \left[\frac{2\omega^2}{1+\beta^2} - {\beta^2+1\over 4\beta^2} \right]
\\
=&\, -4 \repla_k \cos(\omega x)\, \sin(\pi\omega)
\left[ \frac{\pi\omega^2}{2} - {1\over 4\eps}+O(\eps)\right] .
\end{split}
\ee
The contribution from Fig.2(a), involving the open string tachyon contribution to the internal
propagator, cancels the $1/4\eps$ term, and we are left with
\be\label{e217}
h=-2\pi\repla_k\, \omega^2\, \cos(\omega x)\, \sin(\pi\omega)\, .
\ee
This agrees with the order $\repla_k$ contribution to \eqref{ehfin} at the critical points
$x=0$ and $\pi R/k$.

\section{Disc and annulus amplitudes}
\label{ap-CFTampl}

In this appendix we shall summarize the results for the disk two-point function and the annulus one-point function.

\subsection{Disk two-point function}
\label{ap-2pf-disk}

The function $f_E(\omega_1,\omega_2)$, defined in \eqref{deffE} as a ratio of the disk two-point and one-point functions
of the vertex operators $\cU_\omega$ proportional to $e^{i\omega X}$,
comes out of the analysis of \cite{Balthazar:2019rnh,Sen:2020eck}.
It is obtained by analytic continuation from the corresponding correlation function $f(\omega_1, \omega_2)$
with Lorentzian momenta by rotating the arguments by $\pi/2$ in the anti-clockwise direction in the complex plane:
\be\label{edeffE}
f_E(\omega_1,\omega_2) = f(i\omega_1, i\omega_2)\, .
\ee
The Lorentzian two-point function itself has the following integral representation
\ben \label{effinite}
f(\omega_1, \omega_2)
&=& -{1\over 2} \(1 - 2\omega_1\omega_2 \log\beta^2\)
\non\\
&& \hspace{-0.4in} + \frac{2^{-1/4}\pi^{1/2}
2^{(\omega_1^2+\omega_2^2)/2}}{ \sinh(\pi|\omega_1|)\sinh(\pi|\omega_2|)}
\int\limits_0^1 dy
\Bigg[ y^{\, \omega_2^2/2} (1-y)^{1-\omega_1\omega_2}
(1+y)^{1+\omega_1\omega_2}
\left\langle \cV^L_{|\omega_1|}(i) \cV^L_{|\omega_2|}(iy)\right\rangle
\nonumber \\ &&
\qquad
- 2^{-3/4} \pi^{-1/2} \, 2^{-(\omega_1^2+\omega_2^2)/2} \,
\sinh(\pi|\omega_1|)\sinh(\pi|\omega_2|)\,
\frac{1+2\omega_1\omega_2 y}{y^2}
\Bigg]\, ,
\een
which is supposed to hold for real $\omega_1,\omega_2$.
Some features of this equation are as follows:
\begin{enumerate}
\item
$\cV^L_{|\omega|}$ denotes the Liouville vertex operator $e^{(2-|\omega|)\phi}$.

\item
The term given in the last line of \eqref{effinite} is a subtraction
used in \cite{Balthazar:2019rnh}
to make the $y$
integral finite. A detailed analysis from string field theory shows that in order to
compensate for this subtraction, we need to add $-1/2+\omega_1\omega_2\log\beta^2$
to the resulting expression, as given in the first term on the right hand side of
\eqref{effinite} \cite{Sen:2020eck}.

\item
This additional term involves an arbitrary constant parameter $\beta$ labelling the string field theory.
This shows that $f(\omega_1,\omega_2)$ has an ambiguity in the form of an additive term proportional to $\omega_1\omega_2$.
All physical quantities are supposed to be unaffected by this ambiguity.
\end{enumerate}

Using \eqref{edeffE}, \eqref{effinite}, we can now write
\ben \label{effiniteeuclid}
f_E(\omega_1, \omega_2)
&=& -{1\over 2} \(1 + 2\omega_1\omega_2 \log\beta^2\)
\non\\
&& \hspace{-0.4in} - \frac{2^{-1/4}\pi^{1/2}
2^{-(\omega_1^2+\omega_2^2)/2}}{ \sin(\pi|\omega_1|)\sin(\pi|\omega_2|)}
\int\limits_0^1 dy
\Bigg[ y^{\, -\omega_2^2/2} (1-y)^{1+\omega_1\omega_2}
(1+y)^{1-\omega_1\omega_2}
\left\langle \cV^L_{i|\omega_1|}(i) \cV^L_{i|\omega_2|}(iy)\right\rangle
\nonumber \\ &&
\qquad
+ 2^{-3/4} \pi^{-1/2} \, 2^{(\omega_1^2+\omega_2^2)/2} \,
\sin(\pi|\omega_1|)\sin(\pi|\omega_2|)\,
\frac{1-2\omega_1\omega_2 y}{y^2}
\Bigg]\, ,
\een
where $\left\langle \cV^L_{i|\omega_1|}(i) \cV^L_{i|\omega_2|}(iy)\right\rangle$ is obtained from the expression for $\left\langle \cV^L_{|\omega_1|}(i) \cV^L_{|\omega_2|}(iy)\right\rangle$
by replacing $(|\omega_1|,|\omega_2|)$ with $(i|\omega_1|,i|\omega_2|)$. This function satisfies
\be
f_E(\omega_1,\omega_2)=f_E(\omega_2,\omega_1)=f_E(-\omega_1,-\omega_2)\, .
\ee

The function $f_E(\omega_1,\omega_2)$ has been evaluated numerically in \cite{Balthazar:2019rnh}
for a range of values of $\omega_1,\omega_2$.\footnote{We are grateful to Victor Rodriguez
for sending to us the results of this numerical calculation.}
Using these results, we have checked that the following relations are satisfied
\begin{subequations}
\label{relfE}
\bea
& f_E(\omega,\omega)+f_E(\omega,-\omega)= -\pi\omega^2 \cot(\pi\omega), &
\label{epred}
\\
& f_E(\omega,2\omega)+f_E(\omega,-2\omega)= -2\pi\omega^2 \cot(\pi\omega), &
\label{identf12}
\\
& 4f_E(\omega,\omega)-4f_E(\omega,2\omega)+f_E(2\omega,2\omega)=0. &
\label{identf0}
\eea
Note that the left hand sides of all these equations are independent of the parameter
$\beta$ appearing in \eqref{effinite}.
From this it follows, in particular, that
\be
2f_E(\omega,-\omega)-2f_E(\omega,\omega)-f_E(2\omega,-\omega)+3f_E(2\omega,\omega)-f_E(2\omega,2\omega)=0.
\label{identfff}
\ee
\end{subequations}
Finally, numerical results show that $f_E(\omega_1,\omega_2)$ also satisfies a more general relation:
\be\label{eextrareln}
f_E(\omega_1,\omega_2)+f_E(\omega_1,-\omega_2)= -\pi \omega_1\omega_2 \cot(\pi\omega_2) , \qquad \hbox{for $\omega_1>\omega_2>0$}\, .
\ee

In fact, the imaginary part of $f(\omega_1,\omega_2)$ can also be calculated analytically with the following
result \cite{Sen:2020oqr}:
\be
f_{\rm im}(\omega_1,\omega_2)= \frac{\pi i}{2}\,\omega_1\omega_2\(\coth(\pi\omega_1)+\coth(\pi\omega_2)\)\sign(\omega_1+\omega_2).
\ee
This gives
\be
\begin{split}
&f_{\rm im}(\omega,\omega)+ f_{\rm im}(\omega,-\omega)
= \pi i \omega^2 \, \coth(\pi\omega)
\, ,
\\
&
f_{\rm im}(\omega,2\omega)+ f_{\rm im}(\omega,-2\omega) = 2\pi i\omega^2 \coth(\pi\omega)\, ,
\\
&
4 f_{\rm im}(\omega,\omega)- 4f_{\rm im}(\omega,2\omega)+f_{\rm im}(2\omega,2\omega)
= 0\, ,
\\
& f_{\rm im}(\omega_1,\omega_2)+ f_{\rm im}(\omega_1,-\omega_2)
= \pi i \,\omega_1 \omega_2 \coth(\pi\omega_2)\, ,
\end{split}
\label{eimrel}
\ee
for $\omega>0$ and $\omega_1>\omega_2>0$.
If we assume that the real parts of the combinations of $f(\omega_i,\omega_j)$ appearing on the left hand sides of \eqref{eimrel} vanish, then
these relations reproduce the ones in \eqref{relfE}, \eqref{eextrareln}.
However, we do not have an independent argument for
the vanishing of the real parts of  these combinations.

\subsection{Annulus one-point function at $R=\infty$}
\label{ap-1an-disk}

The function $g_E(\omega)$, defined as the $R\to\infty$ limit of the ratio \eqref{e3.17}
of the annulus and disk one-point functions of the vertex operators $\cU_\omega$, similarly to $f_E(\omega_1, \omega_2)$,
is obtained by analytic continuation from the corresponding function $g(\omega)$
with Lorentzian momenta by rotating the arguments by $\pi/2$ in the anti-clockwise direction
\cite{Balthazar:2019rnh,Sen:2020eck}:
\be\label{eG7}
g_E(\omega)=g(i\omega)\, .
\ee
The function $g(\omega)$ has the following representation
\be
\begin{split}
\label{egfinite}
g(\omega)
=&\, \frac{2\, \pi^2}{\sinh(\pi|\omega|)} \int\limits_0^\infty dt \int\limits_0^{1/4} dx
\Bigg[ \eta(it) \left( {2\pi \over \theta_1'(0|it)}\, \theta_1(2\, x|i t)\right)^{\omega^2/2}
\left\langle \cU_{|\omega|} (2\pi x) \right\rangle_A
\\
&\,
- {1\over \pi}\,  \sinh(\pi|\omega|) \, \left({e^{2\pi t}-1\over \sin^2(2\pi x)} + 2\, \omega^2
\right)\Bigg] + {1\over 2}\, \omega^2 \, \log{\beta^2\over 4}\, ,
\end{split}
\ee
where $\beta$ is the same parameter that appeared in \eqref{effinite}.
As usual, in the physical quantities the $\beta$-ambiguity must cancel.
Using \eqref{eG7} and \eqref{egfinite}, we get
\begin{align}
g_E(\omega)
=& -  \frac{2i \pi^2}{\sin(\pi|\omega|)} \int\limits_0^\infty dt \int\limits_0^{1/4} dx
\Bigg[ \eta(it) \left( {2\pi \over \theta_1'(0|it)}\, \theta_1(2\, x|i t)\right)^{-\omega^2/2}
\left\langle \cU_{i|\omega|} (2\pi x) \right\rangle_A
 \nonumber \\
&\,
- {i\over \pi}\,  \sin(\pi|\omega|) \, \left({e^{2\pi t}-1\over \sin^2(2\pi x)} - 2\, \omega^2
\right)\Bigg] - {1\over 2}\, \omega^2 \, \log{\beta^2\over 4}\, .
\label{egfiniteeuclid}
\end{align}

The functions $f(\omega_1,\omega_2)$ and $g(\omega)$ were shown to satisfy some general relations \cite{Sen:2020eck}.
In particular, equating (1.13) and (1.15) in that paper gives the following relation
\be
\begin{split}
& \sum_{i,j=1\atop i<j}^{n+1} f(\omega_i, \omega_j) + \sum_{i=1}^{n+1} g(\omega_i)
= - i\, \sum_{j=1}^n \omega_j \( 1 - \sum_{i=1}^n \pi\omega_i \coth(\pi\omega_i)\), \quad
\\
& \qquad\qquad
\hbox{for $\omega_1,\cdots,\omega_n>0$, \quad
$\omega_{n+1}=-\omega_1-\cdots -\omega_n$}\, ,
\end{split}
\ee
where we have set the constant $C$ appearing in \cite[Eq.(1.13)]{Sen:2020eck} to zero since this is required by
matching the string theory and the matrix model results. Specializing to $n=1$, one finds
\be
f(\omega, -\omega) + 2\, g(\omega)  = -i\, \omega\, (1-\pi\omega\coth(\pi\omega))\, ,
\ee
or its Euclidean version
\be\label{esub1}
f_E(\omega, -\omega) + 2\, g_E(\omega)  = \omega\, (1-\pi\omega\cot(\pi\omega))\, .
\ee
Combining this with \eqref{epred}, one also obtains
\be
2\,g_E(\omega) - f_E(\omega,\omega)= \omega.
\label{2gf}
\ee
We have checked that numerical evaluation of the functions $f_E(\omega_1,\omega_2)$ and $g_E(\omega)$
does confirm this relation.

\subsection{Annulus one-point function for compact time}
\label{ap-anR}

Finally, we calculate the annulus one-point function when the Euclidean time direction is compactified on a circle of radius $R$.

Let us describe the annulus using coordinates $\sigma_1,\sigma_2$ with range
\be
0\le\sigma_1 \le \pi, \qquad 0\le \sigma_2 <2\pi t\, .
\ee
The boundary condition in the Euclidean time coordinate $X$ is given by
\be\label{ebcX}
X(0,\sigma_2)=0, \quad X(\pi,\sigma_2) = 2\pi n R  \quad \hbox{for
$n\in\ZZZ$}, \qquad X(\sigma_1, \sigma_2+2\pi t) = X(\sigma_1, \sigma_2)\, .
\ee
Note that $X$ is not allowed to change by a multiple of $2\pi R$ as $\sigma_2$ changes
by $2\pi$ since the values of $X$ at $\sigma_1=0,\pi$ are fixed.

Our goal will be to compute the one-point function of $e^{i \omega X(\sigma_1=2\pi x,\sigma_2)}$ for
$\omega=k/R$, $k\in {\mathbb Z}$.
We shall evaluate it using the Euclidean path integral approach. For this we decompose
a general fluctuating field $X$ subject to the boundary condition \eqref{ebcX} as
\be
X=\sum_{p,q\in \ZZZ\atop p>0} a_{p,q} \sin (p\sigma_1) e^{ i q\sigma_2/t} +
2n \sigma_1 R\, ,
\qquad n\in\ZZZ,
\quad a_{p,q}^* = a_{p,-q}\, .
\ee
The Euclidean world-sheet action for this background is given by
\be
S_{ws}={1\over 4\pi} \int d\sigma_1 d\sigma_2 \,\Bigl[\(\p_{\sigma_1}X\)^2 + \(\p_{\sigma_2}X\)^2\Bigr]
=\sum_{p,q\in \ZZZ\atop p>0} {\pi\over 4} \(  p^2 t + {q^2\over t} \)|a_{p,q}|^2 +2\pi n^2 R^2 t\, .
\ee
Therefore, the world-sheet path integral over the field $X$ involves integration over $a_{p,q}$
and summation over the integer $n$, with weight factor
\be
\begin{split}
e^{-S_{ws}} e^{i\omega X(2\pi x,\sigma_2)}
=&\, \exp\(-\sum_{p,q\in \ZZZ\atop p>0} {\pi\over 4} \(  p^2 t + {q^2\over t} \)|a_{p,q}|^2
+ i \omega \sum_{p,q\in \ZZZ\atop p>0} a_{p,q} \sin (2\pi x p) e^{ i q\sigma_2/t}\)
\\
& \times\,
 \exp\Bigl( -2\pi n^2 R^2 t +4\pi i n \, \omega\,x\, R\Bigr) \, .
 \end{split}
\ee

Since the term involving $a_{p,q}$ is independent of $R$, the integration
over these variables produces an $R$ independent term that may be identified as part of the integrand
appearing in the first line of \eqref{egfiniteeuclid}.
The effect of compactification is then
to multiply it by the factor
\be
\sum_{n\in\ZZZ} \exp\Bigl( -2\pi n^2 R^2 t +4\pi i n\, \omega \, x\, R\Bigr)
= \sum_{n\in\ZZZ} \exp\bigl( -2\pi n^2 R^2 t \bigr)
\cos\bigl(4\pi n\,  \omega \, x\, R\bigr) \, .
\ee
This gives the unregularized version of $g_R(\omega)$ to be:
\be
\label{egfiniteapp}
-\frac{2i \pi^2}{\sinh(\pi|\omega|)} \int\limits_0^\infty dt \int\limits_0^{1/4} dx\, \mU_\omega(t,x)
\sum_{n\in\ZZZ} \exp\bigl( -2\pi n^2 R^2 t \bigr)
\cos\bigl(4\pi n\,  \omega \, x\, R\bigr),
\ee
where\
\be\label{eanncor}
\mU_\omega(t,x)= \eta(it) \left( {2\pi \over \theta_1'(0|it)}\, \theta_1(2\, x|i\, t)\right)^{-\omega^2/2}
\langle \cU_{i|\omega|} (2\pi x) \rangle_A \, .
\ee
The integral in \eqref{egfiniteapp} diverges from the regions near $x=0$ and $t=\infty$. We have to follow the procedure
described in \cite{Sen:2020eck,Eniceicu:2022xvk}
to subtract the divergent part and replace it by a finite part.
We choose the subtraction term to be
\be \label{eG20}
2\pi  \int\limits_0^\infty dt \int\limits_0^{1/4} dx  \left({e^{2\pi t}-1\over \sin^2(2\pi x)}
\sum_{n\in\ZZZ} \exp\bigl( -2\pi n^2 R^2 t \bigr) - 2 \omega^2
\right) .
\ee
We have checked in subsection \ref{sG4} that for this choice of the subtraction term, the
finite part that replaces this divergent integral has the same form as in the
non-compact theory, leading to
\ben\label{eG21}
g_R(\omega)
&=& -\frac{2i \pi^2}{\sinh(\pi|\omega|)}  \int\limits_0^\infty dt \int\limits_0^{1/4} dx\Bigg[ \mU_\omega(t,x)
\sum_{n\in\ZZZ} \exp\bigl( -2\pi n^2 R^2 t \bigr)
\cos\bigl(4\pi n\,  \omega \, x\, R\bigr)
\\ &&
- {i\over \pi}\,  \sinh(\pi|\omega|)  \left({e^{2\pi t}-1\over \sin^2(2\pi x)}
\sum_{n\in\ZZZ} \exp\bigl( -2\pi n^2 R^2 t \bigr) - 2 \omega^2
\right)\Bigg] - {1\over 2}\, \omega^2 \, \log{\beta^2\over 4}\, .
\non
\een
Comparing this with \eqref{egfiniteeuclid}, we get
\be \label{eDeltagfin}
\begin{split}
\Delta g_E(\omega)
=&\, -\frac{2i \pi^2}{\sinh(\pi|\omega|)} \int\limits_0^\infty dt \int\limits_0^{1/4} dx\Bigg[ \mU_\omega(t,x)
\sum_{n\in\ZZZ\atop n\ne 0}  \exp\bigl( -2\pi n^2 R^2 t \bigr)
\cos\bigl(4\pi n\,  \omega \, x\, R\bigr)
\\
&\,
- {i\over \pi}\,  \sinh(\pi|\omega|) \, {e^{2\pi t}-1\over \sin^2(2\pi x)}
\sum_{n\in\ZZZ\atop n\ne 0 } \exp\bigl( -2\pi n^2 R^2 t \bigr)
\Bigg] .
\end{split}
\ee

The matrix model analysis, described
{\it e.g.} in \refb{Delga}, requires that \refb{eDeltagfin} should
evaluate to
\be
\Delta g_E(\omega)= \hf\,\omega\(\frac{\pi}{R}\, \cot\frac{\pi}{R}-1\) .
\label{Delg}
\ee
The expansion of the right hand side for large $R$
takes the form
\be\label{edelgeexp}
\Delta g_E(\omega) = -{\pi^2\over 6} \, {\omega\over R^2} + \cO(R^{-4})\, .
\ee
In section \ref{sdergexp} we shall give an analytic proof of \refb{edelgeexp} starting from \refb{eDeltagfin} using the fact that for large $R$
the contribution to the integral is dominated by the
small $t$, small $x$ region. We have also tested the
full formula \refb{Delg} numerically by modifying the
code developed in \cite{Balthazar:2019rnh}. However,
unlike in the case of \cite{Balthazar:2019rnh},
a large part of the contribution to the integral comes
from small $t$, small $x$ region even for moderate
values of $R$, and in this region the numerical analysis suffers from significant error due to the fact that terms in both the first and the second line
of \refb{eDeltagfin} become large and only the sum
evaluates to a finite integral. For this reason the
error in the integral is large and we have only been
able to verify \refb{Delg} with about $10\%$ accuracy.

\subsection{Subtraction term for compact Euclidean time} \label{sG4}

\def\fignine{

\def\JPicScale{0.8}
\ifx\JPicScale\undefined\def\JPicScale{1}\fi
\unitlength \JPicScale mm
\begin{picture}(125,90)(0,0)
\linethickness{0.3mm}
\put(30,30){\line(1,0){90}}
\linethickness{0.3mm}
\put(30,30){\line(0,1){60}}
\linethickness{0.3mm}
\put(30,70){\line(1,0){90}}
\linethickness{0.3mm}
\put(50,30){\line(0,1){10}}
\linethickness{0.3mm}
\put(30,40){\line(1,0){90}}

\linethickness{0.3mm}
\put(120,30){\line(0,1){60}}

\linethickness{0.3mm}
\qbezier(35,70)(37.58,59.59)(41.19,52.38)
\qbezier(41.19,52.38)(44.8,45.16)(50,40)
\put(35,35){\makebox(0,0)[cc]{(a)}}

\put(70,35){\makebox(0,0)[cc]{(b)}}

\put(35,50){\makebox(0,0)[cc]{(c)}}

\put(70,55){\makebox(0,0)[cc]{(d)}}



\put(50,24){\makebox(0,0)[cc]{$v\simeq\alpha^{-2}$}}

\put(120,24){\makebox(0,0)[cc]{$v=1$}}

\put(135,41){\makebox(0,0)[cc]{$x\simeq (2\pi\gamma)^{-1}$}}

\put(130,70){\makebox(0,0)[cc]{$x= {1\over 4}$}}

\end{picture}

}

\begin{figure}[t]
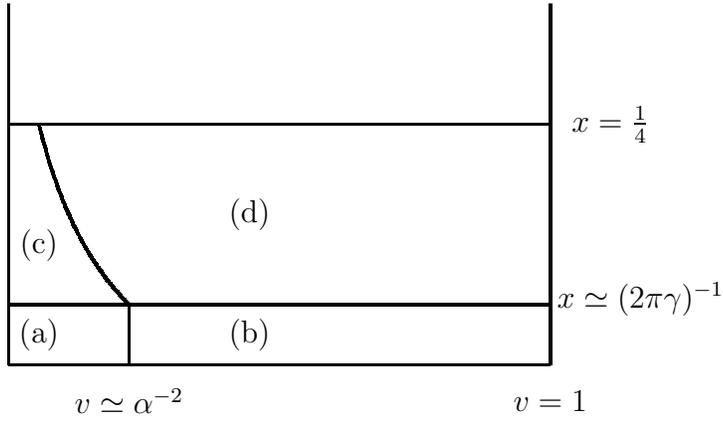

\begin{center}

\hbox{\fignine}

\vskip -.6in

\caption{This figure shows the different regions of the moduli space
covered by different Feynman diagrams contributing to the annulus one-point function.
\label{fignine}
}
\end{center}
\end{figure}

While evaluating the annulus one-point function, we have subtracted from the integrand the term \eqref{eG20}. The $n=0$ term in the sum was present even in the case when the time direction was non-compact, and the effect of this subtraction is compensated by the last term on the right hand side of \eqref{eG21}. Thus, we are left with the term:
\be \label{eG20mod}
  \int\limits_0^1 \frac{dv}{v}\, \int\limits_0^{1/4} dx  \left({v^{-1}-1\over \sin^2(2\pi x)}
\sum_{n\in\ZZZ\atop n\ne 0} v^{n^2R^2}
\right) ,
\qquad v\equiv e^{-2\pi t}\, .
\ee
The general rule for dealing with this term is that we need to first express it as a sum of string field theory Feynman diagrams, extract its finite expression by properly treating the Feynman diagrams and then add this contribution back. The four Feynman diagrams are displayed in Fig.~7 of \cite{Sen:2020eck} and the regions in the $x$-$v$ plane covered by these diagrams have been shown in Fig.~8 of \cite{Sen:2020eck}. We have reproduced the last figure in Fig.~\ref{fignine}.
Here $\alpha$ and $\gamma$ are two parameters of the string field theory. The final results are supposed to be independent of these parameters. The parameter $\beta$ that appeared in the expressions for the functions $f_E(\omega_1,\omega_2)$ and $g_E(\omega)$ is related to $\alpha$ and $\gamma$ as $\gamma=\alpha\beta$. We shall take $\alpha$ and $\gamma$ to be large as in \cite{Sen:2020eck}, and ignore terms containing inverse powers of $\alpha$ and $\gamma$.
To the desired accuracy, the regions (a)-(d) can be described as follows:
\be
\begin{split}
(a)\ :\ &\,
0\le v\le \left(\alpha^{2}-{1\over 2}\right)^{-1},
\qquad
0\le 2\pi x \le \gamma^{-1}\, {2-v\over 2+v}\, ,
\\
(b) \ :\ &\, \(\alpha^{2}-{1\over 2}\)^{-1}<v<1, \qquad
0<2\pi x<  \gamma^{-1}\(1 - \alpha^{-2}\),
\\
(c)\ :\ &\,
{1\over 2\pi\gamma} \,  {2-v\over 2+v}
\le x\le {1\over 4}\, ,
\qquad\quad
0\le v \le L(x),
\\
(d) \ :\ &\,
{\pi\over 2}\ge 2\pi x\ge  \gamma^{-1} (1-\alpha^{-2}),
\qquad
L(x)\le v<1\, ,
\end{split}
\ee
where
\be
L(x)={\left(1 + {1\over 4\gamma^2}\right)^{-2}\over \alpha^2 \gamma^2\sin^2(2\pi x)}
 \Bigg[1- 2 \left\{ \cot^2(2\pi x)- \gamma^2 f^2\right\} \alpha^{-2}\gamma^{-2}
\left(1 + {1\over 4\gamma^2}\right)^{-2}\Bigg]
\ee
and $f\equiv f(\tan(\pi x))$ is another arbitrary function appearing in the definition of string field theory.

We now note that for large $R$, the integrand in \eqref{eG20mod} decays for small $v$. On the other hand, the regions (a) and (c) have $v<\alpha^{-2}$ which is small for large $\alpha$. Therefore, the contributions from regions (a) and (c) will involve negative powers of $\alpha$ which we are ignoring in our analysis. Thus, we only need to concentrate on the contributions from regions (b) and (d).

To evaluate the contribution from region (d), we express the integrand in \eqref{eG20mod} as
\be
d \Omega, \qquad
\Omega\equiv {dv\over 2\pi v} \left({v^{-1}-1}
\right)\, \cot(2\pi x) \sum_{n\in\ZZZ\atop n\ne 0} v^{n^2R^2}
\ee
so that the integral over the region (d) will be given by the integral of $\Omega$ along the boundary of (d) in the anti-clockwise direction. The boundary consists of four parts. Let us consider them one by one.
First, the contribution from the boundary at $v=1$ vanishes since $dv=0$ along this boundary.
Second, the contribution from the boundary at $x=1/4$ is also zero because $\cot(2\pi x)$ vanishes for this value of $x$. Next, along the boundary between the regions (d) and $(c)$, $v$ is bounded above by $\alpha^{-2}$ and, as a result, the integrand is suppressed by positive powers of $\alpha^{-1}$. Thus, we can ignore this contribution as well. This leaves us with the boundary between the regions (b) and (d). To evaluate it, we set
\be
2\pi x = \gamma^{-1} \,
\Bigl(1 - \alpha^{-2}\Bigr)\, ,
\ee
and integrate $\Omega$ over $v$. This gives the contribution from region (d):
\be\label{eregiond}
{\gamma\over 2\pi} \,  (1-\alpha^{-2})^{-1}\,
\int_{\left(\alpha^2 -{1\over 2}\right)^{-1}}^1 \frac{dv}{v}\left({v^{-1}-1}
\right) \sum_{n\in\ZZZ\atop n\ne 0} v^{n^2R^2}\, ,
\ee
where we have ignored terms containing negative powers of $\gamma$.

Finally, let us consider the contribution from region (b). In this case we change variables to $(q,v)$ appropriate to region (b) \cite{Sen:2020eck}:
\be\label{e2.81pre}
2\pi x = \gamma^{-1} q\,
\Bigl(1 - \alpha^{-2}+\cO(\alpha^{-4})+\cO(\gamma^{-2}q^2)\Bigr) \, ,
\ee
and express the integral \eqref{eG20mod} as
\be
\label{eG20modnew}
\int_{\left(\alpha^2-{1\over 2}\right)^{-1}}^1 \frac{dv}{v} \left(v^{-1}-1\right)\sum_{n\in\ZZZ\atop n\ne 0} v^{n^2R^2}
\int_0^{1}{dq\over 2\pi}\,
\frac{\gamma^{-1} \bigl(1 - \alpha^{-2}+\cO(\alpha^{-4})\bigr)}{\sin^2\bigl[\gamma^{-1} q\(1 - \alpha^{-2}+\cO(\alpha^{-4})+\cO(\gamma^{-2} q^2)\)\bigr]} \, .
\ee
The $q$ integral has divergence from the $q=0$ end. The divergent term is proportional to $dq/q^2$, but there are no divergences of the form $dq/q$. The string field theory rule for dealing with these divergences is to replace $\int_0^1 dq q^{-1-\alpha}$ by $-1/\alpha$ \cite{Sen:2020eck}. Operationally this is equivalent to using a lower cut-off $\eps$ on the $q$ integral and then removing all negative powers of $\eps$ from the result of integration before taking the $\eps\to 0$ limit. Using this step, we can replace \eqref{eG20modnew} by
\be\label{eregionc}
-{\gamma\over 2\pi} \, (1-\alpha^{-2})^{-1} \, \int_{\left(\alpha^2 -{1\over 2}\right)^{-1}}^1 \frac{dv}{v}\left({v^{-1}-1}
\right)\, \sum_{n\in\ZZZ\atop n\ne 0} v^{n^2R^2}\, .
\ee
This exactly cancels \eqref{eregiond}, showing that no extra term need to be added to the expression for the annulus one-point function to compensate for the subtraction term given in \eqref{eG20mod}.\footnote{This cancellation can be traced to the fact that for the particular class of string field theories used in the analysis, the relation between $x$ and $q$ given in \eqref{e2.81pre} does not have a term of order $q^2$. In a more general version of string field theory the terms will not cancel, but the difference will be cancelled by another term that involves exchange of the out of Siegel gauge field $c_0|0\rangle$ \cite{Sen:2020eck}.}

\def\figone{

\def\JPicScale{1.5}
\ifx\JPicScale\undefined\def\JPicScale{1}\fi
\unitlength \JPicScale mm
\begin{picture}(110,70)(-5,-5)
\linethickness{1mm}
\put(20,50){\line(1,0){20}}
\linethickness{1mm}
\put(40,50){\line(1,0){20}}

\put(40,50){\makebox(0,0)[cc]{\Large $\times$}}

\put(40,47){\makebox(0,0)[cc]{$A$}}

\put(60,50){\makebox(0,0)[cc]{\Large $\times$}}

\put(60,47){\makebox(0,0)[cc]{$B$}}

\put(50,47){\makebox(0,0)[cc]{$P$}}

\put(30,47){\makebox(0,0)[cc]{$\omega$}}

\end{picture}

}

\begin{figure}[t]
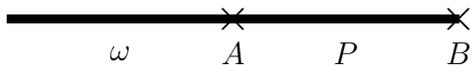

\begin{center}

\hbox{\figone}

\vskip -2.7in

\caption{This is a diagrammatic representation of potentially divergent contribution to the annulus one-point function from the closed string channel. The line labelled by $\omega$ denotes an external on-shell closed string with energy $\omega$, the line labelled $P$ is an internal off-shell closed string with zero energy and Liouville momentum $P$, the vertex $A$ represents a disk two-point function of two closed strings and the vertex $B$ denotes the disk one-point function of a single closed string.
\label{figone}
}
\end{center}
\end{figure}

One can also examine possible divergences from the closed string channel, corresponding to the $v\to 1$ limit. For non-compact Euclidean time and Euclidean external momenta this limit does not produce any divergence \cite{Balthazar:2019rnh}, but we shall now reexamine this for compact Euclidean time coordinate for which the energies are quantized. A diagrammatic representation of the potentially dangerous contribution is shown in Fig.~\ref{figone}.
The contribution from this diagram is an integral over the Liouville momentum $P$ which is taken to be close to 0, with the integrand given approximately by the product of the disk two-point function $f_E(\omega,0)(2T\sin(\pi\omega))(2T\sin(\pi P))\simeq -2\, \omega\, T^2 \sin(\pi\omega) \sin(\pi P)$ from \eqref{eextrareln}, representing contribution from the vertex $A$, a disk one-point function of Liouville momentum $P$ given by $2 T\sin(\pi P)$ from \eqref{eadiskV}, representing contribution from the vertex $B$, and the closed string propagator proportional to $1/P^2$. Thus, the amplitude is proportional to
\be
4\, \omega\, T^3 \sin(\pi\omega)\int  {dP\over P^2} \,\sin^2(\pi P)\, .
\ee
This has no divergence from the $P=0$ end.

As discussed in \cite{Sen:2020eck}, $g_R(\omega)$ receives additional contribution from the exchange of out of Siegel gauge mode $c_0|0\rangle$ and from the Jacobian factors for the change of variables from the string field theory zero mode to the translation zero mode and the string field theory gauge transformation parameter to the conventional gauge transformation parameter. However, all these contributions can be determined in terms of disk amplitudes which do not depend on $R$. Therefore, the computation of $\Delta g_E$ is not affected by these contributions.

\subsection{Analytic study of large $R$ expansion of $\Delta g_E$} \label{sdergexp}

In this section we shall give an analytic derivation of \refb{edelgeexp} starting from
\refb{eDeltagfin}. For this we define the function
\be \label{edefFuv}
F(u,v) = \sum_{n\ne 0} e^{-2\pi n^2 u} \cos(2\pi n v)\, .
\ee
Let us introduce a cut-off $\eps$ on the lower limit of the
$x$ integral in \refb{eDeltagfin}, analyze the unregulated and subtraction terms separately, and
at the end combine the two terms before taking the $\eps\to 0$ limit. The subtraction
term given in the second line of \refb{eDeltagfin} takes the form:
\be\label{eintermed1}
-2\pi \int\limits_0^\infty dt \int\limits_\eps^{1/4} dx\, \, \frac{e^{2\pi t}-1}{\sin^2(2\pi x)}\, F(R^2 t, 0) \, .
\ee
It follows from \refb{edefFuv} that $F(u,v)$ is exponentially suppressed
for $u\gg 1$. Therefore, in \refb{eintermed1} the main contribution to the integral comes from
the region $t\sim R^{-2}\ll 1$ for large $R$, and we can approximate the integral by
\be \label{esubtractfin}
-2\pi \int\limits_0^\infty dt \int\limits_\eps^{1/4} dx \, \frac{2\pi t}{\sin^2(2\pi x)}\,  F(R^2 t, 0)
=- \eps^{-1} \int\limits_0^\infty dt \, t\, F(R^2 t, 0)\, .
\ee

We now turn to the analysis of the unregulated part of $\Delta g_E(\omega)$ given by the
first line of \refb{eDeltagfin}. This can be expressed as
\be -\frac{2i \pi^2}{\sinh(\pi|\omega|)} \int\limits_0^\infty dt \int\limits_\eps^{1/4} dx\,
\mU_\omega(t,x) \, F(R^2 t, 2\, \omega\, R\, x)
 \, .
\ee
Using the exponential fall off of $F(u,v)$ for $u\gg 1$, we see that the integration over $t$
gets significant contribution only from the region $t\sim R^{-2}\ll 1$. Therefore,
we can replace $\mU_\omega$ by its small $t$ approximation, which can be found
in appendix A of \cite{Balthazar:2019rnh} (see eq.(A.3) with $s$ replaced by $1/t$). As a result, we get, up to an overall normalization,
\ben
&&
{1\over \sinh(\pi\omega)} \, \int\limits_0^\infty dt \int\limits_\eps^{1/4} dx \, F(R^2 t, 2\,\omega\, R\, x)\,
 t^{-5/2} \,
{e^{-2\pi x^2\omega^2/t}\over \sinh^2(2\pi x/t)} \\
&& \quad
\times \int\limits_0^\infty dP_1 \, \sinh(2\pi P_1)\,
e^{-2\pi P_1^2/t} (2\tanh(\pi x/t))^{2 + 2 P_1^2} \langle
\cV^L_{|\omega|}(i) \cV^L_{2|P_1|}(i\tanh(\pi x/t))\rangle_{UHP}\, . \non
\een
We could determine the overall normalization by careful comparison between our normalization
conventions and those in \cite{Balthazar:2019rnh}, but instead we shall determine it by an indirect
method later.
Since the integration over $t$ is restricted to a region of order $R^{-2}$, the presence of
$\sinh^2(2\pi x/t)$ in the denominator
restricts the $x$ integral within a range of order $R^{-2}$ and the
presence of the factor $e^{-2\pi P_1^2/t}$ restricts the $P_1$ integral
within a range of order $R^{-1}$. This allows us to simplify the integrand in the large $R$ limit.
We define
\be
y=\tanh(\pi x/t)\, ,
\ee
and express the integral, up to an overall normalization, as
\be
\begin{split}
&
{1\over \sinh(\pi\omega)} \,\int\limits_0^\infty dt  F(R^2 t, 0)\,
t^{-3/2}  \int\limits_0^\infty dP_1 \, \sinh(2\pi P_1)\, e^{-2\pi P_1^2/t}
\\
& \qquad\quad\times \int\limits_{\eps\pi/t}^{1} dy \, (1-y^2) \,
\langle
\cV^L_{|\omega|}(i) \cV^L_{2|P_1|}(iy)\rangle_{UHP}\, .
\end{split}
\ee
Comparing this result with \refb{effiniteeuclid} for small $P_1$, we now see that it can be
expressed in terms of the unregulated part of $f_E(\omega, 2P_1)$ as
\be\label{einterm2}
{C_0\over 2}  \int\limits_0^\infty dt \, t^{-3/2} \, F(R^2 t, 0)
\int\limits_0^\infty dP_1 \, \sinh^2(2\pi P_1)\,
e^{-2\pi P_1^2/t} \Bigl(f_E^{\rm unreg}(\omega, 2|P_1|)+ f_E^{\rm unreg}(\omega, -2|P_1|)\Bigr) ,
\ee
where $C_0$ is an (as yet) unknown normalization constant and $f_E^{\rm unreg}$ denotes
the part of $f_E$ given in the second line of \refb{effiniteeuclid} with
the
lower cut-off on the $y$ integral placed at $\pi\eps/t$. It follows from \refb{effiniteeuclid}
that in order to replace $f_E^{\rm unreg}$ by
$f_E$ in \refb{einterm2},
we need to add to this expression a term of the form
\be
\begin{split}
&
-{C_0\over 2} \, \int\limits_0^\infty dt \, t^{-3/2} \, F(R^2 t, 0)
\int\limits_0^\infty dP_1 \, \sinh^2(2\pi P_1)\,
e^{-2\pi P_1^2/t} {t\over \pi\eps}
\\
\simeq &\, -{C_0\, \eps^{-1}\over 4\sqrt 2}\,
 \int\limits_0^\infty dt\, t\, F(R^2 t,0) \, .
\end{split}
\ee
On the other hand, we know that the actual term that renders the original integral finite is given
by \refb{esubtractfin}. Comparing the two expressions, we get
\be
C_0 = 4\sqrt 2\, .
\ee
Substituting this into \refb{einterm2},
we obtain the full expression for the large $R$ approximation of
$\Delta g(\omega)$
\be
\Delta g_E(\omega) \simeq 2\sqrt 2
\int\limits_0^\infty dt \, t^{-3/2} \, F(R^2 t, 0)
\int\limits_0^\infty dP_1 \, \sinh^2(2\pi P_1)\,
e^{-2\pi P_1^2/t} \Bigl( f_E(\omega, 2|P_1|) + f_E(\omega, -2|P_1|)\Bigr) .
\ee
We now use the observation that $P_1$ is constrained to be small, of order $R^{-1}$, and the result
\refb{eextrareln} to write the above expression as
\be
\begin{split}
\Delta g_E(\omega) \simeq &\,
-2\sqrt 2\, \omega \, \int\limits_0^\infty dt \, t^{-3/2} \, F(R^2 t, 0)
\int\limits_0^\infty dP_1 \, \sinh^2(2\pi P_1)\,
e^{-2\pi P_1^2/t}
\\
= &\,
- \pi \omega  \int\limits_0^\infty dt \, F(R^2 t, 0)\, .
\end{split}
\ee
Using \refb{edefFuv}, this integral can be expressed as
\be
\Delta g_E(\omega)=
-\pi \omega \int\limits_0^\infty dt \,\sum_{n\ne 0} e^{-2\pi n^2 R^2 t}
=-{\omega\over 2R^2} \,  \sum_{n\ne 0} n^{-2}
=- {\pi^2\over 6}\, \frac{\omega}{R^2} \, .
\ee
This reproduces \refb{edelgeexp}.

\providecommand{\href}[2]{#2}\begingroup\raggedright\endgroup


\end{document}